%% file: thesis.tex
\documentclass[11pt,oneside]{book}

\usepackage{latexsym}
\usepackage{comment}
\usepackage{amssymb}
\usepackage{amsmath}
\usepackage{setspace}

\headsep \headheight
\newcounter{subeqn}
\renewcommand{\theequation}{\arabic{chapter}.\arabic{equation}\alph{subeqn}}

\DeclareMathOperator{\sech}{sech}

\begin{document}

\frontmatter

\input{defines}
\parskip6pt

\input{front}


\singlespacing
\tableofcontents
\clearpage

\input{ack} 
%
%
\clearpage
\thispagestyle{empty}
\begin{center}
\vspace*{3.5in} 
To my parents.
\end{center}
\clearpage

\addcontentsline{toc}{chapter}{List of Figures}
\listoffigures
\clearpage
\addcontentsline{toc}{chapter}{List of Tables}
\listoftables
\clearpage

\mainmatter


\include{Intro}

\include{Dynamics}

\include{Coeffs}

\include{Tests}

\include{Results}

\input{Discussion}

\input{Appendix}


\bibliography{thesis}
\addcontentsline{toc}{chapter}{Bibliography}
\bibliographystyle{thesis}

\end{document}

%% file: defines.tex
\parindent0pt

\excludecomment{withecc}
\includecomment{skipecc}

\newcommand{\FP}{Fokker-Planck\ }
\newcommand{\rcore}{r_\mathrm{core}}
\newcommand{\markup}[1]{[{\slshape\sffamily #1}]}
\newcommand{\ellip}{e}
\newcommand{\beq}{\begin{equation}}
\newcommand{\eeq}{\end{equation}}
\newcommand{\bea}{\begin{eqnarray}}
\newcommand{\eea}{\end{eqnarray}}
\newcommand{\bes}[1]{\bea\label{#1}\stepcounter{equation}}
\newcommand{\ees}{\eea\setcounter{subeqn}{0}}
\newcommand{\substep}{\addtocounter{equation}{-1}\stepcounter{subeqn}}
\newcommand{\sublabel}[1]{\substep\label{#1}}
\newcommand{\fcn}[1]{\textsl{#1}}
\newcommand{\sn}[1]{\times10^{#1}}
\newcommand{\rplum}{r_5}
\newcommand{\barmom}{\mathcal{I}_\mathrm{eff}}
\newcommand{\vecI}{{\mbox{\boldmath$I$}}}
\newcommand{\elp}[1]{{\ell{#1}_1\ell{#1}_2\ell{#1}_3}}
\newcommand{\elll}{\elp{}}
\newcommand{\barb}{_{\scriptscriptstyle{\mathrm{B}}}}
\newcommand{\barf}{\mathcal{F}\barb}
\newcommand{\stream}{\mu\barb}
\newcommand{\Cite}[1]{\cite{#1}}
\newcommand{\psilp}[1]{\Psi_{\elp{#1}}}
\newcommand{\Psilp}[1]{\overline{\Psi}_{\elp{#1}}}
\newcommand{\Psilq}[1]{\overline{\Psi^2}_{\elp{#1}}}
\newcommand{\Psilll}{\Psilp{}}
\newcommand{\Psilsq}{\Psilq{}}
\newcommand{\psilll}{\Psilll({I,J})}
\newcommand{\suminf}{\sum^{\infty}}
\newcommand{\reswtp}[2]{(\ell#1_#2w_#2-\omega#1t)}
\newcommand{\reswop}[2]{(\ell#1_#2w_#2-\ell#1_3\Omega_*t)}
\newcommand{\reswt}{\reswtp{}{p}}
\newcommand{\reswp}[2]{(\ell_#2#1\Omega_#2-\omega#1)}
\newcommand{\resw}{\reswp{}{p}}
\newcommand{\resp}[2]{(\ell#1_#2\Omega_#2-\ell#1_3\Omega_*)}
\newcommand{\res}{\resp{}{}}
\newcommand{\opi}{ \frac{1}{\pi} }
\newcommand{\halve}[1]{ \frac{#1}{2} }
\newcommand{\halfpi}{ \frac{\pi}{2} }
\newcommand{\half}{ \halve{1} }
\newcommand{\otwopi}{ \frac{1}{2\pi} }
\newcommand{\oforpi}{ \frac{1}{2\pi} }
\newcommand{\orbint}{\int_{r_p}^{r_a}}
\newcommand{\vrsqp}[3]{\lbrack2(E_{#1}-\Phi{#3})-{#2}^2/r_{#1}^2\rbrack}
\newcommand{\vrsq}{\vrsqp{}{J}{}}
\newcommand{\adhoc}{\textit{ad hoc }}
\newcommand{\etc}{\textit{etc. }}
\newcommand{\etcend}{\textit{etc.}}
\newcommand{\cf}{\textit{cf. }}
\newcommand{\etal}{\textsl{et al. }}
\newcommand{\ie}{\textit{i.e., }}
\newcommand{\eg}{\textit{e.g., }}
\newcommand{\ftd}{{F}}
\newcommand{\vrot}{\varpi}
\newcommand{\di}[2]{\frac{\partial #1}{\partial #2}}
\newcommand{\ddi}[3]{\frac{\partial #1}{\partial #2\partial #3}}
\newcommand{\I}[1]{C_{#1}}
\newcommand{\II}[2]{D_{#1#2}}
\newcommand{\J}{\I{j}}
\newcommand{\IIJ}{\II{i}{j}}
\newcommand{\barIJ}{\mathcal{D}_{ij}}
\newcommand{\summp}[1]{\suminf_{\ell{#1}_3=0}}
\newcommand{\summ}{{\summp{}}}
\newcommand{\sumknp}[1]{\suminf_{\ell{#1}_1,\ell{#1}_2=-\infty}}
\newcommand{\sumkn}{{\sumknp{}}}
\newcommand{\sumalp}[1]{\summp{#1}\,\sumknp{#1}}
\newcommand{\sumall}{{\sumalp{}}}
\newcommand{\Real}{\mathrm{Re}}
\newcommand{\ex}[1]{e^{#1}}
\newcommand{\Chi}{\chi_{\scriptscriptstyle{1}}}
\newcommand{\simle}{\stackrel{<}{_\sim}}
\newcommand{\simge}{\stackrel{>}{_\sim}}
\newcommand{\simgeq}{\; \raisebox{-0.4ex}{\tiny$\stackrel{{\textstyle>}}{\sim}$}\;}
\newcommand{\simleq}{\; \raisebox{-0.4ex}{\tiny$\stackrel{{\textstyle<}}{\sim}$}\;}
\newcommand{\avg}[2]{\left<{#1}\right>_{#2}}
\newcommand{\avgb}[1]{\avg{#1}{h(\beta)}}
\newcommand{\kaplan}[2]{\avg{{#1;#2}}{\theta}}
\newcommand{\vrms}{\sigma_1}

%% file: front.tex
%
%
\begin{titlepage}
\setcounter{page}{1}
\begin{center}
\huge{\textbf{A Fokker-Planck Study of Dense Rotating Stellar Clusters}}
\large
\vfill
A dissertation presented\\
\vfill
by\\
\vfill
\Large{John Andrew Girash}\\
\large
\vfill
to\\
The Department of Physics\\
in partial fulfillment of the requirements\\
for the degree of\\
Doctor of Philosophy\\
in the subject of\\
Physics\\
\vfill
Harvard University \\
Cambridge, Massachusetts
\vfill
November 2009\\
\vfill
\normalsize
\end{center}
\end{titlepage}
\clearpage

\thispagestyle{empty}
\setcounter{page}{2}
\vspace*{3in}

\begin{center} 
\copyright 2009 by John Andrew Girash\\
All rights reserved.
\end{center}
\clearpage

%
%
\setcounter{page}{3}
\addcontentsline{toc}{chapter}{Abstract}
{\noindent Dissertation Advisor: \textbf{Professor George B.\ Field}  \hfill
Author: \textbf{John Andrew Girash}}\\
\begin{center}
\large{\textbf{A Fokker-Planck Study of Dense Rotating Stellar Clusters}}
\end{center}
\begin{center}
\large{\textbf{Abstract}}
\end{center}


~ ~ ~ ~ The dynamical evolution of dense stellar systems is simulated using a
two-dimensional Fokker-Planck method, with the goal of providing a model for
the formation of supermassive stars which could serve as seed objects for
the supermassive black holes of quasars. This work follows and expands on
earlier one-dimensional studies of spherical clusters of main-sequence
stars.  The two-dimensional approach allows for the study of rotating
systems, as would be expected due to cosmological tidal torquing; other
physical effects included are collisional mergers of individual stars and a
bulk stellar bar perturbation in the system's gravitational potential. The
3 Myr main-sequence lifetime for large stars provides an upper limit on the
allowed simulation times.  Two general classes of initial systems are
studied: Plummer spheres, which represent stellar clusters, and ``$\gamma=0$''
spheres, which model galactic spheroids.

~ ~ ~ ~ At the initial densities of the modeled systems, mass segregation and
runaway stellar collisions alone are insufficient to induce core collapse
within the main-sequence lifetime limit, if no bar perturbation is included.
However, core collapse is not a requirement for the formation of a massive
object: the choice of stellar initial mass function (IMF) is found to play a
crucial role. When using an IMF similar to that observed for dense stellar
clusters (weighted towards high masses but with a high-mass cutoff of
$M_\mathrm{max}\lesssim150M_\odot$) the simulations presented here show, in
all cases, that the stellar system forms massive ($250M_\odot$) objects by
collisional mergers of lower-mass stars; in almost all such cases the presence
of a stellar bar allows for sufficient additional outward transport of angular
momentum that a core-collapse state is reached with corresponding further
increase in the rate of formation of massive objects.  In contrast, simulations
using an IMF similar to that observed for field stars in general (which is
weighted more towards lower masses) produce no massive objects, and reach core
collapse only for initial models which represent the highest-density galactic
spheriods.

~ ~ ~ ~ Possible extensions of the work presented here include continuing to track
stellar populations after they evolve off the main sequence, and allowing for
a (possibly changing) nonspherical component to the overall system potential.

\clearpage

%% file: ack.tex
\chapter{Acknowledgements}

The full list of people to whom I am thankful for helping over the course of
this project is too long to mention completely, so first off I would like to
express my great appreciation to everyone in the Department of Physics and at
the Center of Astrophysics while I was there.  Of individual thanks, first and
foremost I'm grateful for my thesis advisor George Field; ever since we started
this project I've been amazed by your knowledge; as it progressed, was also
impressed and aided greatly by your wisdom; and now as we approach the end,
am very thankful for your patience.

To my thesis committee, Professors Christopher Stubbs, Melissa Franklin and
Lars Hernquist, I give my thanks for your unhesitating willingness to join
me in this, and for your wise thoughts and feedback.

No less impressive than George Field's patience is that of my wonderful spouse
Reebee, who without complaint put up with countless late-night coding sessions,
weekends spent debugging, and holidays taken over by writing.  I couldn't have
done this without your support and encouragement.

Both for their friendship and their specific help in writing and reading
this dissertation I am grateful and indebted to Paul Janzen, Warren Brown,
J.D. Paul and Ken Rines.  My long-term officemates along the way -- Dan
Koranyi, Andi Mahdavi and Vit Hradecky -- all provided welcome, supportive
friendship, lessons on the niceties of caffeine, and healthy distractions
both intellectual and not. Thanks for not letting me take myself too
seriously!  When not busy at our respective computers Mike Westover and Dan
Green gave me all I could handle on the tennis court, while David Charbonneau,
Pauline Barmby, Ian Dell'Antonio, Saurabh Jha, Hannah Jang-Condell, Jan Kleyna,
Ted Pyne, Brian Schmidt, Ann Bragg, Dave Latham, Herb Shea and Thomas Dignan
each in their own ways helped keep me sane and balanced along the way.

I thank Josh Grindlay for getting me started in astrophysics, and Jean Collins,
Peg Herlihy, Beth Rigos and Sheila Ferguson for the umpteen things they did to
help keep me going in astronomy and in physics.

Finally to Jim Wilkinson and all my colleagues at the Bok Center goes my
heartful gratitude for your support and enthusiasm as I've led somewhat of a
dual life at Harvard, providing me the opportunity for an substantive professional
career while still completing my graduate work.  I look forward to continuing
the one while building on the other.

This work was supported by NASA through G. Field's research grants, by funding
from the Harvard Deptartment of Physics and the Center for Astrophysics, and
by a Natural Sciences and Engineering Resource Council of Canada ``1967''
Scholarship.

%% file: Intro.tex
\chapter{Introduction\label{chap_intro}}


\begin{quote}``The challenge of simulating a dense star cluster with a million stars is
formidable, because of the enormous ranges in spatial and temporal scales that
have to be modeled simultaneously.'' \Cite{HutReview}\end{quote}


\section{Astrophysical Motivation: Quasar Massive Black Holes}

The observation of high-redshift quasars, \ie those with redshift $z>6$, implies
a problem for understanding how they formed.  The prevailing model for the
underlying astronomical object is that of a supermassive black hole accreting
material gravitationally \Cite{Umemura}.  Each high-redshift quasar thus must
have had a supermassive black hole in place within the first $10^9$ years after
the big bang, which turns out to be a rather strong physical constraint.
Observations of 
high-redshift gamma-ray bursts and massive galaxies place similar limits on
the formation timescales of the first massive stars \Cite{Cusumano} and of the
first massive galaxies \Cite{Naoz}.

Of the two possible basic building blocks for a massive object in the early
universe -- gas, and stars that have already coalesced from the gas --
direct formation from the collapse of a region of gas may seem to have certain
advantages, 
in that it 
avoids the intermediate step of forming stars.  But the required gas cooling
times are barely consistent with current structure formation models (and even
then require reliance on ``extremely rare'' overdensities) and additionally the
centrifugal barrier that arises due to tidally-induced rotation is a ``key
obstacle'' which must be overcome by any model of massive black hole formation
\Cite{KBD}.  As will be described in more detail below, few previous studies have
examined the evolution and possible core collapse of rotating dense stellar
clusters, and none have modeled whether such systems can form massive stars
which could then serve as seeds for quasar black holes.

The task of forming a massive object from stellar-mass objects faces its own
hurdles: even in dense environments stars are less affected by collisions than
is gas (although
stellar collisions have the benefit
of not necessarily causing shock heating) and so energy and angular momentum
transfer due to collisional processes are not as effective {\sl a prioi} for
stars as they are for gas.  However, stellar systems can overcome this
handicap through the formation of a large, coherent dynamical subsystem -- the
most observationally prevalent of which is a ``stellar bar'' --
which can act as a bulk perturbation in the gravitational potential of the
system and so can effectively transport angular momentum outwards, thus allowing
the system's core to contract further.  This mechanism is described further 
in \S\ref{barintro} and developed in detail in \S\ref{barcons}.

The goal of the current project is to generalize the approach of prior
researchers, who developed one-dimensional models which simulated nonrotating
stellar systems (\eg Quinlan and Shapiro \Cite{QS90} and other references given
below), to two dimensions in order to address the questions: What about rotation,
does it inhibit collapse and can it be overcome?  The anticipated answer is that,
despite initial rotational support against collapse, the presence of a
stellar bar will allow sufficient transport angular momentum outwards so
that collisional stellar mergers and mass segregation can produce a massive
object (perhaps $\simeq10^{2-3}M_\odot$) in the core of a dense stellar
cluster such as could be found at the center of a newly-forming galaxy.
Such an object could then undergo growth via accretion to reach supermassive
size ($\sim10^{6-8}M_\odot$) within a Hubble time \Cite{DDC} and be
observed as a quasar or active galactic nucleus, thus bringing together the
formation of early-universe stars, galaxies, and massive collapsed objects.

The remainder of this chapter expands on each of the above ideas, starting with
dense stellar systems, and moving on to how massive black holes may form in such
stellar clusters. How rotation impacts the system is then discussed.  Finally
an overview of the motivation for and development of the numerical
simulation technique used is given.

\subsection{Stellar Systems}

A primary motivation for this study was the work of Quinlan and Shapiro \Cite{QS89}, who
developed a Fokker-Planck model of an idealized nonrotating, spherically
symmetric, dense cluster of compact stars, and found ``rapid buildup of
massive black holes in the cluster core resulting from successive binary
mergers and mass segregation.'' Subsequently, Quinlan and Shapiro studied clusters of
solar-mass main sequence stars and found that it was``remarkably easy for massive
stars to form through multiple stellar mergers in dense galactic nuclei''
\Cite{QS90}.  After performing a complemetary line of simulations using the
specialized GRAPE N-body computing cluster, Portegies Zwart and McMillan
\Cite{ZGrape} speculated that a sufficiently-dense $10^6M_\odot$ stellar
cluster which initially formed $\sim30$\ pc from the Galactic center can spiral
inward due to dynamical friction before being disrupted by the Milky Way's
tidal field.  While a less-dense cluster would be disrupted before reaching
the Galactic center and so merely contribute its stars to the Galactic bulge,
the more-dense cluster
dissolves closer to the center of the Galaxy and leaves behind its central
massive black hole which can then grow to supermassive size by accretion.

In early theoretical work on stellar systems Begelman and Rees \Cite{BR} found
that ``runaway
coalescence'' of stars (\ie stellar mergers combined with mass segregation)
could lead to the formation of a central massive object.  Again, the system's
evolution was a race between coalescence and disruption, the latter in this
case being due to gas released by earlier stellar evolution and collisions.
Lower-density clusters were susceptible to disruption and dissipation, while
more-dense systems retained most of the gas which went to form new stars.
Stellar clusters were thus explicitly linked to active galactic nuclei:
``Dense star clusters may be responsible for some of the low-level
manifestations of activity in galatic nuclei; but they are probably merely
precursor stages of the more spectacular quasar-type phenomena, which
develop after a massive object has formed.''\footnote{This citation also
includes Rees' prototypical supermassive black hole formation-scenarios
flowchart.} The authors note that runaway stellar merging combined with
mass-segregation provides ``one of the quickest routes to the formation of a
massive object in a dense stellar system''.

\subsubsection{Star Cluster Observations}

Dense stellar clusters are fairly common in the nuclei of large galaxies:
75\% or more of late-type (classification Scd-Sm) spiral galaxies have nuclear
clusters, as do at least 50\% of earlier type (Sa-Sc).  These nuclear clusters
are as compact as globular clusters found in the Milky Way, with a typical
(half-light) radius of $2-5$ pc (B\"oker \Cite{Boeker}, and references therein),
but they are massive.  The dynamical mass of a nuclear cluster is in the range
$10^6-10^7M_\odot$, \ie at the very high end of the globular cluster mass
function -- and similar to the cluster masses used here.  B\"oker concludes
that nuclear clusters are ``an intriguing environment for the formation of
massive black holes because of their extreme stellar density''.


An impediment to building massive black holes in stellar clusters is that the
initial mass function (IMF) which describes the initial distribution of stars
across the possible range of masses is observed to have a cutoff at around
150$M_\odot$ \Cite{Figer1}, as described in more detail in \S\ref{subsect_IMF}.
Thus there is a gap between the IMF's upper cutoff and the 250$M_\odot$ minimum
stellar mass required to leave behind a remnant massive black hole after going
supernova \Cite{ONMW}.  If black hole formation is to be feasible this gap
must be overcome by dynamical processes in the cluster, some candidates for
which are considered in the next section.

\subsection{Scenarios for Massive Black Hole Formation in Dense Clusters}

Observationally, Ebisuzaki \etal \Cite{SPZn} note the existence of what they refer to as the
``missing link'' between stars and supermassive black holes: an IMBH seen in
or near a compact stellar cluster in the center of galaxy M82.  They take this
as evidence for supermassive black holes being created through mergers of
IMBHs, which themselves form via the merging of massive stars in compact
clusters.  Coleman Miller and Colbert \Cite{CMC} also note that IMBHs exist in the current universe,
and that inner galactic bars can contribute to the growth of a
$\lesssim1000M_\odot$ black hole; they also postulate SMBHs from early IMBH
mergers, perhaps as stellar clusters percolate to the galaxy center and
merge.


Approaching the problem from the other direction, taking the existence of
high-redshift (\eg $z=6.41$) quasars as a given and examining timescales by
which a supermassive black hole (SMBH) may grow from an earlier IMBH,
Tyler \etal\Cite{TJS} conclude that a supermassive star (whether formed as such or as a
stellar-merger product) was a more likely SMBH progenitor than any of a
primordial IMBH, a large 
Population III star, or the merger product of smaller Population-III black
hole remants.  Each of latter would be required to form very early and at the
extreme upper end of their physically possible mass ranges, followed by
continuous Eddington-limit accretion.  Again, they did not consider rotation.

The overall difficulty of explaining high-redshift quasar black holes is
summarized by D\"uchting \Cite{Duechting}: 
``The existence of sufficient primordial black holes would unaccountably close
the universe'' unless the spectrum of masses is fine-tuned.  It is ``quite
difficult to construct viable physical models.''  Remnants of Population
III stars, or IMBHs formed from the collapse of early star clusters, also have
``serious problems'' accounting for $z=6$ quasars.

In reviewing possible formation processes for massive central black holes in
nonrotating dense stellar clusters with a few thousand to a few million stars,
Rasio \etal \Cite{RFG} found two possible paths.  If core collapse of the
cluster proceeds quickly enough
-- occurring before the typical lifetime of a large main-sequence star,
$\sim3$ Myr \Cite{FAGR}-- stellar collisions can produce a single supermassive
star which then evolves into a intermediate-mass black hole (IMBH) of mass
$10^2-10^4 M_\odot$.  Alternatively, if large individual stars go supernova
before core collapse and leave stellar-mass black hole remnants, the supernova
mass loss may reverse core collapse somewhat for smaller clusters, while larger
systems may still collapse relativistically \Cite{QS89}.  In either case the
initial cluster evolution is dominated by the most massive stars, \ie by the
high end of the stellar initial mass fuction (IMF). 
The former case of forming a massive object from collisions of main-sequence
stars is the object of this dissertation.  

While also plausible, modeling the evolution
of a cluster of relativistic stellar remnants is a sufficiently different as to
require a separate, later study.  Since main-sequence stellar collisions and
cluster dynamics ``depend crucially on each other'' \Cite{LWRSW} it is unclear
what initial conditions to assume for a post-main-sequence cluster.  And
restricting the simulations to a main-sequence stellar population also avoids
the complications of how to model stellar mass loss due to post-main-sequence
stellar evolution in which the liberated gas could do any of: remaining as gas
in the
cluster; evaporate from the system; or coalesce into new stars, all while the
initial stars are transforming into stellar-mass remnants \Cite{QS90}.  While
also interesting, following this stage of the cluster evolution would greatly
increase the complexity of an already quite complicated model.

\subsection{Rotation due to Tidal Torquing}\label{intro_rot}

Rotation is ubiquitous in the universe.  A protosystem's expected rotation from
being spun up by tidal torques due to cosmological perturbations is ``almost
independent of the perturbation spectrum'' and only weakly correlated with the
individual overdensity being considered \Cite{BE87}.  An initial angular
momentum is an ``important piece of realism ${\ldots}$ while the probability
that the cloud from which a star cluster originates, has zero total angular
momentum is very small, practically all models have assumed that.''
\Cite{KELSL}.

Recently, some work has been done on modeling rotating stellar clusters, and
the picture that emerges is not always intuitive.  Rotating isotropic
systems were found to be dynamically stable in the N-body studies performed
by Meza \Cite{Meza}, but the systems were found to be able to rotate ``very
rapidly'' without becoming oblate.  Rotation can accelerate the time to core
collapse in single-component globular clusters
\Cite{KELSL} and was found in 2-component globular clusters to increase the
speed of dynamical evolution via the enhanced outward transfer of angular
momentum, but only if dynamical friction does not dominate.  (However, this
also results in rotation being a check on mass segregation: massive stars'
angular momentum $J$ prevents them from sinking as far in as they would
otherwise, due to the ``gravo-gyro instability'' \Cite{KLS}.)


In general, rotation in dense stellar systems is important for galactic nuclei,
although the vast bulk of previous studies have focused on globular clusters
\Cite{FSK}.  Most prior work on larger systems has either been to model a
specific system and/or to study systems with a pre-existing massive black hole
(\eg the modeling of dwarf elliptical galaxy M32 by Arabadjis \Cite{Arabadjis}, and
references therein).  On galactic scales Barnes and Efstathiou \Cite{BE87} did find that rotation
can inhibit collapse in N-body simulations, but suggest caution with analytic
predictions of angular momentum evolution in complex stellar systems.

As rotation does not necessarily induce flattening of the system,
\begin{withecc}
most of
\end{withecc}
the simulations performed here assume sphericity for simplicity's sake, and the
potential is constructed so that the model is self-consistent with that
assumption \Cite{DJ86}.
\begin{withecc}
Some simulations are performed with non-zero 
ellipticity as a check of its physical effect.
\end{withecc}


\subsubsection{Seeding Black Holes from Low Angular Momentum Material}

One avenue for overcoming the rotational-support problem is to postulate that
only the least-rotating material contributes to the formation of a seed black hole.
Koushiappas \etal \Cite{KBD} modeled SMBH seeds from the gas with the least angular momentum in
``rare-peak'' haloes of the early protogalaxies, under the assumption that
the distribution of the gas has a significant low-angular-momentum tail,
which can be extrapolated to lower masses and smaller radii from the
distribution found
for dark matter in cosmological simulations.  With this assumption they were
able to produce seed IMBHs of $10^5M_\odot$ for any halo larger than
$7\sn7M_\odot$ -- but no black hole at all for smaller haloes, including the
bulges of most disk galaxies. While giving promising results for large
galaxies, they acknowledge that the rotational barrier is a key concept and
that their model still requires another undetermined mechanism for outward transport of
angular momentum in order to achieve the gas's collapse to a black hole, one
which cannot operate for lower-mass haloes and so does not account for any
continuation of the galaxy/central-object scaling relation to lower-mass
galaxies \Cite{Boeker}. 
Thus another solution, which does not depend on the specific requirement of
unusually rare low-rotation initial conditions, is warranted to account for
at least some of the observed massive black holes in disk galaxies.


\subsubsection{Stellar Bars as Bulk Transporters of Angular Momentum}\label{barintro}

Although some angular momentum $J$ is expected to be transported outwards in a
rotating stellar system via the effects of shear between the faster-rotating
inner regions and the slower-rotating outer regions, a bulk perturbation in the
potential can be much more effective than individual stars at transporting
angular momentum.  In general, any nonaxially-symmetric component of the potential
in a rotationally supported gravitational system fosters $J$ transport over large
distances, due to gravitational torques \Cite{Shlos0}.  The amount of transport
can be as much as an order of magnitude greater than that in a purely spherical
stellar system \Cite{Wyse}.  Specifically, a bar-like perturbation in the
potential can provide a much more effective angular momentum transfer
mechanism compared to what independently-orbiting stars can effect \Cite{TW}, with
$J$ being transported from inner to outer regions \Cite{LBK} and eventually to
the stellar system's halo region -- the presence of which does not stabilize the
system against the formation of a bar perturbation in the first place
\Cite{Athan}.  The 
bar excites a gravitational wake in the stellar distribution:  as the bar passes
through a region, other stars experience a shear force along its orbital path.
This has a net drag effect specifically on individual stellar orbits 
in or near resonance with the bar's rotational frequency \Cite{WK}.
This ``resonance drag'' in finite rotating systems is the analogue of classical
dynamical friction in an infinite, homogeneous medium \Cite{MDW85}.


Bar perturbations were first observed within the spiral structure of galaxies
and are still commonly associated with galactic scales.  However, inner bars of
scale $\lesssim 100$ pc are common and form before larger-scale bars \Cite{Shlos}.
While galactic bars may be largely gas, inner bars are mostly stellar and not
good at gas transport \Cite{MTSS}.  There are dynamical arguments for why
having both outer and inner stellar bars should be the standard model, with
the inner bar decoupling from the outer system \Cite{Macie}.  Observationally,
for example, the S0 galaxy NCG 4621 has a 60 pc
counterrotating (\ie decoupled) core, the ``smallest to date'' \Cite{Wernli}.
Thus, subject to tests confirming that a given system  satisfies the dynamical instability
criterion for a bar to form (as given in \S\ref{sect_barimpl}), a stellar bar
is invoked here on the scale of the stellar cluster being modeled in order to
foster angular momentum transport outward from the cluster center.


\section{Motivation for Technique}

\subsection{Comparison of Fokker-Planck and N-body Methods}

The two main techniques for modeling collisional stellar systems such as
globular clusters, open clusters, or galactic nuclei, are through {\sl N-body}
simulations in which particles representing a number of stars are allowed to
interact via relevant gravititational and other physical processes, and through
integration of the {\sl Fokker-Planck} equation which represents the stellar
population statistically as a distribution function $f$.  When compared against
each other on simple (spherical, nonrotating) systems the two techniques agree
to a large extent (\Cite{KELSL}, \Cite{KYLS}) and so the choice of which
method to use depends on the requirements of a given study.

N-body simulations have the advantage of being ``direct'' in that individual
particles are tracked and can be followed through the entire simulated evolution
of the system over time, and the physical effects that dominate their behavior can be
analyzed.  A disadvantage is that it is not always apparent how to incorporate
a given physical effect into the N-body construct in the first place.  Nor do
individual particles in the simulation correspond to individual objects
in astronomy (\ie stars).  The typical $N$ in current N-body studies is still low compared
to the number of stars in real astronomical systems, and how to extrapolate
between ``particles'' and ``stars'' is not apparent \Cite{KELSL}.  Also N-body
methods are noisy in the sense that a given simulation cannot be relied
on to represent the evolution of a system; instead one must run a suite of
simulations in order to get a statistically robust picture, which will be the
case as long as $N$ is not up to realistic particle numbers, preventing
truly star-by-star modeling \Cite{FSK}.

On the other hand, the statistical nature of the Fokker-Planck equation 
``captures the essence'' of continuum mechanics \Cite{Spiegel} but does not
afford direct analysis of how different physical processes contribute to the
observed behavior of the system.  All one knows is how the stellar distribution
function $f$ evolves over time.  However, Fokker-Planck does allow for different aspects
of the physics to be added or removed at will; thus comparing integrations
with and without a given effect included can accomplish the same result.
This ability to include new dynamical aspects in an \adhoc manner
permits us to study a discrete spectrum of stellar masses (\eg as
previously done by Amaro-Seoane \Cite{Amaro}), collisional mergers of
stars (\eg Quinlan and Shapiro \Cite{QS90}) and a bulk perturbation in the
gravitational potential such as the stellar bar.

\subsection{Orbit Averaging and the Third Integral} 

As developed in Chapter \ref{chap_coeffs}, the Fokker-Planck
approximation\footnote{{\sl Approximation} here refers to the idea that the
Fokker-Planck equation can be derived by expanding the collisional Boltzmann
equation under the assumption of weak encounters, \ie $\delta v\ll v$, and
truncating the resulting series after the second-order terms \Cite{BT}.} is
appropriate when small angle, 2-body scattering dominates the global evolution
of the system.  When that is the case, energy transport can be treated as analogous to
heat conduction in a collisional gas -- producing a ``fairly good description
of what happens in N-body simulations'' \Cite{KELSL}.  The full \FP equation
tracks the time-evolution of the system in all 6 dimensions of position and
velocity (or more generally, momentum) which is both computationally
prohibitive and usually not necessary for describing the system's evolved
properties.

In order to simplify the problem one can attempt to eliminate the
positional dependence of the \FP equation, leaving only the momentum
dependence to describe the dynamics.  A typical method used to achieve this is
{\sl orbit averaging} in which the \FP equation and all dynamical quantities
on which it depends are, as the name implies, averaged over a full stellar
orbit before being applied.  These orbit-averaged \FP models are found to
``treat very well the diffusion of orbits according to the changes of their
constants of motion, taking into account the potential and the orbital
structure of the system in a self-consistent way'' \Cite{Amaro}.

One is thus left with a description of a stellar system based on three
constants of motion, corresponding to the three dimensions of momentum.  This
is consistent with the strong Jeans theorem, which requires in the general case
that there be three integrals of the motion, even if only two are well
determined \Cite{KYLS}.  Starting with Goodman's 1993 dissertation
\Cite{Goodman} standard practice has been to use energy $E$ and angular
momentum component $J_z$ as the constants of motion, and to simply ignore
any third integral.  Comparison to N-body results has shown this to be
reasonable if flattening of the system is not extreme \Cite{KELSL}.  For
such systems, however, it is in principle possible to use $J^2$ as a proxy for
the third integral, although this would be ``extremely difficult numerically
and physically'' \Cite{KYLS}.  Strictly speaking, ignoring the third integral
is theoretically valid if and only if the system's gravational potential is
spherically symmetric, even if its velocity distribution is not \Cite{BT}
-- hence the restriction to studying
\begin{withecc}
systems with only small flattening in practice.
\end{withecc}
\begin{skipecc}
rotating, but gravitationally-spherical systems.
\end{skipecc}

\subsection{Choice of Canonical Variables} 

Depending on the choice for constants of the motion, the two-dimensional
orbit-averaged \FP equation can be used to study different dynamical apsects of
a stellar cluster.  With the assumption of isotropy of the velocity dispersion,
the traditional choice of $(E,J_z)$ can represent an axisymmetric rotating
system, \eg Kim \textsl{et al.}'s 
single-mass \Cite{KYLS} and two-mass clusters \Cite{KLS}.  Alternatively
$(E,J^2)$ can be used to study
the somewhat simpler problem of anisotropy in spherically symmetric systems,
as first done by Cohn \Cite{Cohn79} and more recently by Takahashi
(\Cite{TakaI}, \Cite{TakaIII}) and Fiestas \etal \Cite{FSK}.  The latter
problem is simpler in that the distribution can
be averaged over angular momentum $J$ before coefficients of the \FP equation
are calculated, providing a great savings in complexity and calculation time.

\begin{withecc}
When considering a non-spherically symmetric system,
\end{withecc}
\begin{skipecc}
However, once incorporation of \adhoc terms into the \FP equation is desired,
\eg to represent collisional stellar mergers (as originated by Quinlan and
Shapiro \Cite{QS90} in a one-dimensional study)
\end{skipecc}
energy $E$ is no longer necessarily an integral of the
motion.  Any change in the potential will cause changes in the stellar
distribution function when it is expressed as $f(E,J^2)$ or as $f(E,J_z)$.
What is required are the radial action $I$ and the tangential action (\ie the
overall angular momentum) $J$, which are still adiabatic invariants as the
potential evolves, and so also is $f(I,J)$.  Binney and Tremaine \Cite{BT} 
pointed out this exact advantage of using the canonical actions as
coordinates for direct numerical computation of the Fokker-Planck equation,
as a method for stellar dynamics simulations.
\begin{withecc}
And while the amounts of flattening studied here are rarely large enough to
strictly require use of $f(I,J)$ instead of $f(E,J_z)$,
\end{withecc}
\begin{skipecc}
Additional
\end{skipecc}
benefits of this approach are that the \FP equation takes a particularly clean
form and that the procedure of orbit-averaging is then a conceptually simple
averaging of quantities over the $2\pi$ change in (an) orbital angle variable,
as will be seen in Chapter \ref{chap_coeffs}.

With this form it is also possible to straightforwardly incorporate 
terms describing collisional stellar mergers
\begin{withecc}
(as originated by Quinlan and Shapiro \Cite{QS90} in a one-dimensional 
study).
\end{withecc}
into the \FP equation.  The main drawback is that a large number of conversions
are required between the $(E,J,J_z)$ coordinates in which dynamical quantities
are calculated and the $(I,J)$ space in which the system evolution is tracked
by the \FP equation for $f(I,J)$.  Integration of the two-dimensional \FP
equation -- even in its more clean form, as when expressed in term of actions
$I$ and $J$ -- is sufficiently challenging that this tradeoff remains
beneficial.  Effectively the additional complexity of the dynamics has been
separated out of the \FP equation itself and into these coordinate conversions.
The result is a three-step simulation process.  First the Fokker-Planck equation
is used to update the distribution function $f(I,J)$ at a new timestep.  Then
the effects of any additional physical processes -- in this case, collisional
stellar mergers -- are calculated and $f(I,J)$ updated accordingly.  Finally,
given the new $f(I,J)$, consistent values for physical quantities such as the
stellar density and gravitational potential are found in terms of either $E$
and $J$ or the spatial coordinates.

\section{Overview of Resulting Model and of Remaining Chapters}

Given all the above considerations, the basic form of the model developed
for this thesis has the following characeristics:
\begin{enumerate}

\item The system to be studied is a fairly dense stellar cluster, using a
generalization of the initial conditions from Quinlan and Shapiro \Cite{QS90},
with an observationally-motivated initial mass function for the stars
which make up the cluster;

\item Unlike most prior work on dense clusters, an overall rotation of the
system is included, with a value determined by observation;

\item In order to allow for the inclusion of desired additional physical effects
beyond the general gravitional interactions of individual stars in the cluster
-- \ie a stellar bar perturbation, and collisional merging of individual stars --
the Fokker-Planck equation is employed to track the system's evolution; and

\item To capture the physics required while still keeping the problem
numerically tractable the Fokker-Planck model is orbit-averaged, and any
third integral neglected, so that the system is represented in two dimensions.
The canonical actions $(I,J)$ are employed as integrals of the motion in order
for the above-mentioned extra physical effects to be incorporated straightforwardly.

\end{enumerate}

With this setup for the model, the basic question addressed here is under
what circumstances -- using which physical effects, starting with what initial
conditions -- can the system produce a star which can serve as seed object for
an intermediate-mass black hole and eventually a supermassive black hole.  For
this to be possible, the star must have sufficient mass, \ie 250$M_\odot$ or
more, and it must be formed 
before the end of the main-sequence lifetime of large stars, \ie within
the first $3$ Myr or so of the system's evolution.
The simulations will show that a bar-like perturbation can effect sufficient
angular-momentum transfer to allow the creation of a likely seed object
in most rotating clusters whose density and velocity dispersion match what is
seen in the centers of galaxies, but only if the initial mass function is
also set to match that observed in dense clusters -- which is more top-heavy
that the general field-star IMF.


The remaining chapters are organized as follows:

\begin{itemize}
\item Chapter 2 outlines the dynamical and gravitational-potential physics
required for the Fokker-Planck model, lists the numerical considerations
needed to compute it, and outlines the set of possible initial conditions.

\item Chapter 3 derives the diffusion coefficients that incorporate the
gravitational effects of the system's individual stars and of any stellar bar
perturbation into the Fokker-Planck equation for the evolution of the system,
as well as the population effects of collisional mergers.  The
differencing scheme employed to compute the model is developed, and a brief
consideration of the heating and merging effects of binary stars is given.

\item Chapter 4 describes the tests performed on the various aspects of the
model, both to verify its numerical and physical validity and to establish
reasonable values for any numerical input parameters required.

\item Chapter 5 details the results of the simulations for the various possible
choices of initial conditions and of which physical effects to include.

\item Chapter 6 then discusses the simulations' results from an astrophysical
standpoint, gives comparisons to what would be expected from previous studies
and from general considerations such as timescale arguments, and looks forward
to what upcoming observations and future models may be able to provide.

\item Tables of frequently-used symbols and subscripts are given in the Appendix.
\end{itemize}

In the end, the simulations show that dense rotating stellar clusters are a
feasible path to the production of seed objects for massive black holes in
galaxies.  All simulations that were performed with a top-heavy IMF -- such as
is observed in dense stellar clusters -- resulted in at least one sufficiently
massive star (and usually several) being formed.  The result holds for a
range of initial cluster configurations which were chosen to match typical
dense nuclear clusters and galaxy centers. Only for models representing
giant elliptical galaxy cores was a massive object not seen before the
largest stars evolved off the main sequence.  For these systems one would
need to rely either on post-main-sequence dynamics within the core, or on the
presence of a dense nuclear stellar cluster which could evolve independently.

%% file: Dynamics.tex
\chapter{Dynamical and Gravitational Aspects of the Model\label{chap_pot}}

\section{Overview}\label{sect_units}

The first two sections of this chapter describe the simulation model's
dynamical framework, specifically the calculation of orbital parameters \eg
endpoints, frequencies \etcend \ These will be needed in Chapter
\ref{chap_coeffs} to determine the effect of dynamical friction on the
ensemble of stars in the cluster, as described respectively by the diffusion
coefficients and the stellar distribution function in the Fokker-Planck
equation.  The remainder of the chapter then develops the calculations of the
cluster's physical properties, such as the mass density profile $\rho(r)$ and
the gravitational potential
\begin{withecc}
$\Phi(r,z)$,
\end{withecc}
\begin{skipecc}
$\Phi(r)$,
\end{skipecc}
and ends with a summary of the possible initial
conditions used for these quantities.

The fundamental units employed in the model are km/s, $M_\odot$, and pc.  With
these choices the unit of time is $\mathrm{1pc/(km/s)=0.9778\sn6 yr\simeq1Myr}$,
and Newton's constant of gravitation is 
$G\mathrm{=4.299\sn{-3}\,(km/s)^2\,pc/}M_\odot$.

\section{Dynamics}\label{sect_dyn}

The most convenient way to parameterize stellar orbits is to use 
to the canonical radial and angular actions, which together will be
referred to as the action vector $I_j$, with components $I_1$ (the radial action)
and $I_2$ (the angular action).   $I_2$ is equal to the traditional
angular momentum $J$, while $I_1$ can be defined from the radial momentum:
\beq I_1=\otwopi\oint dr\,p_r=\opi\orbint dr\vrsq^{1/2}\label{tw30}\eeq
in which orbital endpoints $r_p$, $r_a$ are the star's periapse and apoapse
distances from the cluster center, $E$ is the star's total orbital energy and
$\Phi(r,\theta)$ is the gravitational potential at distance $r$ and polar
angle $\theta$ from the center.\footnote{Shlosman \Cite{Shlos0} points out that the
Jacobi energy $E$ of (\ref{tw30}) isn't
a true integral of motion, even for individual stars, when there are nested
bar perturbations: the underlying potential is then not a constant.  This is
expected to be the case for clusters such as those modeled here when near the
center of a galaxy \Cite{MTSS}; however, the inner bars decouple dynamically
from the larger-scale galactic bars \Cite{Macie} which will vary on much
longer timescales, and so this nicety is expected to produce only a secular
change in the underlying potential that averages out over long timescales.}
Azimuthal symmetry of the cluster's main stellar distribution (and so also
of the potential) is assumed.

The advantage of using the canonical coordinates is that the stellar
distribution function, when expressed in terms of these actions, is invariant
to adiabatic changes in the potential; hence we do not have to solve for a new
distribution function just because the potential evolves over time \Cite{BT}.
As seen above however it is often necessary to deal with the traditional
variables $r, \theta, \phi, E, J, J_z$ when calculating
orbital parameters or any physical quantity that depends on the value of
the potential.  Alternatively, when the actions $(I_1,I_2)$ are known,
(\ref{tw30}) implicitly defines the stellar orbital energy $E$.

Following Tremaine and Weinberg \Cite{TW}, we can move between action
space and position space via the relations that follow.  The radial and
angular orbital frequencies $\Omega_j$ are found similarly to $I_1$ :
\beq\frac{1}{\Omega_1}=\frac{1}{\pi}\orbint\frac{dr}{\vrsq^{1/2}}\label{tw28}\eeq
\beq\frac{\Omega_2}{\Omega_1}=\frac{J}{\pi}\orbint\frac{dr}{r^2\vrsq^{1/2}}
\label{tw29}\eeq
From these we can relate the dynamical quantities in the two coordinate
systems:
\beq\left(\frac{\partial I}{\partial E}\right)_J=\frac{1}{\Omega_1},~~~
\left(\frac{\partial I}{\partial J}\right)_E=-\frac{\Omega_2}{\Omega_1}\eeq
\beq\left(\frac{\partial E}{\partial I_j}\right)=\Omega_j\label{tw35}\eeq
where the last relation generalizes to the case including a third integral of
motion if we define $I_3=J_z$ and note that $\Omega_3\equiv0$.

Although orbit-averaging will remove any direct dependence of the diffusion
coefficients on the canonical angles $w_1$ and $w_2$, the angles are needed for
intermediate steps in the coefficients' calculation. (The angles are also
needed for some of the tests of the coefficients' validity.) They are given by
\beq w_1=\Omega_1\int_C\frac{|dr|}{\vrsq^{1/2}}\label{tw37}\eeq and
\beq w_2-\psi=\int_C\frac{(\Omega_2-J/r^2)|dr|}{\vrsq^{1/2}}.\label{tw38}\eeq
The azimuthal angle $w_3$ is not of interest except in that it contains a
random initial phase $\phi$.

As shown in Fig. \ref{orbangles},
$\psi$ is the angle swept out by the star's
position vector as measured from the point where it crossed the $\theta=\pi/2$
plane:
\beq\sin\psi\sin\beta=\cos\theta,\label{tw41}\eeq
in which $\beta$ is the orbital inclination, \ie
$J_z=J\cos\beta$.  Integration path $C$ is along the orbit, \ie from $r_p$
to $r_a$ and back to $r_p$ when traversing an entire orbit. For our purposes
the quantity ($w_2-\psi$) is what will be needed most often. The above
relations are intuitive given that $\vrsq^{1/2}$ is the radial component of
the velocity and the canonical angles obey $d^2w_j/dt^2=0$.

\input{epsf}
\begin{figure}
\begin{center}       
\leavevmode  
\epsffile{tw84fig1.epsi}
\end{center}
\hfill\textit{Image source: Tremaine and Weinberg \Cite{TW}.}
\caption{Euler angles for the orbit of a given perturbing potential.}
\label{orbangles}
\end{figure}

\subsection{Numerically Solving for Orbital Endpoints}\label{subsect_endpts}

The general method for finding the orbital endpoints $r_p$, $r_a$ is to solve
numerically for the roots of $v_r^2=\vrsq=0$.  For small values of either
action component $I_1$ or $J$ this may not be numerically robust and other
techniques must be used. If $(E-\Phi)/|\Phi|$ is miniscule (\ie
$\lesssim10^{-8}$), the star is considered to be ``fixed'' at the origin.  For
$(E-\Phi)/|\Phi|\lesssim10^{-3}$ a quadratic approximation to the potential is
employed and the endpoints solved for analytically.  The endpoint-solver
takes as its natural inputs $E$ and $J$, but it also requires at least a
reasonable guess of $I_1$ so it can distinguish between more-radial and
more-circular orbits.

For very-nearly-radial orbits (taken to be $|J|<z_\mathrm{tol}\,I_1$), $r_p$ is
set to a value of $10^{-8}$ in order to avoid numerical problems at exactly $r=0$.
\footnote{Alternatively,
for nearly-radial orbits $I_1/|J|\lesssim10^{-3}$, an epicyclic
approximation for finding $r_p$ and $r_a$ as well as $\Omega_j$ was tried.
This was not found to be an improvement and results with the epicycle
approximation are not reported here.
However, for numerically tiny values of the components of the action (typically 
$I_j\lesssim5\sn{-4}$) \ie for stellar orbits restricted to the cluster center,
an epicylcic approximation \Cite{BT} is employed:
$(\Omega_2)^2\simeq\partial^2\Phi/\partial r^2|_{r=0}$ and $\Omega_1=2\Omega_2$,
although in practice no values of $I_j$ on the grids used trigger this case.}
The input parameter $z_\mathrm{tol}=10^{-5}$ typically, although other values
were tested to make sure the results were robust against such a change.

For moderately-circular orbits (those with $I_1/|J|\lesssim0.17$), first
a value of $r$ for which $v_r>0$ is found;
then $r_p$ and $r_a$ are determined by searching out 
in both directions until $v_r$ changes sign.  
For more radial orbits, the entire allowed range of $r$ is searched
for subranges over which $v_r$ changes sign.  In both cases, the bracketing 
of the ranges over which $v_r$ changes sign is then made increasingly finer
until convergence is reached.  

\subsection{Calculating Orbital Frequencies and other Dynamic Quantities}
\label{subsect_orbfreqs}

Once orbital endpoints $r_p$ and $r_a$ are known then (\ref{tw29}) and (\ref{tw28}) can be used
to find orbital frequencies $\Omega_j$ directly (and similarly (\ref{tw30}) gives the action
components $I_j$ if they are not yet determined). In practice doing integrals
is very expensive in processor time;
thus in general $\Omega_j$ is calculated from (\ref{tw35}).  When values for
$\partial E/\partial I_j$ are not yet known for a given simulation timestep, however,
(\ref{tw29}) and (\ref{tw28}) are employed.

Just as in the orbital endpoint calculation, boundary cases require special
treatment.  When the orbit is highly radial (taken to be when
$J/I_1\lesssim5\sn{-4}$), (\ref{tw29})
becomes unstable and so the limiting case of $\Omega_2/\Omega=\half$ is used.
Finally, for orbits that are very nearly circular ($I_1/J<5\sn{-4}$), even if
the solution for $r_p$ and $r_a$ is obtainable, doing further calculus such as
(\ref{tw29}) and (\ref{tw28}) is numerically difficult and so in this case
the orbit is taken to be exactly circular with $r_p=r_a=r_c$, the radius at
which $\partial v_r/\partial r=0$.  The above $I_1/J$ requirement is sufficiently
strict that this will not affect any dependent calculations.

\input{Potential}

%% file: Potential.tex
\section{Calculation of the Density}
\subsection{Using the Distribution Function in $(E,J^2)$ Space}
\label{subsect_r}

Traditionally, the (smoothed) stellar mass density $\rho$ is given by
the ``full'' 6-dimensional integral of distribution function $\ftd$ over
all velocity space -- which gives the total number density of stars --
multiplied by the per-star mass $m$:
\beq\rho(\mathbf{r})=m\int\,d^3v\,\ftd (\mathbf{r},\mathbf{v})\label{eff}\eeq
which results in a total mass $M$ for the cluster of:
\beq M
=\int\rho(\mathbf{r})\,dV=m\int\,d^3r\,d^3v\,\ftd(\mathbf{r},\mathbf{v}).
\eeq
Previous studies assumed a stellar cluster with spherical symmetry and used
$(E,J^2)$ as the coordinates of choice; using the transformation of Cohn \Cite{Cohn79},
in $(E,J^2)$ space the above expression becomes
\beq M=8\pi^3m\int d(J^2)\int dE\, \Omega_1^{-1} F(E,J).\eeq
With azimuthal symmetry, $2\pi$ of the above comes from integrating
over the velocity-space azimuthal angle $\phi_v$.
Expanding $d(J^2)$ and applying (for fixed $J$) $dE=\Omega_1 dI$ from
(\ref{tw29}), the total cluster mass when integrated over the actions is thus
\beq\label{fejij}M=16\pi^3m\int JdJ\,\int dI\,F(E(I,J),J).\eeq

\subsection{Using the $(E,J^2)$-space Distribution Function in Action Space}
\label{subsect_I}

In order to take advantage of the invariance of the actions, we define a
distribution function analogous to $F$ but expressed in action space:
\beq M=m\int\,dI\,dJ\,f(I,J)\label{oaf}.\eeq
Note that this is by construction an ``orbit-averaged'' distribution
function (\Cite{BT}, \Cite{Cohn79}) and so all angle-dependence has
already been averaged out.  Comparing with (\ref{fejij}) one obtains
\beq f=16\pi^3JF\label{f2f}\eeq
as the conversion between the action-space $f$ and the earlier $F$.

To calculate the spatial density $\rho$, one can either perform the integral
of (\ref{eff}) directly, or first convert it to a 2-dimensional orbit-averaged
form that uses $f$, similar to (\ref{oaf}).  The former requires use of the 
velocity volume element (see \eg Cohn and Kulsrud \Cite{Kulsrud})
\beq \int_{\phi_v=0}^{2\pi} d^3v=4\pi\frac{JdJdE}{r^2v_r}\eeq
and so, if using the two-dimensional $f$ of (\ref{oaf}) and assuming spherical
symmetry,
\bea\label{cohnrho}
\rho(r)&=&4\pi m\int_\Phi^0dE\int_0^{J_\mathrm{max}}\frac{JdJ}{r^2v_r}F(E,J)\\
&=&4\pi m\int_0^{J_\mathrm{max}}JdJ\int dI\frac{\Omega_1}{r^2v_r}F(E(I,J),J)\nonumber\\
&=&\frac{m}{4\pi^2}\int_0^{J_\mathrm{max}}dJ\int dI\frac{\Omega_1}{r^2v_r}f(I,J)\nonumber
\eea
where the endpoints of the $I$ integral are those energetically allowed.
This has the advantage of being a direct method of calculation, but is
otherwise undesirable for this study for several reasons: the $1/r^2$ dependence
means that the $\rho$ calculation is less robust close to the cluster center;
the singularities at the orbit endpoints due to the $1/v_r$ dependence are particularly
difficult to deal with in the $dI dJ$ form of (\ref{cohnrho}),
and spherical symmetry is not an intended assumption for this model.

\subsection{Calculating the Density in Action Space}\label{subsect_rhoI}

In order to deal with the above issues it is necessary to develop a differential
in terms of the action coordinates themselves.  Expanding $d^3v$ in spherical
coordinates we have
\beq d^3v=v^2dv\,d\cos\theta_v\,d\phi_v.\eeq
Defining the ``tangential'' velocity $v_\theta^2+v_\phi^2\equiv v_T^2=(J/r)^2$,
then $dJ=rdv_T=rv\,d\cos\theta_v$ and so
\beq d^3v=2 vdv\,d\phi_v\frac{dJ}{r}\eeq
and 
\beq \int_{\phi_v=0}^{2\pi} d^3v=4\pi\,v\,dv\,\frac{dJ}{r}.\eeq
Note that $2\pi$ in the above again comes from integrating over azimuthal angle
$\phi_v$, and an extra factor of $2$ has been inserted in order to allow use of
the usual convention that $J\geq0$ only.  Finally, with $\Omega_1dI=dE=v\,dv$,
\beq \int_{\phi_v=0}^{2\pi} d^3v=4\pi\,\Omega_1\,dI\,\frac{dJ}{r}.\eeq
Inserting this in (\ref{eff}) and using the conversion from $F$ to $f$,
\beq\rho(r)=\frac{m}{4\pi^2r}\int_r\frac{dI\,dJ}{J}~\Omega_1f(I,J)\label{onerho}\eeq
where as always, the action integral endpoints are determined by which orbits
are energetically allowed at $r$; this is indicated by subscript ``$r$'' on
the integral sign.  This form for the calculation of $\rho$ is
more suited for the overall action-space-based model of this study than is
(\ref{cohnrho}), and is what will be used throughout; the cost of this
choice is that the orbital endpoints are now more complicated to calculate.

In what follows, it will often be convenient to use the notation $(I_1,I_2)$
in place of $(I,J)$.  Generalizing the expression for $\rho$ to the case in
which there are multiple values for the stellar mass,
\beq\rho(r)=\frac{1}{4\pi^2r}\int_r\frac{dI_1 dI_2}{I_2}~\Omega_1\sum_qm_qf_q(I_1,I_2).
\label{homrho}\eeq
where subscript $q$ refers to a particular stellar species
(\eg if the lowest-mass stars in the simulation are of solar mass,
then $m_1=1M_\odot$ and those stars' distribution function is $f_1$).  
The ranges of integration of both of the $I_j$ are the ranges of
energetically allowed values, and so depend on $E(I_1,I_2)$ and on $\Phi(r)$.

\section{Rotational Velocity and Orbital Inclination}\label{sect_vrot}

It is possible to deduce an expression for the rotational velocity $\vrot(r)$
of a given stellar orbit using only the velocity dispersion and the assumption of velocity isotropy.
However, such a relation would depend more heavily on the assumption of isotropy
than does anything else in this study, and in the end it does not yield any
insight into the distribution of orbits over inclination angle $\beta$
(where $\cos\beta=J_z/J$), which will also be needed later.  Instead, we
implement a procedure for first finding the distribution over $\beta$ and
use that to determine $\vrot$.

\subsection{Orbital Inclination}\label{orbinc}

Even in a spherically-symmetric gravitational potential, with an overall
rotational component to the stellar velocity distribution some quantities will
depend on the resulting distribution of orbital inclination.
It is convenient to parameterize the inclination dependence of $f$ as having
factors that are even and odd in $\alpha\equiv\halfpi-\beta$:
\beq h(\alpha)=g(|\alpha|)+\Theta(\alpha)\eeq where
$\int_{-\pi/2}^{\pi/2}d\alpha\,h(\alpha)=\int_{-\pi/2}^{\pi/2}d\alpha\,g(|\alpha|)=1$
and $\Theta(-\alpha)=-\Theta(\alpha)$. Thus defined, $h(\alpha(\beta))$ can
then be applied as a weighting function in averaging any $\beta$-dependent
quantity, which is essentially a way of defining an effective distribution
function:
\beq f_\beta(I,J;\beta)\equiv h(\alpha)f(I,J)=[g(|\alpha|)+\Theta(\alpha)]f(I,J).\eeq

Without any prior knowledge of the rotational state of a stellar system other
than the net amount of rotation, DeJonghe (\Cite{DJ86}, \Cite{DJ87}) has shown that the statistically
most probable form for $\Theta(\alpha)$ is
\beq\Theta(\alpha)=\tanh\left(\frac{bJ_z(\alpha)}{2}\right)\,g(|\alpha|)\eeq
where $b$ is a parameter determined by the total rotational angular momentum.
Other than being an even function of its argument, so far $g(|\alpha|)$ is
unconstrained.  A particularly useful choice is to assume that for a given
$J$, all possible values $0\leq|J_z|\leq J$ are equally likely \Cite{TW},
consistent with the assumption of isotropy.
This is equivalent to setting $g(|\alpha|)=\half\cos\alpha:$
\beq\label{halpha}h(\alpha)=\half\left[1+\tanh\left(\frac{bJ\sin\alpha}{2}\right)\right]\cos\alpha\eeq
or, rewritten in terms of inclination angle $\beta$,
\beq
h(\beta)=\half\left[1+\tanh\left(\frac{bJ\cos\beta}{2}\right)\right]\sin\beta.\label{hbeta}\eeq

It remains to determine the value of $b$.  For notational convenience we
define mass-weighted distribution functions, \ie $\tilde{f}\equiv\sum_qm_qf_q$
and similarly $\tilde{F}\equiv\sum_qm_qF_q$.
Then we define the ``population'' of a given orbital energy as
\beq P(E)\equiv\int_{J_\mathrm{min}}^{J_\mathrm{max}}dJ\,\tilde{F}(E,J)\eeq
and the total angular momentum for a given energy as
\beq J_\mathrm{tot}(E)\equiv\int_{J_\mathrm{min}}^{J_\mathrm{max}}dJ\,J\,\tilde{F}(E,J).\eeq
For a distribution that has no $J$-dependence
(and hence, no rotational streaming),
the expected total angular momentum
is simply
\beq\overline{J}(E)=\half(J_\mathrm{max}+J_\mathrm{min})P(E)\eeq
and so one can deduce the excess of angular momentum at a given $E$:
\beq\Delta J(E)\equiv J_\mathrm{tot}(E)-\overline{J}(E)\eeq
so that $\Delta J(E)=0$ for no net rotation. This allows for the following
scheme for determining $b$: for a given orbital energy $E$, choose $b(E)$ such
 that the resulting amount of net rotation is consistent with $\Delta J(E)$:
\bea\Delta J(E)
&=&\int dJ\,\tilde{F}(E,J)\int^\halfpi_{-\halfpi}d\alpha\,J_z\,h(\alpha;E)\nonumber\\
&=&2\int dJ\,J\,\tilde{F}(E,J)\int^\halfpi_0d\alpha\sin\alpha\,\Theta(\alpha;E)\nonumber\\
&=&\int dJ\,J\,\tilde{F}(E,J)\int^1_0d(\sin\alpha)\sin\alpha\,\tanh\left(\frac{bJ\sin\alpha}{2}\right)\nonumber\\
&=&\int dJ\,J\,\tilde{F}(E,J)\left(\frac{2}{bJ}\right)^2\int^{bJ/2}_0dx\,x\,\tanh x\label{tanhx}\eea
which can be expanded to read
\beq\label{bigrot}\int dJ\,\tilde{F}(E,J)\left[J-\half(J_\mathrm{max}+J_\mathrm{min})
-J\left(\left(\frac{2}{bJ}\right)^2\int^{bJ/2}_0dx\:x\tanh x\right)\right]=0.
\eeq


Root-finding over the above integral equation is a computationally expensive
process; before resorting to that, we first attempt an iterative solution by
expanding $x\tanh x$ for small $x<\halfpi$ and solving the resultant equation
for the leading term in $b$, using the previous iteration's $b$ value in the
higher-order terms.  If this does not converge we do rootfind on the exact
expression above.  For larger values of $bJ/2$, \ie  $bJ/2>\halfpi$ the integral
over $x$ itself asymptotically approaches $\half$, simplifying the process.
In all cases the resulting $b(E)$ is then verified to satisfy (\ref{tanhx});
with this $b$, (\ref{hbeta}) then describes the orbital inclination
distribution of orbits with energy $E$.

\subsection{The Mean Rotational Velocity}

Further defining the population $p$ at a given $r$ as
\beq p(r)\equiv\int_rdIdJ\,\tilde{f}(I,J)\eeq
finding the mean rotational velocity $\vrot$ at a given $r$ is then straightforward:
\bea\vrot(r)&=&\frac{1}{p(r)}\int_r\,dIdJ\,\tilde{f}(I,J)\int^\halfpi_{-\halfpi}d\alpha\,\frac{J_z}{r}\,h(\alpha(E(I,J)))\\
&=&\frac{1}{r\,p(r)}\int_r\,dIdJ\,J\,\tilde{f}(I,J)\left(\frac{2}{bJ}\right)^2\int^{bJ/2}_0dx\;x\,\tanh x\label{tanh}\eea
in which $b=b(E(I,J),J)$ as found in $\S\ref{orbinc}$.  Descriptively,
$\vrot(r)$ is the mean speed at which stars stream in the azimuthal
direction at a given distance from the system center, and is
found by averaging $J_z/r$ over all inclination angles $\beta$ as weighted by
the relative density $h(\beta)$.

\subsection{The Coordinate Grid}\label{subsect_grid}

The self-consistent gravitational potential $\Phi(r)$ needs to be updated after
each time step in the simulation, through the iterative solving of Poisson's
equation.  Up to this point, the dynamics -- including the
calculation of the orbital inclination distribution $h(\beta)$ -- have assumed
a spherical gravitational potential $\Phi(r)$ and (smoothed) density
$\rho(r)$, \ie that the cluster's overall ellipticity $e=0$.
This is done simply to allow the construction of a self-consistent model using
the two-dimensional stellar distribution function $f(I,J)$, as adding a third
integral would be computationally prohibitive.

With the present definitions of dynamical quantities and construction of
$h(\beta)$, setting $e=0$ is mathematically proper \Cite{DJ86} and yields a
self-consistent model. It is possible that $e\neq0$ may be more physically
motivated, although N-body studies do show that ``spherical stellar systems
can rotate very rapidly without becoming oblate'' \Cite{Meza}.
In order to see how much effect the assumption (by construction) of sphericity
has, we can add in allowance for ellipsoidal isodensity surfaces in an \adhoc
manner, using the presumption that the first-order effects of $e(r)\gtrsim0$ on
bulk properties (\eg anything that depends directly on $\rho$ or on $\Phi$) will
outweigh the (small, for small $e$) errors it introduces into the dynamics
proper.\footnote{The only \textsl{inherently} non-spherical aspect of the
model's potential or density, namely a ``bar perturbation'' in the stellar
distribution's potential, will be considered in $\S\ref{barpert}$.}  Thus
unless otherwise specified, ellipticity is pre-defined as $e\equiv0$ for all
results -- although provision for
\begin{skipecc}
the future consideration of the case with
\end{skipecc}
a preset $e(r)\ne0$, which may be allowed to vary along with the coordinate
grid values, is built into the model and is the basis for the development
presented in \S\ref{sect_gravpot} of the gravitational potential and in
Appendix \ref{sect_num} of numerical aspects of the model.

Thus in the general case there will be a ellipticiy $e(a)$ associated with a given
isodensity surface at ellipsoidal radius $a$; this must be taken into account when calculating quantities
that depend on the local position.\footnote{Note that, despite allowing for a
net overall rotation in the stellar population, the procedure for calculating
the non-constant distribution of orbital inclinations $h(\beta)$ does not
alter the overall energy balance.} Equating the exterior value of the potential
$d\Phi_e(a)$ of an ellipsoidal shell of mass $dM$ with the analogous value
$d\Phi_0(r)$ for a spherical shell of identical mass one obtains
\beq-\frac{G\,dM\sin^{-1}e}{ae}=-\frac{G\,dM}{r}\eeq\beq\label{geta2e}
a=r\left(\frac{\sin^{-1}e}{e}\right)\simeq r(1+\frac{1}{6}e^2)\eeq
where the approximation is used throughout for very small ($<2\sn{-3}$)
values of $e$.
From here on, $r$ can be taken as a ``dummy'' variable, used for convenience
to label gridpoints; the shell's semimajor axis length $a$ from (\ref{geta2e})
is the physically-meaningful radial coordinate.

In cylindrical coordinates, $a^2=R^2+\frac{z^2}{1+e(a)}$.  The $a^2$ grid is
allowed to dynamically update with each time step, meaning that the values of
$a^2$ chosen to be grid points can change at each timestep so that regions of
larger $d\Phi/da^2$ and $d\rho/da^2$ are given a greater density of grid points
\Cite{TWM}.  When $e=0$ then $a\equiv r$; this is the case for the majority of
simulations studied here, and for those the coordinate grid is referred to as the
``$r^2$ grid'' and not as the ``$a^2$ grid''.  The numerical consequences of
having a finite number of coordinate gridpoints are discussed in \S\ref{subsect_gridsize}.

\subsection{The Velocity Dispersion}\label{subsect_vdisp}

The one-dimensional root-mean-squared (\textit{``rms''}) stellar velocity
dispersion $\sigma_o$ 
at a given radius $a$ or $r$ is found similarly to $\vrot$:
\beq\sigma_o^2=(1-\delta)\vrms^2=\frac{(1-\delta)}{p(r)}\int_rdIdJ\,\tilde{f}(I,J)
\lbrack2(E-\left<\Phi\right>_a)-\left<v_T^2\right>_a\rbrack\label{sigma2}\eeq
where $\delta$ is an anisotropy parameter that allows for a difference between
the radial velocity dispersion $\vrms$, and $\sigma_o$.
Subscript ``$a$'' on bracket $\left<\cdot\right>_a$ denotes the possible
averaging\footnote{For $\left<v_{_T}\right>_a$, simple geometry yields
$\langle\frac{1}{r^2}\rangle_{_a}=\frac{2}{3a^2}+\frac{1}{3a^2(1-e^2)}
=\halve{a^2}[1+\frac{e^2}{3(1-e^2)}]$. The average value of the gravitational potential
$\left<\Phi\right>_a$ over the ellipsoid surface must be calculated numerically.}
of the contained quantity over the surface of the isodensity ellipsoid
with semimajor axis $a$.
To wit, note that (\ref{sigma2}) assumes when $\delta=0$,
$\sigma_o$ is identical to $\vrms\equiv\langle v_r^2\rangle^{1/2}$.  This is
necessary given that the restriction of working with a two-dimensional
$f(I_1,I_2)$ means there is not full information on the three-dimensional
velocity distribution, and so we must rely on an assumption regarding the
degree of velocity anisotropy in order to calculate $\sigma_o^2$.  It is also
why it is not advisable to depend on this calculation of $\sigma_o^2$ when
determining either the distribution of inclinations $h(\beta)$ or the amount of stellar rotation; those
quantities are of central importance to the overall model whereas $\sigma_o$
is less so.

The expression (\ref{sigma2}) gives the velocity dispersion for all stellar
mass species considered together, but it is just as easy to perform the same
calculation for a particular stellar mass value by restricting the sums in 
$p(r)$ and $\tilde{f}$ to that single species.

\section{Gravitational Potential}\label{sect_gravpot}

As the simulation progresses, new values for the stellar density distribution
and the gravitational potential must be determined at each timestep in turn.
The new potential at each point on the radial-coordinate grid at a given
timestep can be calculated in a manner analogous to that
used by Cohn \Cite{Cohn79}. We write the (inverted) Poisson equation
\newcommand{\new}{\mathrm{new}}\newcommand{\old}{\mathrm{old}}
\beq\Phi^{\new}=\mathcal{L}^{-1}\rho[f^{\new};\Phi^{\new}]\label{cohn24}\eeq
where $\mathcal{L}^{-1}$ is the inverted Laplacian, derived below.
The ``$\new$'' superscript denotes that, starting with a newly-updated stellar
distribution function $f$, (\ref{homrho}) and then (\ref{cohn24}) are used
to find self-consistent values for $\rho$ and $\Phi$.  Note that the allowed
ranges of $I_1$ and $I_2$ in (\ref{homrho}) depend on $\Phi(R,z)$, hence
the implicit dependence of $\rho$ on $\Phi$ in (\ref{cohn24}).
For the first step in the interation $\Phi^{\new}=\Phi^{\old}$ is typically
used in the right hand side of (\ref{cohn24}); if this fails to converge a
guess for $\Phi^{\new}$ based on a comparison of previous timesteps' potential
functions is also attempted.

As done by Cohn \Cite{Cohn79}, we employ the ``Aitken $\delta^2$'' process
\Cite{Henri} to accelerate convergence.  This procedure is iterated until
convergence is achieved, typically to an accuracy of $\sim2\%$
for every point in the Aitken-processed $a^2$ grid.  A final check of
conservation of overall energy is also made.


When the gravitational potential $\Phi$ is taken to be spherically symmetric
-- as is the case for most of the simulations performed -- the procedure is
conceptually straightforward, if numerically complex.  The mass density
$\rho[f^{\new};\Phi^{\old}]$ is updated from the new distribution function
using (\ref{homrho}) and then the solution of the Poission equation is simply
\beq\Phi^{\new}(r)=-4\pi G\int_r^\infty\rho\,\frac{dr}{r}\label{phir}\eeq
after which $\rho[f^{\new};\Phi^{\new}]$ and $\Phi^{\new}$ are iteratively
solved for until convergence is achieved.

For the more general, but rarely needed case in which there is an overall and
variable ellipticity $e(a)$ to the potential, a somewhat more complex procedure
than given above is required.  One still starts with the inverted Poisson
equation as given in (\ref{cohn24}) but now the derivation of the inverted
Laplacian $\mathcal{L}^{-1}$ is more involved.  Details are given in the
Appendix.

\input{ICs}

%% file: ICs.tex
\section{Initial Conditions}\label{sect_ICs}

\newcommand{\evenf}{\bar{F_q}}
\newcommand{\oddf}{\Theta_q}

It is standard in stellar-dynamical work to first consider a Plummer sphere
distribution \Cite{Cohn79} before moving on to other possible
potential-density pairs.
The Plummer sphere is the reference potential for this study: it fits the
light curves of globular clusters \Cite{Spitzer}.  An alternative choice of
initial potential which fits the surface brightness of galactic spheroids is
the ``$\gamma=0$'' model \Cite{Dehnen}.  These two choices have the shared
advantage of providing well-defined values for the initial distribution
function $f(I,J)$ in addition to the potential $\Phi(r)$ and density $\rho(r)$.

To introduce rotation into either of these density-potential pairs requires
an \adhoc alteration of the distribution function as described below in 
\S\ref{subsect_introrot}, which is followed by descriptions of the stellar bar
model and the possible choices for the stellar initial mass fuction (IMF).

\subsection{Potential-Density Pairs}\label{subsect_potdens}

\subsubsection{Base Model: The Plummer Sphere}\label{plummer}

Previous studies of dense-cluster dynamics have often started with a Plummer
sphere (\Cite{QS90}, 
\Cite{Meza}, 
\Cite{Cohn80}, \Cite{Amaro}),
going as far as refering to nonrotating single-mass Plummer spheres as a
``standard testbed'' \Cite{GFR}.
Following Quinlan and Shapiro \Cite{QS89}, the gravitational potential and stellar mass density
for a Plummer model of total mass $M$ are
\beq\Phi(r)=-\frac{GM}{(r^2+\rcore^2)^{1/2}}\label{icphi}\eeq
and
\beq\rho(r)=\frac{3M}{4\pi\rcore^3}(1+r^2/\rcore^2)^{-\frac{5}{2}}.\label{icrho}\eeq
The corresponding energy-space distribution function for mass species $m_q$ is
\beq\evenf(E)=\frac{24\sqrt{2}N_q}{7\pi^3G^5M^5}\rcore^2E^{7/2}.\eeq
where $N_q$ denotes the total number of stars of individual mass $m_q$.
Note that a general non-rotating distribution function is not necessarily
independent of $J$, but the Plummer model's is.

In particular the specific models first employed by Quinlan and Shapiro \Cite{QS89}, such as those
listed in Table \ref{tabics}, have also been commonly used (\eg as done by Rasio \etal \Cite{RFG})
for later studies of nonrotating systems, and so are modified for use here
as described below in \S\ref{subsect_introrot}.
The models in Table \ref{tabics} have two desirable properties:
the initial relaxation time $t_r<1\sn3$ Myr so the total collapse time is
expected to be $\lesssim1\sn{4}$ Myr; and the escape velocity satisfies
$\vrms<v_\mathrm{esc}\simeq600$km/s and so stellar collisions result
in coalescence of the bulk of the combined stellar material, not disruption and dissipation
of it \Cite{QS90}.  The naming convention is that models with the same middle
digit in their name share a common value for initial velocity dispersion
$\sigma_o(0)$, while the final letter indicates the initial central relaxation
time $t_r(0)$.

Relating the models to astronomical systems, model E1B best describes
globular clusters and is not studied extensively here.  E2A and E2B could be
very-dense globular clusters \Cite{McZwart}, or the nuclei of bulgeless spiral
and dwarf elliptical galaxies (as shown in Freitag \etal \Cite{FRB}, \Cite{FGR}).  The
densities of models E4A and E4B match those of massive young stellar clusters
\Cite{FRB}.  The cores of giant elliptical galaxies are observed to have
line-of-sight velocity dispersions $\sigma_\mathrm{LOS}$ near 350 km/s
\Cite{BM}; this is not quite a match for the $\sigma_o=400$ km/s of models E4A
and E4B.  However, near the observed centers of galaxies $\sigma_\mathrm{LOS}$
is more directly comparable to $\vrms$ (as given by \ref{sigma2}) and it will
be shown in \S\ref{subsect_otherqs} that for model E4B, $\vrms(0)\simeq340$
km/s, a good match with the observations of giant elliptical galaxy cores.
Previous simulations of nonrotating systems have found that a very massive
star -- identified as a possible precursor to an intermediate-mass black
hole -- can form in a system with a velocity dispersion of ``many hundreds of km/s''
\Cite{FAGR}, so the high value for the velocity dispersion should not be an impediment.

\subsubsection{Alternative Model: the ``$\gamma=0$'' Sphere}\label{gamma0}

The potential-density pair for the ``$\gamma=0$'' sphere is
\beq\Phi_\gamma(r)=-\frac{GM}{2\rcore}\left[1-\frac{r^2}{(r+\rcore)^2}\right]\eeq
\beq\rho_\gamma(r)=\frac{3M}{4\pi}\frac{\rcore}{(r+\rcore)^4}\eeq
with distribution function 
\beq\evenf(E)=\frac{3MN_q}{2\pi^3(GM\rcore)^{3/2}}\left((2\epsilon)^\half\,
\frac{3-4\epsilon}{1-2\epsilon}-3\sinh^{-1}\sqrt{\frac{2\epsilon}{1-2\epsilon}}
\right),\ \ \ \epsilon\equiv-\frac{E\rcore}{GM}.\eeq
Of other possible models, note that the potential of Jaffe's \Cite{Jaffe} model, and the density of
Hernquist's \Cite{Hern90} model, diverge as $r\rightarrow0$ and so are
inappropriate -- or at least inconvenient -- for use here.

Carrying over values of $M$ and $r_\mathrm{core}$ from Table \ref{tabics} to
the case of the $\gamma=0$ sphere, only one resulting model has central density
$\rho(0)$ and velocity dispersion $\vrms$ in the required ranges: ``G2A'', \ie
a $\gamma=0$ sphere with the same values for $M$ and $r_\mathrm{core}$ as model E2A.
If instead of equating $r_\mathrm{core}$ one uses common values of the
half-mass radius $r_{1/2}$, a potential model ``G3C'' results with core radius
$r_\mathrm{core}=0.20$ pc.  Model G3C has the same total mass $M=1.1\sn7M_\odot$
and the same half-mass radius $r_{1/2}$ as model E2B, albeit with a very large
initial central density of $\rho(0)=1.4\sn9M_\odot/\mathrm{pc}^3$; this model
is likely of most use as a demonstration case rather than as a representation
of a realistic astronomical system.  However an intermediate model with
$r_\mathrm{core}=0.29$pc (\ie half that of model E2B's $r_\mathrm{core}=0.58$pc)
features a less extreme initial density $\rho(0)=5.0\sn8M_\odot/\mathrm{pc}^3$.  This
model has an initial relaxation time similar to that of model G2A but a somewhat
larger initial central velocity dispersion and is thus labelled ``G3A''.

Numerically calculated values for velocity dispersion $\vrms$, density $\rho$
and relaxation time $t_r$ for these $\gamma=0$ sphere models are given with
the overall simulation results in \S\ref{gamm0}.

\subsection{Introducing Rotation}\label{subsect_introrot}

As described in \S\ref{intro_rot}, rotation due to cosmological tidal torques is
ubiquitous in the universe and is only weakly dependent on how a particular system 
was formed.  Barnes and Efstathiou 
\Cite{BE87} considered various models for formation of objects in the early universe,
and in terms of Peebles' dimensionless spin parameter
\beq\label{lamb}\lambda\equiv\frac{J_\mathrm{rot}|W_\mathrm{grav}|^\frac{1}{2}}{GM^{2.5}}\eeq
they determined that the typical value due to cosmological tidal torques is
$\lambda\simeq0.05$, almost independent of the perturbation spectrum.
At the stellar cluster scale required here this is consistent with the measurement
by Cervantes-Sodi \etal \Cite{CSHPK}, who found for a sample of SDSS galaxies that
$\lambda=.04\pm.005$ with a weak trend of increasing $\lambda$ with decreasing
mass. 

In order to introduce rotation into the simulations, we alter the distribution of
$F_q(E)$ so that the total angular momentum of the cluster $J_\mathrm{rot}$ produces
the desired value for $\lambda$, while conserving $N_q$. This results in a ``tilted''
$F(E,J)$ in which the resulting excess total amount of $J$ is attributed to
$J_\mathrm{rot}$:
\beq\frac{1}{(J_\mathrm{hi}-J_\mathrm{lo})}\int_{J_\mathrm{lo}}^{J_\mathrm{hi}}dJ\,JF(E,J)=\left(1+\frac{J_\mathrm{rot}}{J_\mathrm{tot}(E)}\right)\evenf(E)\label{icfej}\eeq
with the tilted distribution function
\beq F_q(E,J)=(1+Q_\mathrm{rot})\evenf(E)\eeq
in which the amount of tilt $Q_\mathrm{rot}(E,J)$ is given by
\beq Q_\mathrm{rot}(E,J)=\frac{(\zeta+1)J_\mathrm{rot}}{\avg{J}{E}J_\mathrm{tot}}
\left(\frac{2}{\Delta J}\right)^{\zeta-1}\left(J-\avg{J}{E}\right)^\zeta.\label{icq}\eeq
In short, this procedure simply replaces $\evenf(E)$ with a new $F_q(E,J)$ that
conserves number and gives the same total angular momentum, but that has a non-flat $J$-dependence of
the form $(J-\avg{J}{E})^\zeta$ which results in a net total rotational angular
momentum $J_\mathrm{rot}$.

Because the $(I,J)$ grid (on which the Fokker-Planck coefficients are calculated)
is square but the corresponding $(E,J)$ grid (representing the space on
which dynamical quantities must be calculated) is not, we must take care in how $\Delta J$ is determined in the rotating case.  In the above,
$\avg{J}{E}=\half(J_\mathrm{hi}(E)+J_\mathrm{lo}(E))$ is the average value of
$J$ on the grid for a given $E$. Similarly, $\Delta J$ is the span of $J$
values over which $\evenf(E)$ is being tilted; to strictly satisfy
(\ref{icfej}) one would take it to equal $J_\mathrm{hi}(E)+J_\mathrm{lo}(E)$
but in practice this results in values of $E$ which do not span a large range
of $J$ being ``overloaded'' with more than their share of rotation.  Thus
instead\footnote{The cost of this change is that the resulting value for
$\lambda$ of the newly-determined distribution $F(E,J)$ is slightly lower than
intended. This would happen in any case however, as the new distribution also
has a lower value for $|W_\mathrm{grav}|$ than the original $\evenf(E)$ did,
due to the additional overall rotation.  All values for $\lambda$ quoted in
Chapter \ref{chap_results} are ``true'' values in that these two effects have been taken into
account by recalculating (\ref{lamb}).} we simply set
$\Delta J=(J_\mathrm{max}-J_\mathrm{min})$.  Polynomial power $\zeta\geq1$ is a
free parameter that sets the shape of the tilt.  We usually apply a linear
tilt $\zeta=1$.

The new rotating distribution $F(E,J)$ is no longer a solution of the Poisson
equation that matches the Plummer potential and density profiles given by
(\ref{icphi}) and (\ref{icrho}).  This ``problem'' is easily overcome by using
the mechanisms of the main simulation itself, as described earlier in this
chapter, to find the $\Phi(r)$ and $\rho(r)$ that do correspond to the new
$F(E,J)$, before starting the simulation proper based on these initial conditions.

\begin{table}\begin{center}
\begin{tabular}{ccccccc}
Model & $\rho(0)$ & $\sigma_o(0)$ & $M$ 
& $r_\mathrm{core}$ & $t_r(0)$ & Corresponding\\
& [$M_\odot/\mathrm{pc}^3$] & [km/s] & [$M_\odot$]
& [pc] & [yr] & Astronomical object\\ \hline\hline
E4A &$3.0\sn8$ & 400 & $1.8\sn7$ & 0.24 &$4.6\sn7$ & core of a\\
E4B &$1.0\sn8$ & 400 & $3.1\sn7$ & 0.42 &$1.4\sn8$ & giant elliptical galaxy \\ \hline
E2A &$4.0\sn7$ & 200 & $6.2\sn6$ & 0.33 &$4.6\sn7$ & nucleus of dwarf elliptical
\\
E2B &$1.3\sn7$ & 200 & $1.1\sn7$ & 0.58 & $1.4\sn8$ &or bulgeless spiral galaxy
\\ \hline
E1B &$1.8\sn6$ & 100 & $3.6\sn6$ &0.79  &$1.4\sn8$ & globular cluster\\
\end{tabular}\end{center}
\caption{The set of initial condition models. 
For purposes of the numerical simulation, the quantities taken as fundamental
parameters are total mass $M$ and core radius $r_\mathrm{core}$.  The central
density $\rho(0)$, velocity dispersion $\sigma_o(0)$ and relaxation time
$t_r(0)$ listed here are the resulting analytic values for a Plummer sphere
distribution.  Model names are taken from Quinlan and Shapiro \Cite{QS89}.}
\label{tabics}\end{table}

\subsection{Initial Mass Function}\label{subsect_IMF}

The spectrum of masses which is input to the model is expected to play a large role
in its dynamical evolution \Cite{GFR}.  A simple power-law
$dN/dm\propto m^{-\alpha}$ represents perhaps the simplest form for the
initial mass function (IMF).  The traditional Salpeter IMF has $\alpha=2.35$.
Other observationally-determined IMFs include the Miller-Scalo \Cite{GFR}:

\beq\label{MSIMF}\frac{dN}{dm}\propto\left\{\begin{array}{ll}
     m^{-1.4},&\mbox{$0.1<m/M_\odot<1.0$}\\
     m^{-2.5},&\mbox{$1.0<m/M_\odot<10$}\\
     m^{-3.3},&\mbox{$10<m/M_\odot$}
\end{array}\right.\eeq
and the Kroupa IMF \Cite{Kroupa}:
\beq\label{KrIMF}\frac{dN}{dm}\propto\left\{\begin{array}{ll}
     m^{-1.3},&\mbox{$0.08<m/M_\odot<0.5$}\\
     m^{-2.2},&\mbox{$0.5<m/M_\odot<1.0$}\\
     m^{-2.7},&\mbox{$1.0<m/M_\odot$}
\end{array}\right.\eeq

A problem with the above IMFs is that none was determined for the specific
case of dense clusters near centers of galaxies.  However conditions in the
centers of galaxies are sufficiently different from those elsewhere that all but
the highest-density molecular clouds (number density $n>10^4/\mathrm{cm}^3$)
will be shredded by tidal forces \Cite{Figer1}.
Recent observational work on such IMFs -- most
notably for the Arches cluster located 30pc from the center of the Milky Way, as well
as the Quintuplet and Central clusters, and R136 in the Large Magellanic
Cloud, has led to the following picture of the IMF in the centers of galaxies
(Figer \Cite{Figer1}, \Cite{Figer2}, Oey and Clarke \Cite{Oey}, and references therein, but see
Elmegreen \Cite{ElmeNot} for a dissenting opinion):
\begin{itemize}
\item galaxy-center IMFs are top-heavy, with $\alpha\simeq1.7-1.8$ for
$M\gtrsim\mathrm{(6-10)}M_\odot$, or $\simeq1.8-1.9$ when differential
extinction is taken into account;
\item stellar masses below $\sim2M_\odot$ do not add appreciably to the total
mass of the stellar cluster;
\item there is a fundamental, initial upper mass cutoff of $\sim150M_\odot$;
\item central densities can reach $\simeq10^6M_\odot/\mathrm{pc}^3$.
\end{itemize}
Notable exceptions to the high-mass cutoff include the Pistol Star and its
``twin'' FMM362 in the Quintuplet cluster; despite a cluster age of at least
4 Myr these stars have masses in the $150-200M_\odot$ range with an expected
stellar lifetime of $2.5-3$ Myr.  One explanation is that these stars are the
product of stellar mergers of stars in the $100M_\odot$ range \Cite{Figer1}.  

Thus an IMF for the Arches cluster, representative of dense clusters near
the centers of galaxies, can be expressed as
\beq\frac{dN}{dm}\propto\left\{\begin{array}{lrl}
     m^{-1}           ,&2\lesssim&\!\!\!\!m/M_\odot<8\pm2\\
     m^{-1.8\pm0.1}   ,&8\pm2<&\!\!\!\!m/M_\odot\lesssim150
\end{array}\right.\eeq
and a simplified version using somewhat overly-conservative choices of
parameters is
\beq
\frac{dN}{dm}\propto\begin{array}{ll}m^{-1.9},&2<m/M_\odot<150.
\end{array}\label{ArchesIMF}\eeq
It is the above IMF that is used for the ``Arches'' simulations in this study.
Some other cases here use the Salpeter IMF of $\alpha=2.35$ but most start
with the Kroupa IMF which for a lower-mass bound of $1M_\odot$ corresponds
simply to $\alpha=2.7$; both of these were also used by Amaro-Seoane \Cite{Amaro} for an
analogous study of dense but nonrotating clusters which had already formed
massive central objects, \ie black holes or supermassive stars.







With all relevent dynamical quantities now calculable, the Fokker-Planck
equation's coefficients for the system can now be determined; the validity
of the overall simulation technique and its various components tested; and
the actual simulations performed using the initial conditions given above.
These steps comprise the next three chapters respectively.

%% file: Coeffs.tex
\chapter{Derivation of the Fokker-Planck Diffusion Coefficients\label{chap_coeffs}}

\newcommand{\fpf}{f}
\section{The Orbit-Averaged Fokker-Planck Equation}\label{sect_oafp}

As described in Chapter \ref{chap_pot},
it is not advantageous to use energy $E$ and angular momentum $J$ (or its vertical
compenent $J_z$) for the integrals of the motion here as prior two-dimensional,
spherically-symmetric studies have done \Cite{Cohn79}; any change in the
gravitational potential of the system will cause changes in the distribution function when it is expressed as
$F(E,J)$ or as $F(E,J_z)$.  What is required are the radial action $I_1$ and
tangential action $I_2\equiv J$, which are adiabatic invariants as the potential
evolves -- and so also is $f(I_1,I_2)$ given the assumption of weak encounters.
As given by Binney and Tremaine \Cite{BT}, 
orbit-averaging is then a simple averaging of quantities over the
$2\pi$ change in (an) angle variable.  It is customary to average over the
radial angle.  The orbit-averaged Fokker-Planck equation then takes a
particularly simple form:
\beq\label{simplefp}\di{}{t}\fpf(I_i,t)=-\di{}{I_j}[\fpf\J\,]+\half \ddi{}{I_i}{I_j}[\fpf\IIJ\,]\eeq
where $f(I_i,t)=f(I_1,I_2,t)$ is the distribution function 
and $\J, \IIJ$ are the drift and diffusion coefficients. 
Summation over repeated indices $i$ and $j$ is implied.

Physically, the Fokker-Planck equation can be understood as a collisionless
Boltzmann equation $df/dt=0$ with a collisional term
added\footnote{``Collisional'' here refers to the relatively distant 
gravitational encounters that dominate the overall dynamical scattering, and
is not to be confused with the close-range ``collisional mergers'' of stars
which are treated in \S\ref{mergerrates}.} to account for particles (here, stars)
scattering in and out of a given volume of phase space.  When the collisional
term is expanded in a Taylor series and truncated after the second order term,
the remaining two terms are the drift and diffusion coefficients; as stated by
Binney \& Tremaine \Cite{BT}, they describe the expected rate of change in
$I_i$ or $I_iI_j$ for a given test star with actions $(I_i,I_j)$.
When the coefficients satisfy the relations
\beq\IIJ=\II{j}{i}\label{symdij}\eeq
and
\beq\J=\half\di{}{I_i}\IIJ\label{fpfc}\eeq
then the Fokker-Planck equation can be recast in a flux-conservative form
\beq\di{}{t}\fpf(I_i,t)=\half \di{}{I_j}[\IIJ\,\di{}{I_i}\fpf]\label{fcfp}.\eeq
It will be seen below that this flux-conservative form has a distinct
advantage when it comes to finite differencing.

The Fokker-Planck approximation remains valid as long as number 
$N=\int dI_1dI_2f\gg1$, and requires that the time step $\Delta t$ satisfy
$t_{dyn}\ll\Delta t\ll t_{r}$. We have a dynamical time (as defined by
Tremaine and Weinberg \Cite{TW}) of $t_{dyn}\equiv\sqrt{3\pi/16G\rho}\lesssim10^4$ yr, and a
relaxation time of $t_{r}>10^6$ yr or more, and so in practice finding an
appropriate size of timestep $\Delta t$ is not difficult.\footnote
{For a given test star, the relaxation time is given by
$t_r=\max_{ij}(\frac{I_iI_j}{D_{ij}})$.  In Chapter \ref{chap_results}
the median relaxation time \Cite{QS90} of
$t_{rq}=\vrms^3/[4\pi\sqrt{3/2}G^2m_q^2\ln(0.4N)]$ is used instead,
in order to indicate the relaxation time of stars of a particular stellar
mass in position space, instead of in action-space as considered here.}

Generalizing to the case of multiple distribution functions $f_q$ and
adding \adhoc terms $L_q, G_q$ to account for losses and gains due to
stellar mergers, as well as analogous parameters $B_q$ and $R_q$ for
stellar births and deaths (``remnants'') due to stellar evolution,
the most-general Fokker-Planck equation in this study is
\beq\di{}{t}\fpf_q(I_i,t)=\half\di{}{I_j}[\IIJ\,\di{}{I_i}\fpf_q]-L_q+G_q-B_q+R_q
\label{fullfp}\eeq
in which subscript $q$ refers to a given subpopulation of stars labeled
``$q$'', usually distinguished by being of mass $m_q$.

The remainder of this chapter develops the physics of the various terms on the
right hand side of (\ref{fullfp}).  First the general perturbation potential
is converted into action space for use in the diffusion coefficients $\IIJ$.
The specific case of field-star perturbers is developed, which is then
used to derive expressions for $\IIJ$.  The other relevent perturbation
potential, that of a stellar bar, is then constructed and its diffusion
coefficients also developed.  How to finite-difference the Fokker-Planck
equation using these diffusion coefficients follows, and expressions
for the stellar merger loss and gain terms $L_q$ and $G_q$ are then derived.
The chapter closes with a discussion of the possible effects of binary heating
on the model and the reasons for not including it in the calculation; despite
its omission from the Fokker-Planck model, the amount of possible binary
heating that could have occured is still tracked througout the simulations
in order to confirm that the reasons for omitting it remain valid.

\section{The Perturbing Potential}\label{coeffs_pert}

\subsection{General form of the expansion}\label{subsect_pert}

The drift and diffusion coefficients describe the rate of transfer of action
(momentum) to a star due to interactions with other individual stars or with
a bulk stellar bar perturbation.  To do so, first it is necessary to
express the perturbing potential $\Phi_*$ in action space.  This has been done
by Tremaine and Weinberg \Cite{TW}; the following summarizes their results.  Starting in traditional
spherical coordinates $\Phi_*$, when expanded in spherical harmonics is
\beq\Phi_*=\sum_{l=0}^\infty\sum_{m=-l}^l\Phi_{lm}(r)Y_{lm}(\theta,0)\,
               \ex{im(\phi-\Omega_*t)}.\label{tw47}\eeq
(Label ``$*$'' can refer to any perturbing mass, \ie either to a stellar
bar as a whole or to a single field star.)
Expressed in action space, (\ref{tw47}) becomes
\beq\Phi_*=\sum_{m=0}^\infty\sum_{k,n=-\infty}^{\infty}\Psi_{knm}(I_1,I_2)\,
               \ex{i(kw_1+nw_2+mw_3-\Omega_*t)}\label{tw55}\eeq
where
\beq\label{tw56}\Psi_{knm}=\sum_{l=0}^\infty\left(\frac{2}{1+\delta_{m0}}\right)
            V_{lnm}(\beta)W_{klnm}(I_1,I_2)\eeq
and in which $\beta$ is the perturber's orbital inclination (thus $\beta=0$ for
a bar in the rotational plane of the system). Quantity $V_{lnm}(\beta)$ contains the effect of rotating the frame of
reference so that it is aligned with the orbital plane, and $W_{klnm}(I_1,I_2)$
is the strength of the interaction of the $(k,n,m)$ resonance for multipole
expansion term $l$, as will be seen below\footnote {For completeness' sake, we
note here that $V_{lnm}(\beta)=(-)^{(m-n)/2}r_{lnm}(\beta)Y_{ln}(\halfpi,0)$,
where $r_{lnm}$ is the Slater rotation matrix $r_{lnm}=\sum_t(-)^t
\frac{\sqrt{(l+n)!(l-n)!(l+m)!(l-m)!}}{(l-m-t)!(l+n-t)!t!(t+m-n)!}
\tan^{2t+m-n}(\halve{\beta})\cos^{2l}(\halve{\beta})$ \Cite{TW}.  The sum is
taken over all $t$ such that the factorials' arguments are all positive 
semidefinite, and similarly $r_{lnm}\equiv0$ if any of $(l\pm n)$ or $(l\pm m)$
is negative.  
}.  In particular, $W_{klnm}$ has the form
\beq W_{klnm}(I_1,I_2
)=\opi\int_0^\pi dw_1\,\cos[kw_1-n(\psi-w_2)]\Phi_{lm}(r)\label{tw53}\eeq
in which one may note that angles $w_1$ and $(\psi-w_2)$ are functions of $r$ 
as given by (\ref{tw37}) and (\ref{tw38}).  By inverting
(\ref{tw37}) one obtains $r(w_1)$ and by extension $(\psi-w_2)(w_1)$ as
required for use in (\ref{tw53}). Various symmetries of spherical harmonics
and in the expansion are of relevance:
\begin{itemize}
\item $\Phi_*$ is real, and all perturbers studied have real $\Phi_{lm}$ for
all $(l,m)$, so $\Phi_{l-m}=(-)^m\Phi_{lm}$;
\item the system is symmetric about the $z=0$ plane, so $(l+m)$ must be even;
\item $V_{lnm}(\beta)=0$ unless $(l+n)$ is even;
\item the above two items result in $V_{lnm}$ being real for all $(lnm)$,
and so $\Psi_{knm}$ is real for all values of $(knm)$;
\item $V_{l-n-m}=(-)^mV_{lnm}$ and $W_{-kl-n-m}=(-)^mW_{klnm}$, and so
$\Psi_{-k-n-m}=\Psi_{knm}$.
\end{itemize}


Note that the choice of radial basis functions $\Phi_{lm}(r)$ is not
unique \Cite{Meza}; the current method is most convenient for dealing
with inclined orbits while still allowing development of the diffusion
coefficients in terms of orbital resonances, below.

\subsection{Effect of Orbital Inclination\label{coeffs_orbinc}}

Many of the physical quantities that determine the diffusion coefficients
depend either on orbital inclination $\beta$ and/or the polar angle of the
curent position $\theta$.  For the former dependence we invoke the orbital
inclination distribution $h(\beta)$ developed in $\S\ref{orbinc}$.
Quantities such as $V_{lnm}(\beta)$ are then averaged over
$(0,\beta_\mathrm{max})$ with weighting function $h(\beta)$; notationally this
will be denoted by angle brackets with subscript $h(\beta)$.  Thus for any
quantity $y$
\beq\avgb{y}\equiv\frac{\pi}{\beta_\mathrm{max}}\int_0^{\beta_\mathrm{max}}d\beta\,h(\beta)\,y(\beta).\eeq
Normally $\beta_\mathrm{max}=\pi$, which accounts for both prograde and
retrograde orbits. 
As the $\beta$ dependence of $V_{lnm}(\beta)$
is strictly trigonometric, if the form of $h(\beta)$ allows it and if no
part of $\Phi_{lm}$ as contained in $W_{klnm}$ depends on $\beta$, the
averaging of (\ref{tw56}) can be done analytically; else it must be done
numerically. 
This will be of help later, when weak approximations will be all that are
needed to satisfy the requirements for doing analytical averaging.

A further example is given by the square of (\ref{tw56}):
\beq\label{avgpsq}\overline{\Psi^2}_{knm}\equiv\left(\frac{2}{1+\delta_{m0}}\right)^2
\avgb{\left[\sum^\infty_{l=0}V_{lnm}(\beta)W_{klnm}(I_1,I_2)\right]^2}.\eeq
The above $\overline{\Psi^2}_{kmn}$ is what will be used in the diffusion
coefficients below.  To calculate the average, the squared sum is expanded
and evaluated term by term; this is more efficient than averaging over the
entire squared sum at once.  For convenience, the triple $(knm)$ will be
written as $(\elll)$ in most sums.

The direct calculation of (\ref{avgpsq}) takes an
unacceptably long time, largely due to the complexity of the calculation of
$\Phi_{lm}(r)$ as will be seen below; it is sped up by evaluating the
integrands at preset values of $\theta$ and $w_1$, and interpolating from
those values when performing the integrations.  Similarly, solving (\ref{tw37}) for
$r(w_1)$ and then integrating (\ref{tw38}) for $w_2-\psi$ each time they are
needed in (\ref{tw53}) would be computationally prohibitive, and so they are
also calculated on a grid and interpolated for.

The ability to transform away the $\theta$ dependence of $Y_{lm}(\theta,0)$ in
going from (\ref{tw47}) to (\ref{tw55}) notwithstanding, other quantities that
depend not on $\beta$ but instead on $\theta$ directly require a slightly more
subtle treatment.  The approach here is to average any such quantity over its
allowed range of $\halfpi-\beta\leq\theta\leq\halfpi$ before doing the
averaging over $\beta$ itself.  Allowing there to also be a purely
$\beta$-dependent factor $y$:
\beq\kaplan{y}{x}=\frac{\pi}{\beta_\mathrm{max}}\int_0^{\beta_\mathrm{max}}d\beta\,h(\beta)y(\beta)
\,\frac{2}{\pi}\int_0^\halfpi d\psi\,x(\theta(\psi,\beta))\label{prekap}.\eeq
The inner averaging is actually performed over $\psi$ instead of $\theta$
because, averaged over all stellar orbits of a given $(I,J)$, $d\psi/dt$ is a
constant whereas $d\theta/dt$ is not.  (See (\ref{tw38}) and note that the
phases of $w_1$ and $w_2$ are independent; Figure \ref{orbangles} gives a
visual depiction.)  To convert between $\theta$ and $\psi$, (\ref{tw41}) is used.

\section{Field Star Perturbers}\label{sect_fieldpert}

For an individual star of mass $m_*$ (not to be confused with spherical
harmonic index $m$) at position 
$(r_*,\,\theta_*,\,\phi_*=\Omega_*t+\phi_o)$ in the field, the potential 
at $(r,\theta,\phi)$ is
\begin{eqnarray}\Phi_*(\mathbf{r};\mathbf{r_*})&=&-\frac{Gm_*}{|\mathbf{r}-\mathbf{r_*}|}\\
&=&-Gm_*\sum_{l=0}^\infty\frac{r_<^l}{r_>^{l+1}}\frac{4\pi}{(2l+1)}
\sum_{m=-l}^l(-)^mY_{lm}(\theta,\phi)Y_{lm}(\theta_*,\phi_*)\nonumber\\
&=&-Gm_*\sum_{l=0}^\infty\frac{r_<^l}{r_>^{l+1}}\frac{4\pi}{(2l+1)}
\sum_{m=-l}^l(-)^mY_{lm}(\theta,0)Y_{lm}(\theta_*,0)
e^{im(\phi-\phi_*)}\nonumber
\eea
with the traditional $r_<$ and $r_>$ respectively being the smaller and
larger of $r$ and $r_*$ .  Initial phase $\phi_o$ can be absorbed into the
initial phase of canonical angle variable $w_3$ and so has already been dealt with in the previous
section.  Comparison with (\ref{tw47}) thus yields
\beq\label{philm}\Phi_{lm}(r;\mathbf{r_*})=
-Gm_*\frac{r_<^l}{r_>^{l+1}}\frac{4\pi}{(2l+1)}(-)^mY_{lm}(\theta_*,0).
\eeq
The field star distribution is defined as the overall cluster distribution,
excluding those stars specifically assigned to a stellar bar (which is labeled
``$B$'' to denote that it is a bulk perturber, different from individual stars)
and so is expressed in terms of the canonical actions:
$f_*(I'_1,I'_2)=\sum_qf_q(I'_1,I'_2)-f\barb(I'_1,I'_2)$.  (Here, primes
indicate actions of the perturbing field star, and not of the stellar orbit
being perturbed.)

To deal with the undesired perturber radial coordinate $r_*$, we perform an
additional orbit-averaging of the field star's 
contribution to the overall perturbing potential as given by (\ref{philm}),
this time averaging over the perturbing field star's orbital range.
It is more straightforward to directly average over $r_*$ weighted by
$1/|v_{r*}|$ than to convert to the corresponding canonical angle $w'_1$
first, which would accomplish the identical task but require an additional
step.  Both $Y_{lm}(\theta_*,0)$ and $v_{r*}$ depend on $\theta_*$ (the latter
via $\Phi(\mathbf{r_*})$) which can at the same time be averaged over using the
technique of (\ref{prekap}):
\bea\lefteqn{\avg{1;\avg{\Phi_{lm}(r;\mathbf{r_*})}{w'_1}}{\theta_*}
=\avg{1;
\int_{r_{p*}}^{r_{a*}}dr_*\frac{\Phi_{lm}(r)}{|v_{r*}|}
  \Bigg/
\int_{r_{p*}}^{r_{a*}}\frac{dr_*}{|v_{r*}|}
}{\theta_*}}
\label{dblavg}\\
&=&-\frac{4\pi Gm_*}{(2l+1)}(-)^m\avg{1;Y_{lm}(\theta_*,0)
\int_{r_{p*}}^{r_{a*}}
  \frac{dr_*\ r_<^l/r_>^{l+1}}{\vrsqp{*}{I'_2}{\mathbf{(r_*)}}^\half}
\Bigg/
  \int_{r_{p*}}^{r_{a*}}\frac{dr_*}{|v_{r*}|}
}{\theta_*}.\nonumber
  \eea
Of course for orbits that are very nearly circular ($r_{p*}\simeq r_{a*}$),
$r_*$ is well-defined for any given value of $I'_1$ (noting $I'_2\gtrsim0$)
and so the above averaging over $r_*$ is unnecessary; one can then simply
take $\Phi_{lm}(r)=\Phi_{lm}(r;r_*)$.  Otherwise the full form of
(\ref{dblavg}) is required, and is what is used for the $\Phi_{lm}(r)$
factor in (\ref{tw53}).

\section{The Diffusion Coefficients}\label{coeffs_coeffs}

With knowledge of the perturbation coefficients $\Psilll$ we can now proceed to
compute the drift and diffusion coefficients $\J$ and $\IIJ$.  The former were
derived by Van Vleck \Cite{VV} in a quantum mechanical context, although a more related
calculation is that of Tremaine and Weinberg \Cite{TW} who derived the dynamical friction exerted on
an external satellite by a stellar system. Our situation is analogous, but
requires the 
generalization of considering both components of the action instead of only
$I_2$.  Thus the derivation of $\IIJ$ here follows a similar form.

Denoting first-order approximations by $\Delta_1$, Hamilton's equations for
the system are
\beq\Delta_1I_j=\di{\Chi}{w_j},\ \ \Delta_1w_j=-\di{\Chi}{I_j}\eeq
with generating function
\newcommand{\xfrac}[2]{\frac{\ex{i\reswtp{#1}{#2}}}{i\reswp{#1}{#2}}}
\beq\Chi=-\Real\left[\sumall\psilp{}\xfrac{}{p}\right]\eeq
in which $\Omega_p=\frac{dw_p}{dt}$, and
$\omega=\ell_3\Omega_*+i\eta$ is the frequency of the perturbation potential
including a slow ``turning on'' of the perturber in the distant past effected
by $\eta>0$.
Throughout this section repeated index $p$ will indicate summation over all
possible values 1,2,3 of the index, and the same character $p$ when used
in different factors indicates distinct sums, \eg
$(\ell_p\Omega_p)(\ell'_p\Omega_p)=
(\ell_1\Omega_1+\ell_2\Omega_2+\ell_3\Omega_3)
(\ell'_1\Omega_1+\ell'_2\Omega_2+\ell'_3\Omega_3)$.
So, to second order
\newcommand{\fcomp}[3]{\frac{#1}{\resp{#2}{#3}^2+\eta^2}}
\newcommand{\freal}[2]{\fcomp{\resp{#1}{#2}}{#1}{#2}}
\newcommand{\fimag}[2]{\fcomp{\eta}{#1}{#2}}
\newcommand{\fbase}[5]{\fcomp{\eta#4\reswop{#1}{#2}
                            #5\resp{#1}{#2}#3\reswop{#1}{#2}}{#1}{#2}}
\newcommand{\fboth }[2]{\fbase{#1}{#2}{\cos}{\sin}{-}}
\newcommand{\fderiv}[2]{\fbase{#1}{#2}{\sin}{\cos}{+}}
\bea
&&\Delta_1I_i\,\Delta_1I_j=\di{\Chi}{w_i}\ \di{\Chi}{w_j}\\
 &&=\Real\left[\sumall\ell_i\Psilll\xfrac{}{p}\right]
    \Real\left[\sumalp{'}\ell'_j\Psilll'\xfrac{'}{p}\right]\nonumber\\
 &&=e^{2\eta t}\sum\sum\ell_i\Psilll\left[\fboth{}{p}\right]\nonumber\\
 &&\ \times\sum{}'\sum{}'\ell'_j\Psilll'\left[\fboth{'}{p}\right]\nonumber
\eea
where for notational convenience, a prime on a quantity indicates that any
indices it takes are primed, \ie $\omega'\equiv(\ell'_3\Omega_*+i\eta)$ \etcend.
To second order, the rate of change of ($\Delta I_i\,\Delta I_j$) is
\newcommand{\Psibar}{\overline{\Psi}}
\bea
\label{cosums}\frac{d}{dt}\lefteqn{[\Delta_1I_i\,\Delta_1I_j]=
2\eta\left[\Delta_1I_i\,\Delta_1I_j\right]+e^{2\eta t}}\\
&\times&\left[\sum\sum\ell_i\Psibar\resp{}{p}\fderiv{}{p}\right.\nonumber\\
&&\times\left.\sum{}'\sum{}'\ell'_j\Psibar'\fboth{'}{p}\nonumber\right]\nonumber\\
&+&e^{2\eta t}\left[\sum\sum\ell_i\Psibar\fboth{}{p}\right.\nonumber\\
 &\times&\left.\sum{}'\sum{}'\ell'_j\Psibar'\resp{'}{p}\fderiv{'}{p}\nonumber\right]
\eea
in which for brevity $\Psibar\equiv\Psilll$ and $\Psibar'\equiv\Psilll'$.

The diffusion coefficient for an individual perturber is $\frac{d}{dt}(\Delta_1I_i\,\Delta_1I_j)$, averaged
over all initial phases of the $w_p$; as these are assumed to be distributed
randomly, most of the terms in (\ref{cosums}) vanish or cancel out.  Considering
each of the three ``grand sums'' (\ie each quantity contained in outer sets of
brackets in \ref{cosums}) separately, this happens by the following reasons:
\begin{enumerate}
\item all terms without either $\ell_p=\ell'_p$ for all $p$, or else
$\ell_p=-\ell'_p$ for all $p$, vanish upon the averaging;
\item all cross terms of the form $[\pm\sin\reswop{}{p}\cos\reswop{}{p}]$ also
vanish upon averaging over (any) initial phase angle;
\item $\ell_p=-\ell'_p$ requires that $\ell_3=0$; thus in each grand sum,
for each term with a given $\ell_1=-\ell'_1$ and $\ell_2=-\ell'_2$ there
is an equal-magntitude, opposite signed term with $\ell_1=\ell'_1$ and
$\ell_2=\ell'_2$, and so all terms with $\ell_3=0$ cancel out;
\item in the second and third grand sums, the remaining terms can all be
grouped into compound terms having angle dependence of the form
$[\cos^2\reswop{}{p}-\sin^2\reswop{}{p}]$, which also
vanishes upon averaging over initial phase.
\end{enumerate}

With the above, the only remaining terms are those from the first grand sum
having $\ell_p=\ell'_p$ and $\ell_3>0$, and excluding cross-terms.
Using angle brackets to denote averaging over initial phases, the reduced 
form of the averaged perturbation is thus
\bea\label{predij}
\lefteqn{\left<\frac{d}{dt}[\Delta_1I_i\,\Delta_1I_j]\right>_w
=2\eta\,e^{2\eta t}\times}\\
&&\suminf_{\ell_3=1}\suminf_{\ell_1,\ell_2=-\infty}
\ell_i\ell_j\Psilll^2\left<\left[\fboth{}{p}\right]^2\right>_w\nonumber\\
&&=\eta\,e^{2\eta t}\sum\sum\ell_i\ell_j\Psilll^2
\left[\frac{1}{\resp{}{p}^2+\eta^2}\right]\nonumber\\
&&=\eta\,e^{2\eta t}\sum\sum\ell_i\ell_j\,\frac{\Psilll^2}{|\resp{}{p}-i\eta|^2}
\nonumber\eea
in which $\left<\sin(\ell x+b)\cos(\ell x+b)\right>_x=0$ and
$\left<\sin^2(\ell x+b)\right>_x=\left<\cos^2(\ell x+b)\right>_x=\half$
have been used to eliminate the trigonomic factors.

The analogous derivation of the drift coefficients
performed by Tremaine and Weinberg \Cite{TW} found that the effect of the
first-order perturbation vanishes, and that to lowest order
\beq\left<\frac{d}{dt}[\Delta_2I_j]\right>_w=\half\eta\,e^{2\eta t}\label{tw63}
\sum\sum\ell_i\ell_j\,\di{}{I_i}\frac{\Psilll^2}{|\resp{}{p}-i\eta|^2}.
\eeq
From (\ref{predij}) and (\ref{tw63}) it can be seen that (\ref{symdij}) and
(\ref{fpfc}) do indeed hold; from here on the flux-conservative form
of the Fokker-Planck equation (\ref{fcfp}) will be assumed and no further explicit
consideration of the drift coefficients is required.

In order to be used in the Fokker-Planck equation, the limit
$\eta\rightarrow0$ must be taken, which allows us to use the relation
$\delta(x)={\opi}\lim_{\eta\to0}\,\eta|x-i\eta|^{-2}$.  Thus, one obtains
\beq\left<\frac{d}{dt}[\Delta_1I_i\,\Delta_1I_j]\right>_w=\pi\suminf_{\ell_3=1}
\suminf_{\ell_1,\ell_2=-\infty}\ell_i\ell_j\Psilsq\delta\resp{}{p}.
\label{bardij}\eeq

To obtain the full diffusion coefficient one must integrate over the full
phase space of the perturber.  With 
$f_*(I'_1,I'_2)$ denoting the
distribution function of the perturbers under consideration
and invoking the $\beta$-averaged expansion coefficients of the perturbing
potential from (\ref{avgpsq}) one obtains
\footnote{The expression here differs from the one derived by Tremaine and Weinberg \Cite{TW}
who additionally had
an integral over $dJ_z$, but in our case that's already been taken care of
through the averaging over $\beta$.  Also, $f_*$ here is already ``orbit
averaged'' and no summing over the ranges of $w_1$ and $w_2$ is needed.}
\bea\label{dij}
\IIJ&=&\int d^2I'\,f_*(I'_1,I'_2)
\left<\frac{d}{dt}[\Delta_1I_i\,\Delta_1I_j]\right>_w\\
&=&\pi\int d^2I'\,f_*(I'_1,I'_2)\suminf_{\ell_3=1}
\suminf_{\ell_1,\ell_2=-\infty}\ell_i\ell_j\Psilsq\delta\resp{}{p}.\nonumber
\eea


Note that in this formalism the effect of the dynamical friction which
underlies the interactions is expressed in terms of resonances between
the orbital frequencies of any star
and the secular frequency of the perturbing potential (which is
either that of the bar, or the sum of all potentials of field stars of a
given orbital frequency $\Omega_*$).  In practice, for a field star perturber,
$\IIJ$ is evaluated by using the delta-function that describes the resonance
to eliminate one of the action integrals and then
integrating over the other.  For the perturbation due to
a stellar bar (for which $\Omega_*$ is taken to be a common value for all stars
comprising the bar), the resonance delta-function does not integrate out in
(\ref{dij}) and will have to be handled separately as will be seen below.

\subsection{Mass Segregation}

Because $\Psilll$ scales linearly with $m_*$, and $f_*$ is the number-weighted
distribution function, one can see that $\IIJ$ also scales with $m_*$;
this results in the same mass
segregation effect as described by Quinlan and Shapiro \Cite{QS90} in which the tendency towards
equipartion of energy causes lighter stars to receive kinetic energy from
heavier stars, which then preferentially sink towards the center.

Mass segregation is observed in globular clusters, and can be qualitatively
described using energy equipartition.  The segregation can be unstable
in that a population of more-massive stars decouples dynamically from the
lower-mass population if no equipartition is possible; whether this happens
depends on $m_*'/m_*$ and the ratio of total masses of the two populations
\Cite{FJZR}.
Mass segregation is one example of the complex dynamics that can arise when
there is a spectrum of masses in the system.

As a more specific example, McMillan \etal \Cite{MBZHM} calculate that if $m_*'\leq20M_\odot$,
then a subpopulation forms in a globular cluster core in $\simeq0.2$ of the
half-mass relaxation time $t_\mathrm{rh}$, and the central density of the
larger-mass stars increases by 2-3 orders of magnitude for ``typical IMFs'';
in denser clusters it is possible for this to occur before many stars evolve
to supernovae.  Even a modest range of masses can produce enough mass
segregation to greatly increases the rate of core collapse \Cite{GFR}).
The minimum central density determined by McMillan \etal as leading to 
mass-segregation-driven core collapse in globular clusters is similar
to that used in some of the cases studied here.

\subsubsection{Is Mass Segregation Primordial, or Dynamic?}

Whether mass segregation in stellar clusters is primordial or is the result of
dynamical evolution is a matter of some debate: N-body simulations of globular
cluster dissolution performed by Baumgarft \etal \Cite{BdMK} on the specialized
GRAPE6 computing engine matched observed properties of globular clusters when
an initial mass segregation was assumed and did not not match when no initial
segretation was included; they also deduced a ``near-universal'' mass function
for 
low-metalicity star formation environments. 
However, using the same ``NBODY4'' code McMillan and Portegies Zwart \Cite{McZwart} found no firm evidence
for {\sl a priori} mass segregation in young dense clusters.  In Fokker-Planck
(nonrotating) models of individual clusters, Amaro-Seoane \Cite{Amaro} found no
evidence of initial mass segregation while acknowledging that other models for
young clusters had predicted it.  While none of the above studies looked at exact
analogues of the early-universe stellar clusters studied here, the conservative
approach is still not to assume any initial mass segregation, so that any mass
segregation observed in the simulations is the result of the dynamical friction.
Collisional mergers of large stars can also be a source of effective mass
segregation, in that mergers should preferentially occur in the more-dense
central region of the system.



\section{The Stellar Bar: General Considerations}\label{barcons}

As stated in the Chapter \ref{chap_intro}, although some angular momentum is
expected to be transported outwards via the effects of shear between the
higher-$\Omega_2$ inner regions and the lower-$\Omega_2$ outer regions, a
bulk perturbation in the potential is predicted to be much more effective
than individual stars at transporting angular momentum, as indicated by the
appearance of the square of the perturbation coefficient $\Psilsq$ in the
overall Fokker-Planck diffusion coefficient (\ref{dij}).

Most work on stellar bars, both observed and modeled, has been on the scale
of entire galaxies.  But ``inner bars'' are common in observed samples of
galaxies even though the formation and dynamics of nested bars is still
poorly understood \Cite{Shlos0}.  Even for galactic-scale bars few studies
address the issue of ``strength'' of the bar, in terms of the overall mass
associated with the bar perturbation.  Typical values for the strength are in
the range of 5-15\% of the overall mass available in the relevant volume: for
their N-body simulations Weinberg and Katz \Cite{WK} assumed the bar
comprised 15\% of the disk and halo mass within the bar's radius (which they
took to be half the corotation radius, although their results were
not sensitive to this choice); and Athanassoula \Cite{Athan} found in N-body
simulations that a stellar bar quickly grew to $\simeq5\%$ of the mass within
the radius of their stellar disk, regardless of whether the potential within
that radius was dominated by the disk or by the halo.

In terms of orbital composition, bars are elongated along constituent orbits'
long axes \Cite{CE}, and chaotic orbits contribute to the bar's overall
makeup \Cite{Athan}.  Chaotic orbits are not considered here, but this is in
line with stellar dynamics as practiced
since the 1940s in which the chaotic effects of N-body interactions
are ignored in general \Cite{Gurz}.\footnote{Gurzadyan \Cite{Gurz} goes on to
state that it would be expected from ergodic theory that N-body interactions and
the ensuing chaotic effects would dominate over 2-body interactions -- however,
it is emphasized that this applies for the case containing a central massive
object, unlike the situation here.}
Preliminary 3-dimensional modeling of stellar bars by Athanassoula \Cite{Athan}
shows orbits similar to, but more general than, those of two-dimensional models.
He finds that a halo component in the galaxy's overall gravitational potential
does not stabilize the system against forming a stellar bar (\cf also \Cite{Athan1}),
and infers that (as yet undetermined) bar models with non-isotropic distribution
functions would be expected to lead to even stronger bar growth.

Dynamically, a ``bar perturbation'' has typically been taken as being synonymous
with the presence of an additional quadrupole ($m=2$) component in the potential
(\eg Binney and Tremaine \Cite{BT}), and it will be taken as so here.  It is worth
noting that Athanassoula \Cite{Athan} has recently found some bars to have
relatively strong $m=4$, and in some cases even $m=6$ or $m=8$, components.
These modes represent a possible avenue for further study but to lowest order
would be expected to simply enhance the rate of transport of angular momentum
within the cluster.

\section{The Stellar Bar: Implementation}\label{sect_barimpl}

Early in the study of bar dynamics, a high central density $\rho(0)$ was argued
to preclude bar formation due to the disruptive presence of an inner Linblad
resonance (ILR), and the ``Toomre'' $T_\mathrm{rot}/|W|\gtrsim0.14$ instability
criterion \Cite{Toomre} was formulated as a test for whether or not such a
non-axisymmetric perturbation could form in a system.  However, high-$\rho(0)$
barred galaxies with ILRs are observed, and in general bars are observed to be
much more ubiquitous than the Toomre criterion allows.  Thus the true trigger
of bar formation remains unknown \Cite{Sellwood}.   Of particular interest is
the finding of Gadotti and de Souza \Cite{Gadotti} that bars can form in Plummer
sphere potentials which are embedded in nonspherical halos but which possess no
disk component.  Many studies (\eg Athanassoula \Cite{Athan}) note that bars are
very common features.

That being said, all simulations performed here possess
$T_\mathrm{rot}/|W|\gtrsim0.17$ and thus satisfy not only the Toomre criterion
but also the $T_\mathrm{rot}/|W|$ value of $0.171$ calculated by
Christodoulou \etal \Cite{CST} (which was for stellar Maclaurin spheroids, but
was the strictest bar-forming criterion found in the literature).
This is not by design, but is a function of using an observationally-motivated
value for the overall rotation parameter of $\lambda\simeq0.05$ as discussed in
\S\ref{subsect_introrot}.  Thus the bar is simply assumed to exist.

N-body studies show that bars form over 1--2 rotational periods but last for
many more ($\gtrsim10$) \Cite{CS}, and that the stars trapped in the bar do
not undergo much individual dynamical evolution \Cite{Ceverino}.  Some N-body
simulations show the bar losing strength over time, but this effect has been
found not to be due to angular-momentum transfer weakening the bar directly but
is instead due to a vertical-buckling instability of the bar in the presence
of background gas \Cite{Berentzen}, which does not pertain here.  This has led
us to build the nonaxisymmetric potential from a fraction of the lowest-mass
component of the stellar population, and to force those stars to orbit in
``lock step'', sharing a common orbital frequency and phase (\ie it is assumed
that they all orbit at a frequency $\Omega\barb$ whenever the diffusion
coefficient calculation requires knowledge of the orbital frequency).
We assume there is no transfer of stars from bar to field or vice versa,
consistent with studies showing the same stars remain trapped within the bar
(\Cite{Linblad}, \Cite{Ceverino}).  N-body studies show that
the bar frequency is a compromise of the orbital frequencies of its component
stars \Cite{CE}, so our $\Omega\barb$ is set by either conserving the total
angular momentum of those stars, or by simply averaging over their individual
orbital frequencies; each possibility is dicussed below.

Informed by the considerations of \S\ref{barcons}, the most straightforward
way to incorporate a stellar bar into our model is to assume that a fraction
of the stars are trapped in the bar from the outset, that the resulting
nonaxial perturbation in the potential is dominated by the quadrupole term,
and that the structure rotates in bulk at a common angular speed.  With these
assumptions the stars comprising the bar can be considered as simply a subset
of the overall stellar population for the purposes of calculating updated
values of the distribution function $f$ or stellar density $\rho$, while for
determining the bar's contribution to the diffusion coefficients $\IIJ$ the
bar is treated as a semi-solid object with a single overall rotation speed
$\Omega\barb$.  Allowing the bar distribution to evolve like the field stars
avoids any problem of inserting the bar binding energy by hand.

The mass fraction assigned to the bar here is typically 1\% of the total system
mass, for simplicity taken from the lowest stellar mass in the IMF being used
(although the choice of stellar mass in the bar was found to have little or no
effect on the simulation's results).  Using the definition of bar strength
suggested by Weinberg and Katz \Cite{WK} as the ratio of the bar's mass to the
total mass within half of the corotation radius, this gives a bar strength of
5-10\% depending on the initial cluster model being considered, well within the
5-15\% range of previous studies.

One possible concern is that the general formalism of resonant interactions
represented by the diffusion coefficients (\ref{dij}) is only valid when the
bar's rotation speed $\Omega\barb$ does not change too slowly \Cite{TW};
presumably this is because if $d\Omega\barb/dt$ is too small, the
back-reactions of the resonant interaction on the bar itself will not be
smoothly distributed over frequency space and may disrupt the stellar orbits
comprising the bar.  In this study a typical value for the timescale of the
rate of change of the bar frequency $\Omega\barb/(d\Omega\barb/dt)$ is
$\gtrsim10-200$ bar rotation times;
this timescale is consistent with the range of Weinberg's \Cite{MDW85}
calculation for a general bar interacting with a halo and with Athanassoula's
\Cite{Athan1} N-body bar simulations.

The back-reaction issue may explain why, when a bar mass-fraction of
$\gtrsim2\%$ (corresponding to a bar strength much greater than $10\%$) was
tested, the simulations were unstable to numerical divergences in which the
results were not consistent for different choices of numerical stepsize
$\Delta t$.  Backing off to a bar fraction of $1\%$ of the total system mass
allowed for numerical stability, at the cost of being a slightly conservative
choice for the strength of the bar perturbation.




\subsection{Bar Speed Determined by Angular Momentum Conservation}

Being made up of point masses, the bar's effective moment of inertia $\barmom$
about the polar axis can be expressed as simply
\beq\barmom=\stream\int\rho\barb\,dV\,r^2\sin^2\theta\eeq
in which $\rho\barb$ represents the smoothed mass distribution of stars in the
bar.  (Throughout this study, all quantities $\rho$ are smoothed stellar
densities.) The factor $\stream$
allows for \adhoc tuning of the moment of intertia, and is intended to account
for the fact that the bar is not a solid object but is a collection of
individual stars streaming through the bar's overall structure; $\stream$ can
even in principle be negative but in general is of order 
$\stream\simeq\rho\barb/\rho$ \Cite{WT},
and so normally we will set it as an input parameter.  (Note that Athanassoula
\Cite{Athan} and others treated the bar as a rigid object, but they did not
``build'' their bar models out of the consituent stars of the system.)


Quantities depending on the radial coordinate, \eg $\rho\barb$ and $e$, are
stored in terms of ellipsoidal coordinate $a^2$ and so it is convenient to peform the integration
over $a$ instead of $r$.  The mass element is then
$\rho\barb dV=4\pi\rho\barb a^2(1-e^2)^{1/2}\,da$, and one achieves
\bea\barmom&=&2\pi\stream\int_{-1}^1d\cos\theta\int da\,\rho\barb\,a^2(1-e^2)^{1/2}a^2
\left(\sin^2\theta+\frac{\cos^2\theta}{1-e^2}\right)\sin^2\theta\label{bari}\\
&=&4\pi\stream\int da\,a^4\rho\barb\,(1-e^2)^{3/2}
\int_0^1d\cos\theta\,\frac{1-\cos^2\theta}{1-e^2(1-\cos^2\theta)}\nonumber\\
&=&2\pi\stream\int da^2\,a^3\,\rho\barb\,\left[\frac{1-e^2}{e^2}\left(\frac{\sin^{-1}e}{e}
-\sqrt{1-e^2}\right)\right].\nonumber\eea
The above uses standard tables and trigonometric identities to do the angular
integration\footnote{Substitution $z=1-e^2(1-\cos^2\theta)$ aids in the
evaluation of the term with $\cos^2\theta$ in the numerator of the second line
in (\ref{bari}).} \Cite{crc}.  The quantity enclosed in square brackets in
(\ref{bari}) approaches a value of $\frac{2}{3}$ in the limit of $e\rightarrow0$
as is expected when finding the moment of inertia of a spherical shell; we set
it to be exactly $\frac{2}{3}$ for tiny values of $e$ ($\lesssim0.001$, which
gives an accuracy in $\barmom$ of better than one part in $10^6$) in order to
avoid any numerical problems of dividing by a vanishlingly small denominator.

The bar frequency is then
\beq\Omega\barb=\frac{J\barb}{\barmom}\label{Jbar}\eeq
where $J\barb$ is the total angular momentum of the stars in the bar,
$J\barb=\int dI\,dJ\,f\barb(I,J)J$.  Note that the integral for $J\barb$ should
properly be of $J_z$ and not of the full $J$.  We do not invoke an isotropy
argument to approximate $J_z/J$ because it is more straightforward to simply
modify $\stream$ accordingly, increasing it by a factor of $\simeq2$ or
possibly $\simeq\sqrt{2\pi}$ over the value $\stream\simeq\rho\barb/\rho$ given above.

\subsection{Bar Speed Determined by Angular Frenquency Conservation}

Alternatively, one may assume that the bar pattern speed is simply the average
over all stars comprising the bar of those star's orbital frequencies, as has
been found to be the case in N-body simulations of barred systems \Cite{CE}.
This case is very straightforward:
\beq\Omega\barb=\frac{1}{M\barb}\int d^2I\,m\barb f\barb\Omega_2\label{obar}\eeq
As stated above, for simplicity in this study the bar is assumed to consist
only of stars of a single mass, \ie $m\barb=m_1$.  The total mass in the bar
is simply $M\barb=m\barb \int d^2I\,f\barb$.  This method of determining
$\Omega\barb$ has the advantage of not requiring knowledge of the bar's moment
of inertia and so involves no assumptions about $\stream$.  In general we
employ (\ref{obar}) by default; when (\ref{Jbar}) is used it will be
specifically noted.

\subsection{Bar Perturbation}\label{barpert}

As the physical effect of the bar is largely to add an additional quadrupole
term to the overall potential, the bar can be modeled numerically by the
following three terms in the expansion (\ref{tw47}), with all other terms zero:
\begin{itemize}
\item the $\Phi_{2\pm2}$ term that describes the quadrupole interaction;
\item the $\Phi_{20}$ term that incorporates the ellipsoidal nature of the
bar \Cite{arfken}; and 
\item the $\Phi_{00}$ term that sets the zero point of the bar's potential.
\end{itemize}
Only the $m=0$ terms in (\ref{tw47}) contribute to the value of $\Phi\barb$
averaged over phase angle $\phi$; because of this they are useful in
determining how large the $m=2$ term is, despite the fact that they do not
directly contribute to any transfer of angular momentum themselves. The
``bar potential'' $\Phi\barb$ is simply the fraction of the overall potential
attributable to the subpopulation of stars that comprise the bar (as found by
using $\rho\barb$ in place of $\rho$ in \ref{phir}). The requirement that
the expansion sum, averaged over $\phi$, matches the known value of the bar
potential $\Phi\barb(a,\theta)$ implies that at a given semimajor radius
value $a$,
\beq\Phi_{00}(a)Y_{00}+\Phi_{20}(a)Y_{20}(\theta,0)=\Phi\barb(a,\theta).\label{c20}\eeq
Evaluating (\ref{c20}) at the pole and equator ($\theta=0$ and $\theta=\halfpi$
respectively), and solving for $\Phi_{20}$ and then $\Phi_{00}$ gives 
\beq\Phi_{20}(a)=\frac{2}{3}\sqrt{\frac{4\pi}{5}}
  \left[\Phi\barb\left(a,0\right)-\Phi\barb\left(a,\halfpi\right)\right]\eeq and 
\beq\Phi_{00}(a)=\sqrt{4\pi}\left[\frac{1}{3}\Phi\barb\left(a,0\right)
  +\frac{2}{3}\Phi\barb\left(a,\halfpi\right)\right].\eeq
For the strength of the quadrupole, $\Phi_{22}$ is chosen so that the more
elliptical the orbits of the stars in the bar are, the more the $\Phi_{22}$
term contributes.  By the symmetry of the bar, the $m=\pm2$ terms in 
(\ref{tw47}) are equal, leading to
\beq\label{c22}\Phi_{2\pm2}(a)Y_{2\pm2}\left(\halfpi,0\right)=\half\left(
    1-\frac{\avg{r_p}{}}{\avg{r_a}{}}\right)\Phi\barb\left(a,\halfpi\right)\eeq
\ie circular orbits contribute nothing to the bar quadrupole, and purely
radial orbits make for a maximally strong bar.  $Y_{22}$ is at its largest
magnitude at $\theta=\halfpi$, so this guarantees that the bar potential as
expressed by this expansion remains negative at all $\phi$ while its average
over $\phi$ vanishes. The averages in (\ref{c22}) are over all bar stars
whose orbits cross radial coordinate $a$.  When $a$ is greater than the
outmost point $a_\mathrm{max}$ contained in the numeric radial grid, each
$\Phi_{lm}$ is assumed to drop off as $1/a^l$ (or $1/r^l$ when the underlying
potential is spherical symmetric) from the value at $a_\mathrm{max}$:
$\Phi_{lm}(a>a_\mathrm{max})=\left(\frac{a_\mathrm{max}}{a}\right)^l\Phi_{lm}(a)$.

Now that the strength of the bar perturbation has been characterized, the
remaining difficulty is in evaluting the integrals of (\ref{dij}). Note
that, by incorporating $\Phi\barb$ as defined for the entire bar as a whole
in the calculation of the bar's $\Phi_{lm}$ (and hence $\Psi_{knm}$)
coefficients, the integration over the bar's distribution function $f\barb$
has effectively already been performed; this contrasts with the field-star
case in which the $\Phi_{lm}$ (and $\Psi_{knm}$) are defined for individual
stars and only afterwards is the distribution function integrated over.  Thus 
(\ref{bardij}), with its ``bare'' $\delta$-function, is actually the proper 
expression to use for the bar's contribution to the diffusion coefficients.


The problem then is that, for the bar, the $\delta$-function is then no longer
a function of an integration variable.  This can be thought of as being due
to that, by construction, the stars comprising the bar's gravitational potential
orbit in ``lock step'', sharing a common bulk orbital frequency, and so the
$\delta$-function no longer has a distribution of frequencies on which to act.
Fortunately, the fact the calculation is being done on a numeric grid imposes a
finite scale over which any bulk resonance effect must be ``smeared'' anyway --
namely the difference $\Delta\Omega(I_1,I_2)$ between the nearest points on
the action-space grid -- and so the resonance effect must be widened from
$\delta$-function-width to at least this size.
Thus to take the place of the $\delta$ function in the bar's diffusion
coefficient, we insert a Gauusian window function $\barf$:
\beq\barf(\Omega)\equiv\mathcal{A}e^{-(\Omega-\ell_3\Omega\barb)^2/2(\varepsilon\ell_3\Omega\barb)^2}\eeq
so that for the $(\ell_1,\ell_2,\ell_3)$ resonance, the closer the frequency
$\Omega=\ell_1\Omega_1(I_1,I_2)+\ell_2\Omega_2(I_1,I_2)$ for any given
gridpoint $(I_1,I_2)$ is to $\ell_3\Omega\barb$, the
more effect of the resonance it experiences.  For the width of the resonance
we compare the bar pattern speed to the timescale of the bar's slowing down
due to dynamical friction:
\beq\varepsilon=\dot\Omega\barb/\Omega\barb^2.\eeq
According to Tremaine and Weinberg \Cite{TW}, $\varepsilon$ as defined above is of order the
fractional mass density of the bar
and so it is set accordingly as an input parameter.  The normalization factor
$\mathcal{A}=1/(\sqrt{2\pi}\varepsilon\ell_3\Omega\barb)$.
With this window function we can define an effective bar diffusion coefficient
\beq\label{effdij}\barIJ=\pi\suminf_{\ell_3=1}
\suminf_{\ell_1,\ell_2=-\infty}\ell_i\ell_j\Psilsq\barf(\Omega).\eeq
The above $\barIJ$ is used in place of $D_{ij}$ in (\ref{bardij});
this scheme allows the bar to act as a single, bulk perturber of frequency
$\Omega\barb$ whose effects are felt only on the ``cells'' in the action
space grid over which the line of resonances 
$\delta(\ell_1\Omega_1+\ell_2\Omega_2-\ell_3\Omega\barb)$ falls.
The model's computer code issues a warning if the smearing width
$\varepsilon\ell_3\Omega\barb$ is smaller than the inter-gridpoint spacing of 
$\Delta\Omega(I_1,I_2)$ values for any action-space grid cell in which the resonance does fall.

\input{Finite}
\input{Mergers}

%% file: Finite.tex
\section{Finite Differencing Scheme}



\subsection{Numerical Stability}

Any finite differencing must be numerically stable to be useful, and so here
we employ the \textsl{Crank-Nicholson} method, which averages implicit and
explicit schemes and is guaranteed to be numerically stable for any
stepsize -- although the diffusion coefficients, which are very weakly
nonlinear and are very expensive to calculate, are treated explicitly.  The
resulting sparse matrix is solveable using standard methods as shown below.

\newcommand{\upr}{{x_+y}}
\newcommand{\dnr}{{x_-y}}
\newcommand{\upt}{{xy_+}}
\newcommand{\dnt}{{xy_-}}
\newcommand{\ctr}{{xy}}
\newcommand{\fff}[2]{\fpf_{{#1}{#2}}}
\newcommand{\fxy}{\fff{x}{y}}
\newcommand{\fuy}{\fff{x+1}{y}}
\newcommand{\fxu}{\fff{x}{y+1}}
\newcommand{\fdy}{\fff{x-1}{y}}
\newcommand{\fxd}{\fff{x}{y-1}}
\newcommand{\stepp}[1]{{(\tau#1)}}
\newcommand{\stepa}{\stepp{}}
\newcommand{\stepb}{\stepp{{+\frac{1}{3}}}}
\newcommand{\stepc}{\stepp{{+\frac{2}{3}}}}
\newcommand{\stepd}{\stepp{{+1}}}

The Fokker-Planck equation (\ref{fcfp}) expressed in Crank-Nicholson form is
\bea\label{fd}\frac{1}{\Delta t}(\fpf_{xy}^\stepd-\fpf_{xy}^\stepa)
=\frac{1}{4}\sum_\stepa^\stepd\bigg[
\frac{1}{\Delta I_{x_+x_-}}\left(D_{11\upr}\frac{\fuy-\fxy}{\Delta I_{x+1\,x}}
      -D_{11\dnr}\frac{\fxy-\fdy}{\Delta I_{x\,x-1}}\right)\ \ &&\\
+\frac{1}{\Delta J_{y_+y_-}}\left(D_{22\upt}\frac{\fxu-\fxy}{\Delta J_{y+1\,y}}
       -D_{22\dnt}\frac{\fxy-\fxd}{\Delta J_{y\,y-1}}\right)\nonumber\ \ &&\\
+\frac{1}{\Delta\nonumber
I_{x_+x_-}}\left(D_{12\upr}\frac{\fpf_{x_+y_+}-\fpf_{x_+y_-}}{\Delta J_{y_+y_-}}
       -D_{12\dnr}\frac{\fpf_{x_-y_+}-\fpf_{x_-y_-}}{\Delta J_{y_+y_-}}\right)\ \ &\\
+\frac{1}{\Delta\nonumber
J_{y_+y_-}}\left(D_{21\upt}\frac{\fpf_{x_+y_+}-\fpf_{x_-y_+}}{\Delta I_{x_+x_-}}
    -D_{21\dnt}\frac{\fpf_{x_+y_-}-\fpf_{x_-y_-}}{\Delta I_{x_+x_-}}\right)\bigg]&&
\eea
The notation in (\ref{fd}) requires explanation: superscripts denote the
current already-solved-for timestep $\stepa$ and the to-be-solved-for timestep
$\stepd$; they are implicit for all $\fpf$ values on the right hand side.
The averaging over $\stepa$ and $\stepd$, \ie over the explicit and implicit
differencings of (\ref{fcfp}), constitutes the Crank-Nicholson scheme.
Subscripts $xy$ denote points on the two-dimensional grid in action space,
along the radial (first subscript) and tangential (second) directions.
A trailing ``$\pm$'' indicates the position partway between the labeled
gridpoint and the one $\pm1$ gridpoints away (\eg $D_{22\dnt}=D_{22x(y-1)_+}$
if ``partway'' is taken to be ``halfway'', as will be discussed in
\S\ref{subsect_cc}). Thus (\ref{fd}) is an example of differencing the
flux-conservative diffusion equation ``as it stands'' \Cite{NR}.  Finally, 
$\Delta I_{zz'}\equiv(I_z-I_{z'})$ and similarly for $\Delta J_{zz'}$.

Note that the diffusion coefficients are treated purely explicitly; as are
$\Phi$, $\rho$ \etcend, they are only dependent on the overall ensemble $\fpf$ and
not upon any particular $\fxy$, and so are similarly considered to be part of
the ``snapshot'' of the system at timestep $\tau$ over which the evolution of
$\fpf$ is calculated.

\subsection{Time Splitting}

\textsl{Operator splitting} \Cite{NR} (also called ``time splitting'' when
referring to a time-evolution operator such as used here) is used to evaluate
each of the lines in (\ref{fd}) individually.  Using fractional increases
in $\tau$ to conceptually denote this splitting, this can be written as
\bes{opsplit}
\frac{\fpf_{xy}^\stepb-\fpf_{xy}^\stepa}{\Delta t}&=&\frac{1}{4}
\sum_\stepa^\stepb
\frac{1}{\Delta I_{x_+x_-}}\left(D_{11\upr}\frac{\fuy-\fxy}{\Delta I_{x+1\,x}}
-D_{11\dnr}\frac{\fxy-\fdy}{\Delta I_{x\,x-1}}\right)\sublabel{onedii}
\ \ \ \ \ \ \ \ \ 
\\
\frac{\fpf_{xy}^\stepc-\fpf_{xy}^\stepb}{\Delta t}&=&\frac{1}{4}
\sublabel{onedjj}
\sum_\stepb^\stepc\frac{1}{\Delta J_{y_+y_-}}\left(D_{22\upt}\frac{\fxu-\fxy}
{\Delta J_{y+1\,y}}-D_{22\dnt}\frac{\fxy-\fxd}{\Delta J_{y\,y-1}}\right)
\ \ \ \ \ \ \ \ \ 
\eea\bea
\frac{\fpf_{xy}^\stepd-\fpf_{xy}^\stepc}{\Delta t}=\frac{1}{4}
\sum_\stepc^\stepd
\frac{1}{\Delta_\pm^2}\bigg(\fpf_{x_+y_+}(D_{12\upr}+D_{21\upt})-\fpf_{x_+y_-}
(D_{12\upr}+D_{21\dnt})\sublabel{crossd}\ \ &&\\
-\fpf_{x_-y_+}(D_{12\dnr}+D_{21\upt})+\fpf_{x_-y_-}(D_{12\dnr}+D_{21\dnt})\bigg)&&
\ \ \nonumber\ees
in which for convenience 
$\Delta_\pm^2\equiv(\Delta I_{x_+x_-}\Delta J_{y_+y_-})$ has been used and
terms rearranged in (\ref{crossd}). Collecting terms for $(\tau+\frac{1}{3})$
on the left and for $(\tau)$ on the right, (\ref{onedii}) becomes
\bea
-\frac{1}{4}\frac{D_{11\dnr}}{\Delta I_{x\,x-1}}\fdy^\stepb
+\left(\frac{\Delta I_{x_+x_-}}{\Delta t}
  +\frac{1}{4}\frac{D_{11\upr}}{\Delta I_{x+1\,x}}
  +\frac{1}{4}\frac{D_{11\dnr}}{\Delta I_{x\,x-1}}\right)\fxy^\stepb
-\frac{1}{4}\frac{D_{11\upr}}{\Delta I_{x+1\,x}}\fuy^\stepb
&&\label{stepx}
\\=\frac{1}{4}\frac{D_{11\dnr}}{\Delta I_{x\,x-1}}\fdy^\stepa
+\left(\frac{\Delta I_{x_+x_-}}{\Delta t}
  -\frac{1}{4}\frac{D_{11\upr}}{\Delta I_{x+1\,x}}
  -\frac{1}{4}\frac{D_{11\dnr}}{\Delta I_{x\,x-1}}\right)\fxy^\stepa
+\frac{1}{4}\frac{D_{11\upr}}{\Delta I_{x+1\,x}}\fuy^\stepa
&&\nonumber\eea
which can be seen to form a tridiagonal linear set of equations in $x$ that
can be solved for the unknowns $\fxy^\stepb$, for a given stepsize $\Delta t$
and (fixed) $y$.  Routine \fcn{TRIDAG} from Press \etal \Cite{NR} is used to solve
(\ref{stepx}) for each $y$ value in turn.  Given knowledge of the full
$\fxy^\stepb$, the solution of (\ref{onedjj}) for $\fxy^\stepc$ is entirely
analogous, with the radial and tangential coordinates swapping roles.

Time splitting is not mandatory for setting up differencing, but it greatly
simplifies the scheme both conceptually and mathematically.  It will be shown
in \S\ref{sect_diffscheme} that time splitting does result in a more accurate
and robust differencing.

\subsection{Ensuring a Positive-Definite Distribution}\label{subsect_cc}

In addition to numerical stability, when tracking the evolution of a
distribution function of real objects it is greatly advantageous to apply a
differencing scheme which guarantees a positive-definite solution for every
gridpoint and at each timestep.  For the one-dimensional Fokker-Planck equation
there exists the \textsl{Chang-Cooper} spatial differencing scheme which (when
combined with the Crank-Nicholson time-differencing described above) has both
these qualities \Cite{CC}.  The Chang-Cooper scheme consists of a method to formulate a working prescription for
the ``half-grid'' points such as $\fpf_\upr$: for example, if one defines a
parameter $\delta_{x(y)}$ using
$\fpf_\upr\equiv(1-\delta_{x(y)})\fuy+\delta_{x(y)}\fxy$ then Chang-Cooper
provides a calculation for $\delta_{x(y)}$ which guarantees $\fpf$ will remain
positive-definite.

A one-dimensional differencing scheme does not necessarily generalize to two
or more dimensions, however: Chang-Cooper does not give a general form for
calculating both $\delta_{x(y)}$ and $\delta_{(x)y}$.  Fortunately when the
Fokker-Planck equation is specifically cast in its flux-conservative form
(\ref{fcfp}), the Chang-Cooper method reduces trivially to the case of
``centrally weighting'' all derivatives that involve the distribution function:
$\delta_{x(y)}=\half$ and $\delta_{(x)y}=\half$ for all $x$ and $y$.
Thus $\fpf_\upr=\half(\fuy+\fxy)$ \etc for the half-grid values of $\fpf$;
substituting these into (\ref{crossd}) and recollecting for the various values
of $\fpf$ on the gridpoints (\ie $\fxy$ and its eight immediate neighbors) yields
a difference equation for $\fxy^\stepd$ analogous to that of (\ref{stepx}) for 
$\fxy^\stepb$.  The ensemble of such equations for all values of $x$ and $y$ forms a 
sparse matrix system which is solved using the routine \fcn{LINBCG} from
Press \etal \Cite{NR}.


\subsection{Numerical Boundary Conditions}

At the edges of the $(I,J)$ grid the differencing schemes as described above
require knowledge of $\fxy$ values beyond the grid proper.  Denoting the
highest values of $x$ and $y$ on the grid as $X$ and $Y$, for $x=1$ or $x=X$
and for $y=1$ or $y=Y$ numeric boundary conditions must replace (\ref{stepx})
and its analogues for $\stepc$. 
Using the $x=X$ case as an illustration, possibilities include:
\newcommand{\iratio}{\frac{\Delta I_{XX-1}}{\Delta I_{X-1X-2}}}
\bes{striks}\fpf^\stepb_{Xy}&=&\fpf^\stepb_{X-1y}\sublabel{strika}\\
            \fpf^\stepb_{Xy}&=&\left(1+\iratio\right)\fpf^\stepb_{X-1y}-
                        \left(\iratio\right)\fpf^\stepb_{X-2y}\sublabel{strikb}\\
            \fpf^\stepb_{Xy}&=&\fpf^\stepa_{X-1y}\sublabel{strikc}.
\eea
These are all straightforward extrapolations; cases (a) and (c) are taken
directly from Strikwerda \Cite{strik}, while case (b) has been generalized for a
non-uniform grid. 

An option which does not require any extrapolation consists of employing a
one-sided (towards the interior) differencing on the boundary instead:
\bea\frac{\Delta \fpf(I_X)}{\Delta I}=\frac{\fpf_{Xy}-\fpf_{X-1y}}{\Delta I_{XX-1}}
\sublabel{natspl}\ees
which can then be used in place of Chang-Cooper's centered differencing to
derive analogues of equations (\ref{fd}) and (\ref{opsplit}) for use on the
boundaries.  This scheme effectively causes $\Delta^2f/\Delta I^2$ and
$\Delta^2f/\Delta J^2$ to vanish on the boundaries, and so is referred to as 
the \textsl{natural-spline} method (although the full differencing of the
Fokker-Planck equation only
vanishes if \eg $D_{11X_+y}=D_{11X_-y}$ as well).  It is not a numerical
boundary condition \textit{per se}, but rather a new differencing scheme for 
the boundary points that avoids the need for any numerical boundary condition.

Experimentation with all four possible boundary condition schemes showed that
only the natural-spline treatment of (\ref{natspl}) accurately tracked an
analytically-solvable test case, as shown in Figure \ref{plot1xy} and
described in \S\ref{sect_diffscheme}.

%% file: Mergers.tex
\newcommand{\ti}{\mathcal{I}}
\newcommand{\tj}{\mathcal{J}}
\newcommand{\tr}{\mathcal{R}}
\newcommand{\tw}{\scriptstyle{\mathcal{W}}}
\newcommand{\tww}{\scriptscriptstyle{\mathcal{W}}}
\newcommand{\tv}{\vartheta}
\newcommand{\tvr}{\tv_\pm}
\newcommand{\tvt}{\vec{\tv}_T}
\newcommand{\tm}{{\scriptscriptstyle\mathcal{Q}}}
\newcommand{\tM}{{\scriptstyle\mathcal{Q}}}

\section{Stellar Mergers}\label{mergerrates}

\subsection{Rates of Loss due to Mergers}\label{lossrate}

\newcommand{\collgen}[2]{\dot{\mathcal{N}}_{#2q'}#1}
\newcommand{\collrate}[1]{\collgen{#1}{q}}
\newcommand{\avgint}[3]{\left<#3\collgen{#1}{#2}\right>_\gamma}
\newcommand{\avggen}[2]{\avgint{#1}{#2}{}}
\newcommand{\avgrate}[1]{\avggen{#1}{q}}
Modeling stellar mergers due to collisions requires the transfer of stars from
one $f_q$ to another, \eg if a star of mass $m_\tm$ with actions $(\ti,\tj)$
and one of mass $m_{q'}$ with actions $(I',J')$ combine into a new star of mass
$m_q=m_\tm+m_{q'}$, then $f_\tm(\ti,\tj)$ and $f_{q'}(I',J')$ must be
decreased, and $f_q$ increased at a dynamically appropriate value of the
action $(I,J)$.\footnote{In practice there isn't usually a mass bin $q$ such
that $m_q=m_\tm+m_{q'}$ exactly, and so interpolation over the two bins nearest
$(m_\tm+m_{q'})$ is required; this is discussed in \S\ref{subsect_bookkeep}.}

This alteration of $f_q$, $f_\tm$ and $f_{q'}$ is accomplished via the loss
($L_q$) and gain ($G_q$) terms added to the Fokker-Planck equation, as in
(\ref{fullfp}).  Note that these terms are not part of the differencing
scheme, as they are simple rates of change and so can be calculated directly.
The terms $L_q$ and $G_q$ are similar to those developed by Quinlan and
Shapiro \Cite{QS89}, who used a one-dimensional
distribution function $F(E)$ only.  We adapt them to the two-dimensional
$f(I,J)$, calculating the probability of collision given cross section
$\sigma$.
(Context should prevent any notational confusion between the cross section
and the components of the velocity dispersion, \eg $\sigma^2_\phi$.)
Expressing the cross section as a sum of powers of the collisional
speed $|\Delta\vec v|$ with mass-dependent coefficients, \ie
\beq\sigma\equiv\sum_\alpha\sigma_\alpha(q,q')|\Delta\vec v|^\alpha\eeq
then the rate of collisions of a given ``target'' star of mass $m_q$ and
actions ($I,J$) at radius $r$ with all stars of mass $m_{q'}$ is
\beq \label{mrij}\collrate(r,I,J)=\frac{\rho_{q'}}{m_{q'}}\,\sigma(q,q')
|\Delta\vec v|=\sum_\alpha\frac{\sigma_\alpha(q,q')}{4\pi^2r}
\int_r\frac{dI'\,dJ'}{J'}\Omega_1'f_{q'}(I',J')|\Delta\vec v|^{\alpha+1}\eeq
where primes are used to indicate quantities dependent on the actions being
integrated over, and the only dependence upon the target star's actions
$(I,J)$ is in $\Delta\vec v=|\vec v-\vec v'|$.  The above makes use of
the density (\ref{onerho}) as an operator, but in doing so introduces an
ambiguity due to the lack of full information about $r$, $\vec v$ and
$\vec v'$ in the orbit-averaged distribution function $f_{q'}$.
This ambiguity will be dealt with below.

Given values for the actions, the magnitudes of the radial ($|v_r(r)|$,
$|v_r'(r)|$) and tangential ($|\vec{v}_T|=J/r$, $|\vec{v}_T'|=J'/r$)
components of the stellar velocities are well-defined, but there is ambiguity
in the latter's subcomponents due to lack of information regarding $J_z/J$.
A conservative approach is to assume that the relative orientation of the
tangential parts of the velocity vectors is random, \ie if $\gamma$ is the 
angle between $\vec{v}_T$ and $\vec{v}_T'$, so that
\beq\label{dvpm}|\Delta\vec{v}|^2_\pm\equiv(|v_r|\mp|v_r'|)^2
  +(|\vec{v}_T|-|\vec{v}_T'|\cos\gamma)^2+|\vec{v}_T'|^2\sin^2\gamma\eeq
then $\gamma$ is randomly distributed.  This is conservative because it
assumes no ``collimation'' of $\vec{v}_T$ and $\vec{v}_T'$ due to the effect
of overall cluster rotation.  Taking the average over random angle $\gamma$
and over the possible relative signs of the radial components one can define
for any given function $g(|\Delta\vec{v}|)$,
\beq\left<g(|\Delta\vec{v}|)\right>_\gamma\equiv\otwopi\int^\pi_0d\gamma\sum_\pm
g\left(|\Delta\vec{v}|_\pm\right)\label{gammavg}.\eeq

The $r$ dependence is easily eliminated by integrating over the orbit of the 
target star.  Thus the orbit-averaged rate of collisions for a single star is
\beq\dot{N}_{qq'}(I,J)=\int_0^\pi\frac{dw_1}{\pi}\avgrate{(r(w_1),I,J)}=
\int_{r_p}^{r_a}\frac{\Omega_1dr}{\pi\vrsq^{1/2}}\avgrate{(r,I,J)}\label{mij}\eeq
where $\collrate$ is given by (\ref{mrij}), in which
the endpoints of the action integrals therein are those
energetically allowed at $r$.  In practice the integral over $dr$ above is brought
inside the one over $dIdJ$, in which case one integrates over all $(I,J)$ and
restricts the integration over $r$ to the range, if any, over which the orbits
overlap.  A possible simplifying assumption for the 
integral in (\ref{mij}) is due to the fact that the target star spends the bulk
of its time at or near its orbital endpoints $r_p=r(0)$ and $r_a=r(\pi)$.)

With this in hand the total loss rate due to stellar collision of all
stars with given values of action $(I,J)$ and mass $m_q$ is simply
\beq L_q(I,J)=f_q(I,J)\sum_{q'}\dot{N}_{qq'}(I,J).\label{lm}\eeq
Of course, in reality there is a continuum of masses and so the sum in
(\ref{lm}) should properly be an integral, but in this study stellar masses
are defined on a finite grid of values and so the sum is over those values.

\subsection{Rates of Gain}

The calculation of the gain rate $G_q(I,J)$ is conceptually similar but
slightly more cumbersome, owing to the difficulty in determining the new velocity
vector of a star which is the product of a merger of two smaller stars.
In $G_q$, $I$ and $J$ now denote the actions of the star of mass $m_q$ produced
by the collision, and so we introduce the notation $(\ti,\tj)$ for the target
star's actions.  Thus at any given $r$, conservation of momentum requires
\beq m_q^2J^2=[(m_q-m_{q'})\tj+m_{q'}J'\cos\gamma]^2+m_{q'}^2J'^2\sin^2\gamma\label{j1}\eeq
and
\beq m_qv_r(r)=(m_q-m_{q'})\tvr\pm m_{q'}v_r'\label{i1}\eeq
in which $\tvr$ has been used for the target star's radial velocity and the
choice in relative sign accounts for the possibilities that the stars have
like or opposite radial velocity directions at the time of collision.
Note that this is the \textit{same} ``$\pm$'' as in (\ref{dvpm}), although
of opposite sign, and does not need to be independently averaged over.  After 
solving (\ref{j1}) for $\tj=r|\tvt|$ and (\ref{i1}) for $\tvr$, the target
star's orbital energy is simply $\mathcal{E}_\pm=\half(\tvr^2+|\tvt|^2)+\Phi(r)$,
and then $\ti_\pm$ can be found directly from (\ref{tw30}).

The analogue of (\ref{mij}) in the merger-gain case is the rate of producing
stars of mass $m_q$ and actions $(I,J)$ from those of masses
$m_\tm\equiv m_q-m_{q'}$ and $m_{q'}$ and with respective actions
$(\ti,\tj)$ and $(I',J')$.  Because we are
``back-constructing'' what $\ti$ and $\tj$ are from knowledge $(I',J')$ and
$(I,J)$, the orbit-averaging is only over that portion of the orbit of a star
with actions $(\ti,\tj)$ that overlaps with the product star orbit as
determined by its actions $(I,J)$.
(Consideration of the overlap of the orbit having actions $(I',J')$ with that of $(I,J)$
is already implicit in the use of $\rho_{q'}(r)$ in (\ref{mrij})).  Labeling
the target star's radial frequency as $\Omega_{\tww}$, its associated
canonical angle variable as $\tw_1$, and its orbital endpoints $\tr_p$ and
$\tr_a$,
\bes{nmmw}\dot{N}'_{\tm q'}(I,J)&=&\avgint{(r(w_1),\ti,\tj)}{\tm }
{\int_{\tww_\mathrm{min}}^{\tww_\mathrm{max}}{d\tw_1\ }}\substep\\
&=&\avgint{(r,\ti,\tj)}{\tm }{\int_{\max(\tr_p,r_p')}^{\min(\tr_a,r_a')}
\frac{\Omega_{\tww}dr}{|\tvr|}}\sublabel{nmmr}.\ees
Note that in this case we integrate over $d\tw_1$ so that contributions 
to $\dot{N}'_{\tm q'}(I,J)$ over the entire orbit of the merging stars are
considered, whereas in the merger-loss case we orbit-averaged over $dw_1$
in order to eliminate the $r$ dependence.

To determine ${\tw_1}(r)$ (or $\tr_p(\ti,\tj)$ and $\tr_a(\ti,\tj)$)
one must solve (\ref{i1}) and (\ref{j1}) -- which is only possible within
the integral over $d\tw_1$ (or over $dr$), and which in turn requires
knowledge of ${\tw_1}(r)$ in order to determine the integral's
endpoints.  To avoid the impasse, the form
\beq\dot{N}'_{\tm q'}(I,J)
=\avgint{(r,\ti,\tj)}{\tm }{\int_{\max(r_p,r_p')}^{\min(r_a,r_a')}
\frac{\Omega_{\tww}dr}{|\tvr|}}\label{nmmrr}\eeq
is used instead, with the additional requirement that only dynamically-allowed
solutions of (\ref{j1}) are allowed: \ie for a given value of $\gamma$ we define
$\collgen{}{\tm }(r,\ti,\tj)\equiv0$ if $\tj<0$ or is complex; this
is equivalent to only considering orbital ranges that do indeed overlap.

The rate of gain due to the mergers is thus
\beq G_q(I,J)=\half\sum_{q'}\dot{N}'_{\tm q'}(I,J)f_\tm(\ti,\tj)\label{gm}\eeq
with an extra factor of $\half$ inserted to counteract the effect of
double-counting;\footnote{Other studies, such as \Cite{QS90}, have included
a factor of $1/(1+\delta_{\tm q'})$ instead of $1/2$ to prevent 
double-counting. The distinction is that here we sum over all possible
collisional pairs and attribute their products to the appropriate
$m_q=m_\tm+m_q'$, as opposed to predetermining the product $m$ value being
considered and invoking a delta function $\delta(m_q-m_\tm-m_q')$ to restrict
the mass values of the colliding stars as needed.  The present method is more
efficient computationally as all $G_q$ for a given $(I,J)$ are found
simultaneously.} it is arbitrary which is called the ``target'' star, and it
takes one of each to create a merger product.  The right side of (\ref{gm})
contains a notational sleight of hand: since $(\ti,\tj)$ depend on $I'$,
$J'$, and $\gamma$, the integrals from (\ref{mrij}) and (\ref{gammavg}) now
act as operators on $f_\tm(\ti,\tj)$.  For clarity, the full expression for
the gain rate is
\beq\label{fullgm}G_q(I,J)=\sum_{\alpha,q'}\frac{\sigma_\alpha(\tM,q')}{16\pi^3}
\int\frac{dI'\,dJ'}{J'}\Omega_1'f_{q'}(I',J')\int^\pi_0d\gamma
\int_{\tww_\mathrm{min}}^{\tww_\mathrm{max}}\frac{d\tw_1}{r({\tw_1})}
\sum_{\pm,\tj}f_\tm(\ti_\pm,\tj)|\Delta\vec v|_\pm^{\alpha+1}\eeq
in which the final sum is over the two possible sign choices in (\ref{dvpm}),
and over (one or both) dynamically-allowed values of $\tj$ found from solving
(\ref{j1}).  In practice it is required to do the orbit-averaging integral
directly over $r$ and not over $\tw_1$, as in (\ref{nmmr}) and (\ref{nmmrr}).

An artifact of this approach is that the symmetry between ``target'' and
``object'' stars is not obvious in (\ref{fullgm}).  However as the numerical
calculations for $L_q$ and $G_q$ are performed by independent sections of
computer code, a match between overall merger losses and gains shows the
validity of the method, as will be discussed in \S\ref{sect_LG}.

\subsection{Mass Bookkeeping}\label{subsect_bookkeep}

The above derivation made use of the manifest fact that any star which is the
product of a purely-inelastic stellar collision and merger will have a mass equal
to the sum of the colliding stars' masses: $m_q\equiv m_\tm+m_{q'}$. But the
model only tracks stars of a finite number mass values, and so it's possible
that $m_q$ does not correspond to a stored value.  To account for such
intermediate mass values, merger products are interpolated across the nearest
values below and above $m_q$ on the mass grid \Cite{Lee}.
Referring to these values as $m_{q_-}$ and $m_{q_+}$ respectively, one has
\beq G_{q_-}=\frac{m_{q_+}-m_q}{m_{q_+}-m_{q_-}}G_q\label{gminus}\eeq
and
\beq G_{q_+}=\frac{m_q-m_{q_-}}{m_{q_+}-m_{q_-}}G_q.\label{gplus}\eeq
Note that because actions are conserved in collisions, no interpolation in $I$
or $J$ is required.  Thus the merger-gain rates from both (\ref{gminus}) and
(\ref{gplus}) for a given mass value contribute to the full Fokker-Planck
equation (\ref{fullfp}).

One may ask why do we not merely round down the merger-product's mass to the
next lowest mass bin value, in order to account for mass loss during a
not-completely inelastic stellar collision which produced the merger.  The main
reasons are (1) that it would also inadvertently reduce the total mass of the
cluster, which while it would be a reasonable effect in an ordinary globular cluster would
not be expected to occur in these more massive systems; and (2) that there is no way
to account for diffuse gas in the calculation of the cluster's gravitational
potential, nor in associated dynamical quantities.  Thus all collisional mass
is assumed to go into the newly-merged star, a similar approach to that taken
by Quinlan and Shapiro \Cite{QS90} who surveyed results of hydrodynamical
simulations of stellar collisions -- most of which find a maximum mass loss per
collision of $\lesssim11\%$ -- and concluded that the average mass loss will
be much less than that and thus also assumed complete coalescence into the
produced larger-mass star.  It is also consistent with the finding of
Freitag \etal \Cite{FAGR} that assuming coalescence
is ``fully justified'' for velocity dispersions $v_{rms}\lesssim300$ km/s -- a
condition largely satified by the simulations here, the vast bulk of which have
a calculated initial velocity dispersion of $\vrms\leq310$ km/s.


\subsection{The Delta-function Approximation}

The full merger-gain coefficient calculation (\ref{fullgm}) is by far the most
computationally intensive part of the model, and so calls for a simplifying approximation.
The ``delta-function approximation'', in which all stars at a given of radial
distance are assumed to share a common $|\vec v'|=v_\textsl{rms}(r)$, was found
by Quinlan and Shapiro \Cite{QS89} to provide large performance gains with only a marginal loss in
accuracy in most cases.  The key to the approximation is to not replace the
density $\rho_{q'}$ in (\ref{mrij}) with its distribution-function integral form
($\ref{onerho}$) and instead to substitute $|\vec v'|$ with $v_\textsl{rms}(r)$
throughout, thus avoiding two computationally-expensive integrals over the
action components.  However the present two-dimensional study requires velocity
components $v'_r\equiv|\vec v'|\cos\theta_v=v_\textsl{rms}\cos\theta_v$ and
$v'_T\equiv|\vec v'|\sin\theta_v=v_\textsl{rms}\sin\theta_v$, and so the
approximation comes at the cost of an averaging over $\sin\theta_v$:
\beq\label{dfagm}G_q(I,J)\simeq\sum_{\alpha,q'}\frac{\sigma_\alpha(\tM,q')}{4\pi}
\int^1_0d\sin\theta_v\int^\pi_0d\gamma\int_{\tww_\mathrm{min}}^{\tww_\mathrm{max}}
{d\tw_1}\frac{\rho_{q'}(r({\tw_1}))}{m_{q'}}
\sum_{\pm,\tj}f_\tm(\ti_\pm,\tj)|\Delta\vec v|_\pm^{\alpha+1}.\eeq
The actions $(\ti_\pm,\tj)$ at a given position $r({\tw_1})$ are determined by the new conservation rules
\beq
m_q^2v_T^2=[(m_q-m_{q'})|\tvt|+m_{q'}v'_\textsl{rms}\sin\theta_v\cos\gamma]^2+m_{q'}^2v'^2_\textsl{rms}\sin^2\theta_v\sin^2\gamma\label{j2}\eeq
and
\beq m_qv_r(r)=(m_q-m_{q'})\tvr\pm m_{q'}v'_\textsl{rms}\cos\theta_v\label{i2}\eeq
which, as before, may not have a solution for either or both $(\ti_+,\tj)$ or
$(\ti_-,\tj)$ (in which case $f_\tm(\ti,\tj)\equiv0$). The notation
$v'_\textsl{rms}(r)$ indicates the $\textsl{rms}$ velocity specifically of the
subpopulation of stars of mass $q'$, making this an even less-severe approximation than
in the one-dimensional case.  Note that, consistent with the conservative
assumption of a randomly-varying $\gamma$ in \S\ref{lossrate}, $\vec v'$ is
also taken to be evenly distributed over $\theta_v$.  The averaging is
restricted to the range $0\leq\theta_v<\halfpi$ because the ``$\pm$'' in
(\ref{i2}) already accounts for $\halfpi\leq\theta_v<\pi$.

\subsection{The Cross Section}

Also following Quinlan and Shapiro \Cite{QS90}, for the stellar mergers we take a 
hard-sphere cross section with gravitational focusing correction:
\beq\sigma=
\pi(r_*^2+r_*'^2)\left[1+\frac{2G(m+m')}{(r_*+r_*')|\Delta\vec v|^2}\right]\eeq
in which $m$ and $m'$ are the individual stellar masses, and $r_*$ and $r'_*$
are the physical stellar radii of main-sequence stars of those masses.

In practice, either term in the above may dominate in a given collision,
depending on the size of the relative speed of collision $|\Delta\vec v|$;
Ebisuzaki \etal \Cite{SPZn} has found observational evidence that the gravitational
focusing term can be important in compact stellar clusters, and Freitag \etal 
\Cite{FAGR} claim that the gravitational focusing term dominates when
the velocity dispersion $\vrms<300$ km/s, which is the case for most but
not all simulations presented here (the exception being the ``E4A'' models
-- which, as mentioned above, almost satisfy that criterion).

\section{Binary Mergers, Binary Heating}

It will be seen in Chapter \ref{chap_results} that two-body mergers, while in
some cases fairly frequent, do not dominate the dynamics in our simulations.
Given that 3-body collisions are much more likely to form binaries than to
result in a direct merger, that Quinlan and Shapiro \Cite{QS90} found that few
hard 3-body binaries formed during most of their simulations, and that in the
absence of a central massive object Freitag \etal \Cite{FAGR} found that 2-body
mergers start before 3-body binaries can form, we ignore the effects of
3-body collisions for stellar mergers as well as for the effect of binary
hardening on the overall energy budget of the system. In addition, following
the lead of prior studies (\Cite{FRB}, \Cite{KELSL}) the effect of primordial
binaries is also not considered for simplicity's sake.  For determining the
rate of stellar mergers this is a conservative approach, as when
there is no massive central object primordial binaries are likely to foster
collisions (even if once a massive black hole does form the binaries then
serve to ``grind down'' stars instead of growing them) \Cite{FAGR}.
In their N-body simulations Portegies Zwart and McMillan \Cite{ZGrape} also found that 3-body encounters
(binary+star) increase the rate of mergers, in contrast to prior assumptions.

Binaries can also serve to heat the overall distribution of field stars as the
binaries harden \Cite{QS90}, with the potential to eventually halt and reverse
core collapse.  However, in practice for dense clusters this effect is not
important prior to the late stages of core collapse and so can be ignored in
earlier stages \Cite{KELSL}.  Also, stellar collisions between binaries and
other stars lead to a significant reduction in the amount of heating produced
\Cite{FRB}.  Finally, binary + field star interactions are most likely for very
small ($\delta v\simeq 5$ km/s) relative velocities \Cite{Freitag}, whereas the
models employed here feature velocity dispersions measured in the hundreds of
km/s.

For all the above reasons, binary heating has not been incorporated into
the simulations performed.  However, a test of the maximum possible amount
of binary heating the system could have achieved was performed for a variety
of models, the results of which are given in \S\ref{sect_binheat}.






%% file: Tests.tex
\chapter{Validity Tests and Model Parameter Choices\label{chap_tests}}

\section{Overview}

There are five major calculational components to the simulation method:
\begin{enumerate}
\item updating the gravitational potential $\Phi(r)$ and stellar density
$\rho_q(r)$ given a new distribution function $f_q(I,J)$ at each timestep;
\item calculating the resulting new values for stellar-dynamical quantities
such as the orbital frequencies $\Omega_j(I,J)$, as well as the conversion
between orbital angle
$w_1$ and radial position $r(w_1)$ for a given actions $(I,J)$ \etcend);
\item calculating the full set of diffusion coefficients $D_{ij}(I,J)$;
\item calculating the rates of mass loss ($L_q(I,J)$) and gain ($G_q(I,J)$)
for each stellar mass value $m_q$ within the distribution functions $f_q(I,J)$; and
\item finite-differencing the Fokker-Planck equation to update each
$f_q(I,J)$ using $D_{ij}$, $L_q$ and $G_q$.
\end{enumerate}

Also requiring consideration are choices for the purely-numerical parameters
of each simulation:
\begin{itemize}
\item the size and range of the grid of actions $(I,J)$ on which many quantities are calculated;
\item the size and range of the radial-coordinate $r$ grid for $\rho(\vec r)$, $\Phi(\vec r)$
\etcend;
\item the choice of timestep $\Delta t$;
\item the number of terms used in the spherical-harmonic expansion of $\Phi$;
and \item how to ``bin'' the range of stellar masses studied into discrete
values $m_q$.
\end{itemize}

Tests of each of these aspects of the overall calculation are presented in turn
in this chapter, as is verification that the effect of binary heating is
sufficiently small that it need not be included in the model.

\section{Potential-calculating Tests}\label{potests}

\begin{figure}
\input{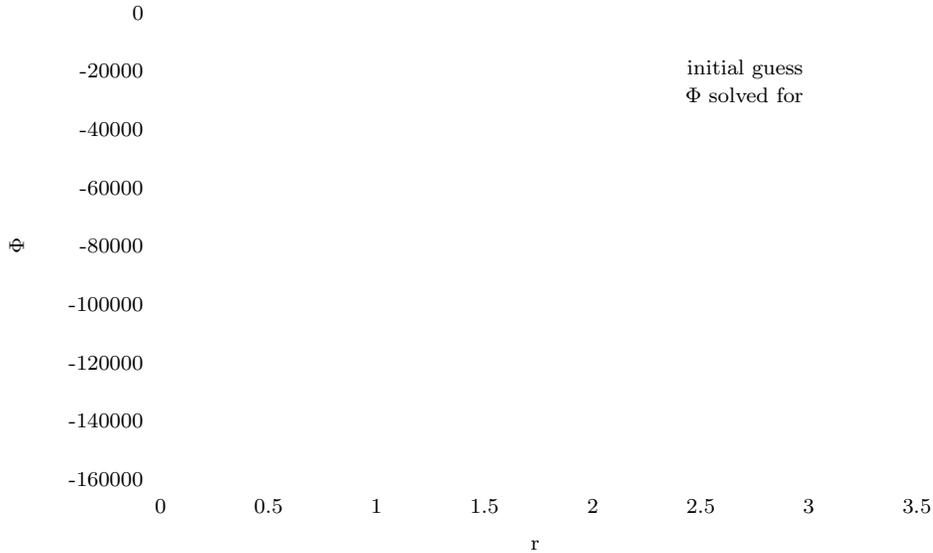}
\caption{Demonstration of the ability of the potential-solver to converge even
when given an initial guess for $\Phi$ and $\rho$ that are extremely in error.
The solid line is the analytical curve for $\Phi(r)$ for a Plummer sphere,
the long dashes are the initially input values, and the short dashes are the
iteratively solved-for potential.
This convergence was obtained after only 4 iterations.
The units in this figure are arbitrary for testing purposes.}
\label{phitest.e2a}\end{figure}

\begin{figure}
\input{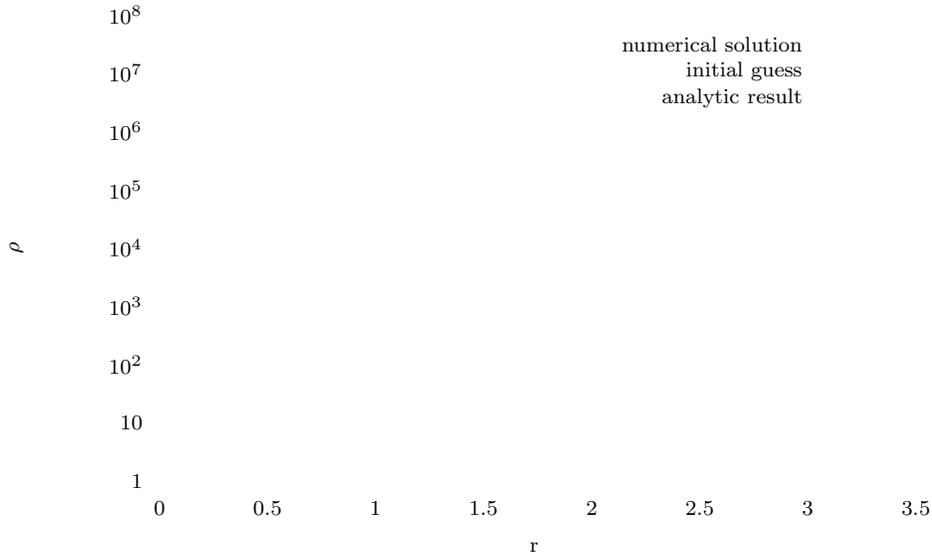}
\caption{The same test of the potential solver as shown in Figure
\ref{phitest.e2a}, but now plotting $\rho(r)$, again in arbitrary units.  The
initial guess was chosen to be much smaller and flatter than the true solution.
}\label{rhotest.e2a}\end{figure}

The first part of the model iteratively calculates the new density $\rho$ and
gravitational potential $\Phi$ given an updated $f$; it is most easily tested
by giving it an artificially bad (\ie far from correct) initial guess for
$\Phi(r)$ and $\rho(r)$, and verifying that it does converge on the true
values.  That it does can be seen in Figures \ref{phitest.e2a} and
\ref{rhotest.e2a}.  Of note is that the input functions were close to
being ``maximally'' bad, \ie any much larger discrepency from the correct
values produced unphysical results in intermediate calculations (\eg $\rho<0$).
Yet in these tests cases the solver converged on the proper solutions
in approximately the same number of iterations as it does when used in the
actual model calculation. 

\section{Dynamical Tests}
\subsection{Orbital Frequencies}

The orbital-dynamic quantities described in \S\ref{sect_dyn} appear in
almost every higher-level calculation in the model.  In order to test the
accuracy of the calculation of the dynamical quantities' values, two different
potential-density pairs for which all quantities are also analytically calculable were
used as test cases: the two-dimensional 
Simple Harmonic Oscillater (SHO), and the isochrone potential \Cite{BT}.

\begin{figure}
\input{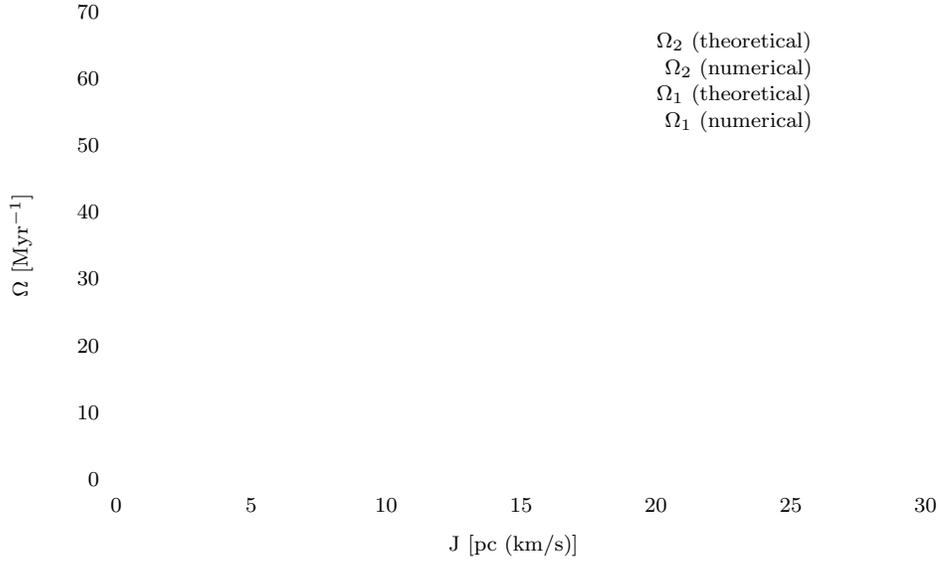}
\caption{Orbital frequencies $\Omega_1$ (upper plot) and $\Omega_2$ vs. $J$ for a Simple
Harmonic Oscillater potential, at a constant energy.  Straight lines are the
analytic values, varying ones are the numerical-found solutions.  Numerical
values of other dynamical quantities (energy, orbital endpoints) are accurate
to within
$10^{-3}$ or better of the analytic values.  As explained in the text,
numerical curves in this plot were found using the simulation code's
calculations of the potential and density;
substituting analytically-known potential and density values into the
calculation produced identical curves.  Similarly, whether the $\Omega_j$ were
found by integrating over the orbit (using (\ref{tw28}) or (\ref{tw29})) or
by taking $\Omega_j=\di{E}{I_j}$ (\ref{tw35}) made no difference.
}\label{harm13fp}\end{figure}

Figure \ref{harm13fp} shows a sample result of the SHO case; the
upper curves are of $\Omega_1$, lower are of $\Omega_2$.  The results do not
depend on whether the $\Omega_j$ were found by directly integrating over the
orbit (\ie using (\ref{tw29}) or (\ref{tw28})) or by calculating the partial derivative of
the energy (using (\ref{tw35})); this is consistent with the fact that the
numerically-determined values for $E$ and for the orbital endpoints matched
the analytic values to within $10^{-3}$ or better. As can be seen, there is
a slight systematic offset in the value found for $\Omega_1$, but otherwise
the match to the analytic curves is good; this is a typical result.

\begin{figure}
\input{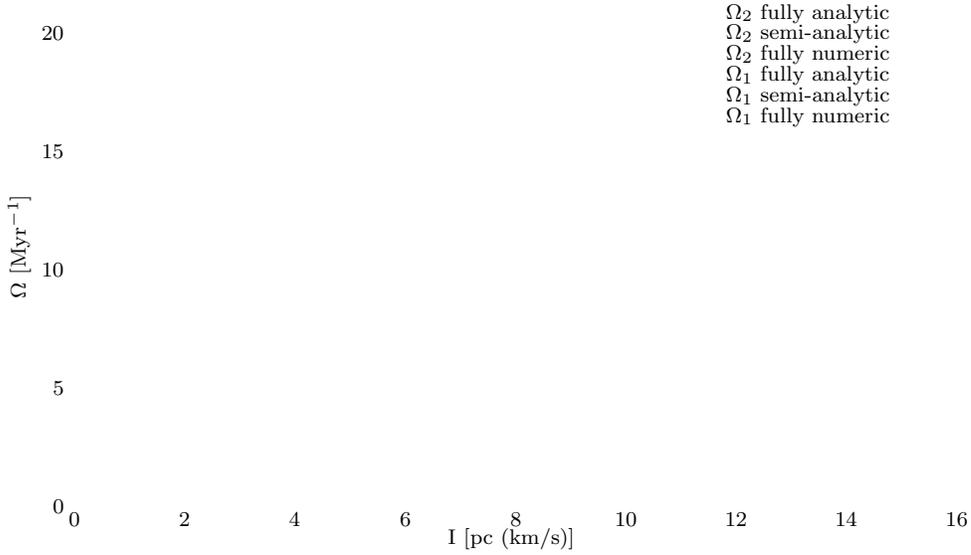}
\caption{$\Omega_1$ (upper plot) and $\Omega_2$ vs. $I$ for an isochrone
potential, at a constant energy.  Similar to Fig. \ref{harm13fp} but here
the numerically-found $\Omega$ curves show both the semi-analytic- and
fully-numeric-potential cases.
}\label{chron33}\end{figure}

When testing the calculation of dynamical quantities, there was a choice in
what values to use for the potential $\Phi(r)$: either numerically-determined
values (\ie a ``fully-numeric'' test which employed the entire calculational
machinery), or the analytically-known potential (a ``semi-analytic'' test
which isolated the dynamical calculations only).
Figure \ref{chron33} shows results for both, in the case of the isochrone
model.  Again there is a slight offset between the fully-analytic
$\Omega_1$ and the numeric curves.  Of note is the close match between
the semi-analytic and fully-numeric calculations; this is not surprising
given how well the potential-finding algorithm described in Chapter
\ref{chap_pot} works, as seen in \S\ref{potests}.

\subsection{Orbital Angles}

As stated in \S\ref{sect_dyn}, the diffusion coefficients do not depend on
the canonical angle variables $w_j$, nor is knowledge of them needed for any
of the other calculations in the model; this is largely due to the fact that
the sinusoidal dependence of the multipole components of the potential in
action space (\ie as shown in (\ref{tw55})) is eliminated upon
orbit-averaging.\footnote{Although as shown \eg in (\ref{nmmw}), if
desired it is always possible to change variables from $r$ to $w_1$ by using
(\ref{tw37}).}.  However, $r(w_1)$ and $[\psi-w_2](r)$ are explicitly
required to perform the integral of (\ref{tw53}) in the diffusion coefficient
calculation, and $w_2(r)$ is needed for
many of the tests described in this chapter. 

\begin{figure}
\input{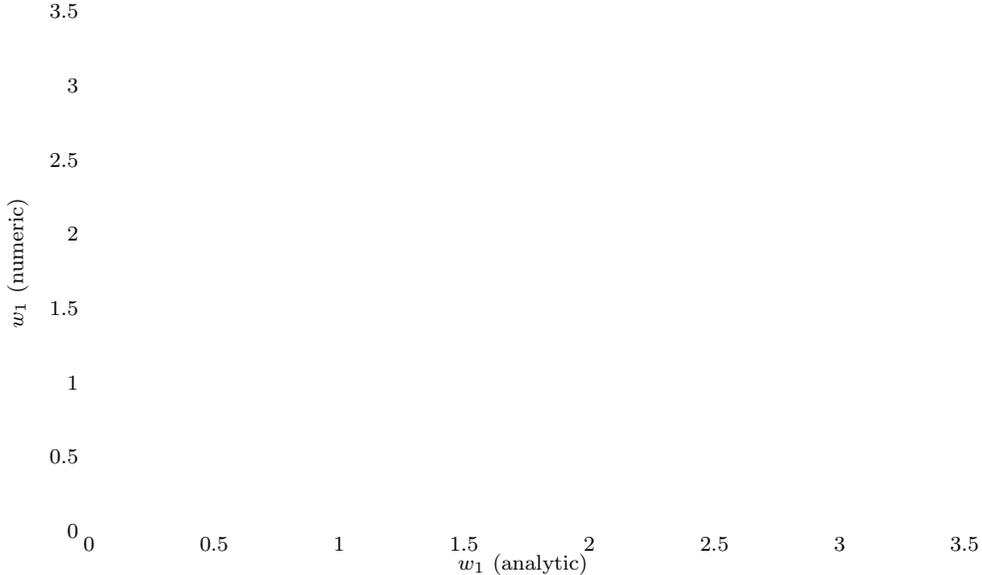}
\caption{Orbital angle $w_1(r)$ interpolated from values calculated using
(\ref{tw37}) v. the analytically-known value of $w_1$ in an SHO potential is
plotted as a dashed line.  Results of a semi-analytic test  were identical to
the  fully-numeric results shown here.  The solid line shows the diagonal, for
comparison.}
\label{angtest.r}\end{figure}

Figure \ref{angtest.r} shows the numerically-calculated $w_1(r)$ versus the
analytically-known function $w_1(r)$ in an SHO potential.  For the
calculation of the curve in Fig. \ref{angtest.r}, first $w_1(r)$ was integrated
using (\ref{tw37}) at a finite number of values of $r$ and then the shown
curve was interpolated at other points from that grid of values.  The
technique of calculating a complicated function on a finite grid and then
interpolating to determine the function's value at an arbitrary point was
employed throughout this study in order to reduce computation time, and so
this serves as a test of that procedure as well.

\begin{figure}
\input{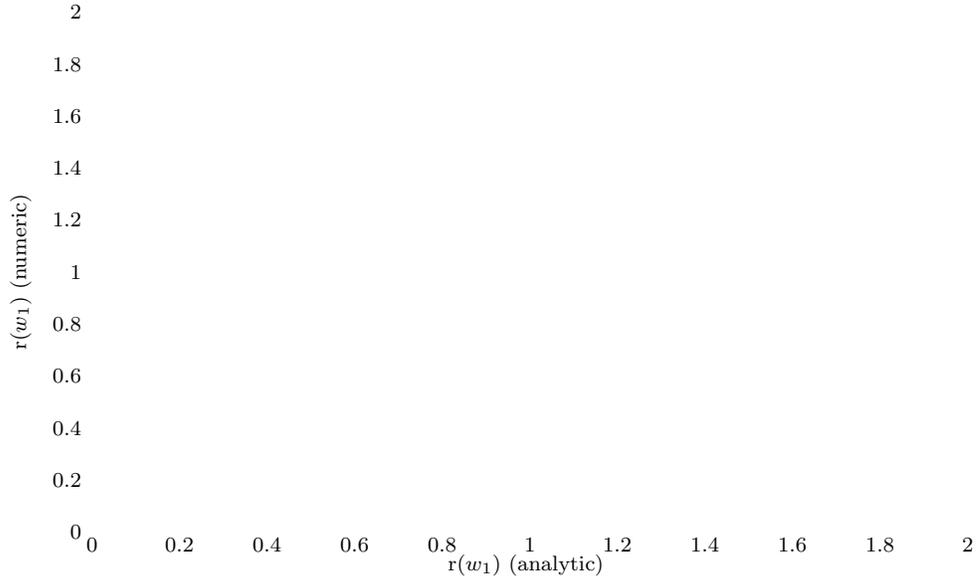}
\caption{As in Fig. \ref{angtest.r}, but here the interpolated inverse
function $r(w_1)$ is plotted v. the analytic values of $r$.}
\label{angtest.w1}\end{figure}

Figure \ref{angtest.w1} is the analogous plot of $r$, \ie in which $w_1(r)$
was calculated on a grid and then the inverse funtion $r(w_1)$ found by doing
a reverse interpolation.  In this case the interpolation does more than merely
reduce the required computation time, it is also required to actually invert the
function.  Figures \ref{angtest.w1} and \ref{angtest.r}
show that the results of the numeric calculation of $w_1$ and $r$ from $\Phi$
and $\rho$ match the true values of those orbital angles.

\newcommand{\wtp}{(w_2-\psi)}
The other interesting orbital angle term is $\wtp$ as described in
\S\ref{sect_dyn}.  Figure \ref{angtest.w2} plots $\wtp$ calculated
from (\ref{tw38}) versus the analytically-known values in an SHO potential,
for two different values of the actions $(I,J)$ (\ie across two different
orbits). The orbit plotted with ``+'' in that figure is a typical result,
while the one shown with ``$\times$'' is one of the more extreme cases. It
can be seen that although $\wtp$ isn't calculated quite as accurately as
$w_1$ (perhaps because of the delicacy of the [$\Omega_2-J/r^2$] factor in
(\ref{tw38})) neither orbit's values are far from the expected diagonal.

\begin{figure}
\input{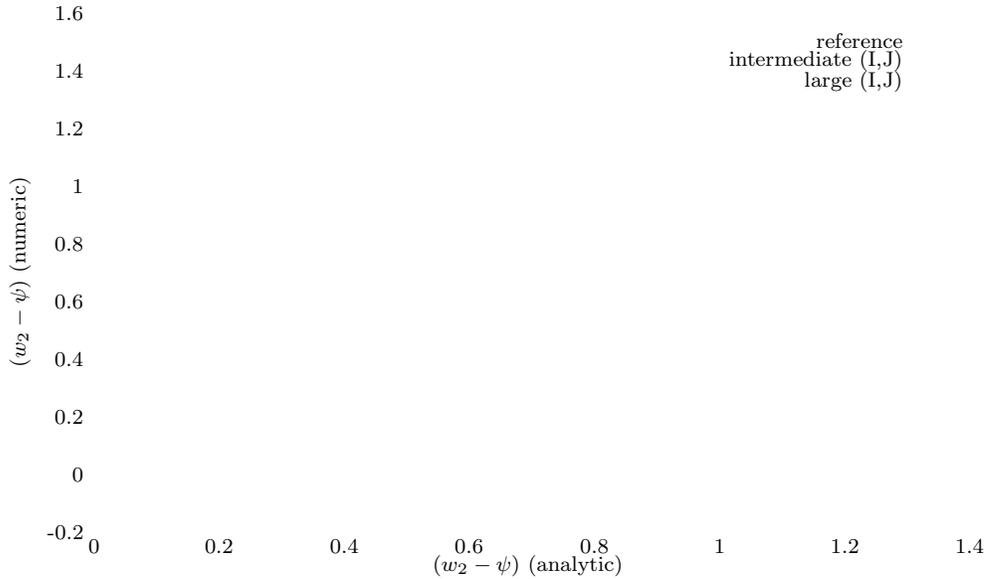}
\caption{Angle term $\wtp(r)$ plotted as numerically-calculated vs.
analytically-known values.  Points denoted by ``+'' are for an orbit with
intermediate values of actions $(I,J)$ for the SHO potential used, while
``$\times$'' are for an orbit with $(I,J)$ near the upper end of the range of
actions that are still bound in the potential.  This figure is done as a
scatter plot because, unlike $w_1(r)$, $\wtp(r)$ is not a monotonic function
of $r$.  As in Figs. \ref{angtest.r} and \ref{angtest.w1}, $\wtp$ is first
integrated from (\ref{tw38}) for a finite number of $r$ points, and then the
values plotted here interpolated are from that grid in $r$.}
\label{angtest.w2}\end{figure}


\section{Diffusion Coefficient Tests}
\subsection{Low-level Calculations}


The field-star-orbit averaging of the potential's expansion terms as given by
(\ref{dblavg}) is one of the more complex steps required in calculating the
diffusion coefficients.  But for some values of the expansion index $l$, and
over some ranges of $r/R_c$, the integral in (\ref{dblavg})
is analytically solveable in the test potentials
-- specifically, for $l=0$ or $2$ in the isochrone potential, and for
$0\leq l\leq6$ in the SHO potential.  In both cases a consistent good
match, within $10\%$ or better, was found between the numerically-integrated
values and the analytic one, and a match within $2\%$ or better between
numeric integrations using analytic or fully-numeric integrands.
(Here ``fully-numeric'' means using only code from the model itself,
with the test potential values as inputs.)  While encouraging,
this test only verifies that the integrals in (\ref{dblavg}) are being
performed accurately, and not that they are correct in their form.
For that, see below.


\subsection{Diffusion Coefficients: Reproducing the Potential}

As shown in (\ref{dij}) the major part of the diffusion coefficient calculation
consists of determining the expansion coefficients $\Psilll$ of the potential
in action space, whose relation to the (position-space) potential $\Phi_*$
of a given perturber is given by (\ref{tw55}).  Thus a fundamental test of the
validity of the diffusion coefficients is to check whether they can be used to
reconstruct the overall gravitational potential of a field star or of the bar,
effectively reversing the change of coordinates from position-space to
action space from which they were calculated.  To reconstruct the potential,
first we have to add in the $l=0$ and $l=1$ terms to (\ref{tw55}) that don't
contribute to the dynamical friction:
\beq\Phi({\bf r})=\suminf_{l_3=0}\suminf_{l_1,l_2=-\infty}\psilll\cos\reswt.
\label{TW55}\eeq
Not all of the above terms are needed: as stated in \S\ref{sect_dyn}, the
angle $w_3$ contains a random phase $\phi_o$ and so only terms with $l_3=0$
contribute to the sum, due to the cosine factor.  Angle $w_2$ includes $\psi$,
which is measured from the ascending node to the current orbital position
and is unconstrained by where in the allowed range of orbital radius $r$ the
star currently is; however, this does not imply that terms with $l_2\neq0$
do not contribute: even though $\psi$ and $w_1$ are uncorrelated, $\psi$
and $w_2$ may be and so $w_2$ does not act as a random phase as $w_3$ does.

So upon orbit-averaging, the cosine factor restricts contributions to only
those terms with $l_3=0$. Note that the cosine factor does not appear in the
actual dynamical friction calculation, and so non-zero $l_3$ terms do
contribute there -- in fact, their presence is what allows for the resonant
interaction between field star and perturber. All values of $l_1$ and $l_2$
must be included here however, as $w_1$, $w_2$ and the $\Psi$ are all
dependent upon $r$ for a given value of the orbital action vector $(I,J)$,
and so the only terms that contribute to (\ref{TW55}) are
\beq\Phi({\bf r})=
  \suminf_{l_1,l_2=-\infty}\overline\Psi_{l_1l_20}(I,J)\,\cos(l_1w_1+l_2w_2-\omega t)
  \label{phidecomp}\eeq
which can then be averaged over any object-star orbits being considered.

\begin{figure}
\input{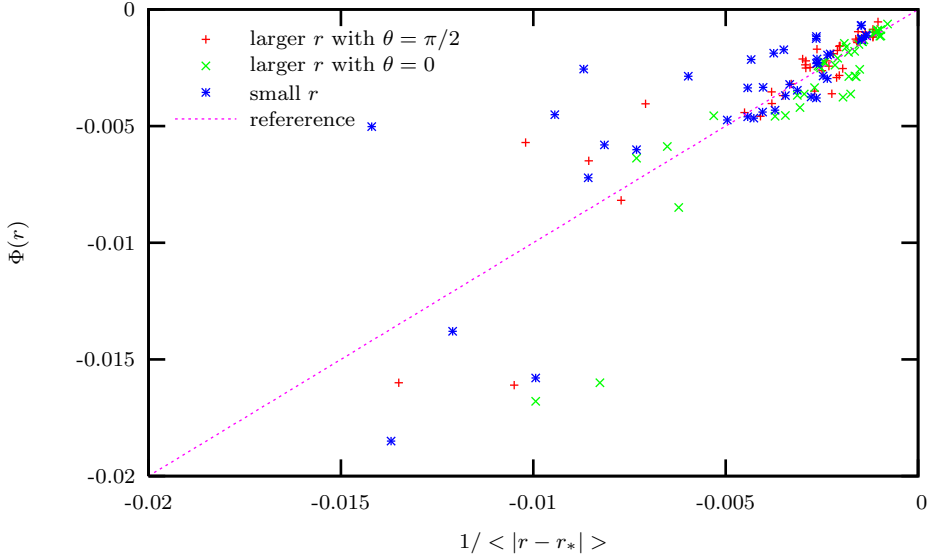}
\caption{Scatter plot showing the results of using (\ref{phidecomp}) to
reconstruct the potential $\Phi(r)=1/\langle|r-r_*|\rangle$ of a {\sl single}
star within a larger cluster from its diffusion coefficients $\Psi_{l_1l_20}$.
The various points plotted represent different choices both of test-point
location ${\bf r}$ and of object-star action values $(I,J)$. Units are arbitrary
but equivalent along both axes.}
\label{decomp0}\end{figure}

\begin{figure}
\input{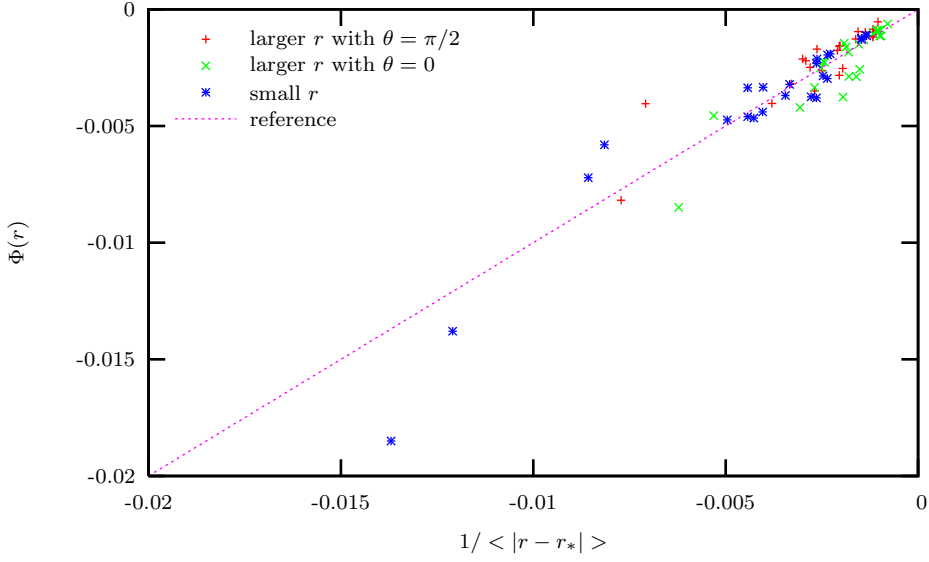}
\caption{Similar to Fig. \ref{decomp0}, but with circular and purely-radial
object-star orbits removed.
}
\label{decomp1}\end{figure}

The simplest test case to consider is that of a single object star's
potential as given by $\Phi(r)=1/|r-r_*|$.  Figure \ref{decomp0} shows
the values of $\Phi(r)$ reconstructed using (\ref{phidecomp}) from the
diffusion coefficients of object stars with various orbital actions $(I,J)$,
calculated at several locations ${\bf r}$ within the overal cluster.  In
general the match between the reconstructed $\Phi(r)$ and the expected value
$1/\langle|r-r_*|\rangle$ is fairly close, with some exceptions at larger
values of $1/\langle|r-r_*|\rangle$, \ie when the gravitational effect of the
object star is greatest.  This is not a large cause of concern however: Figure
\ref{decomp1} shows that the majority of the outliers in Fig. \ref{decomp0}
are due to object stars on nearly-circular ($I\simeq0$) or nearly-radial
($J\simeq0$) orbits; as such orbits are nearly unpopulated in the actual
distribution functions $f(I,J)$ used in the simulations they do not contribute
strongly to the overall diffusion coefficient values.  For the non-extreme
orbits that make up the bulk of $f(I,J)$, the diffusion coefficients are seen
to be able to reproduce the perturbing potential as expected.

\subsection{The Bar Perturbation}

\begin{figure}
\input{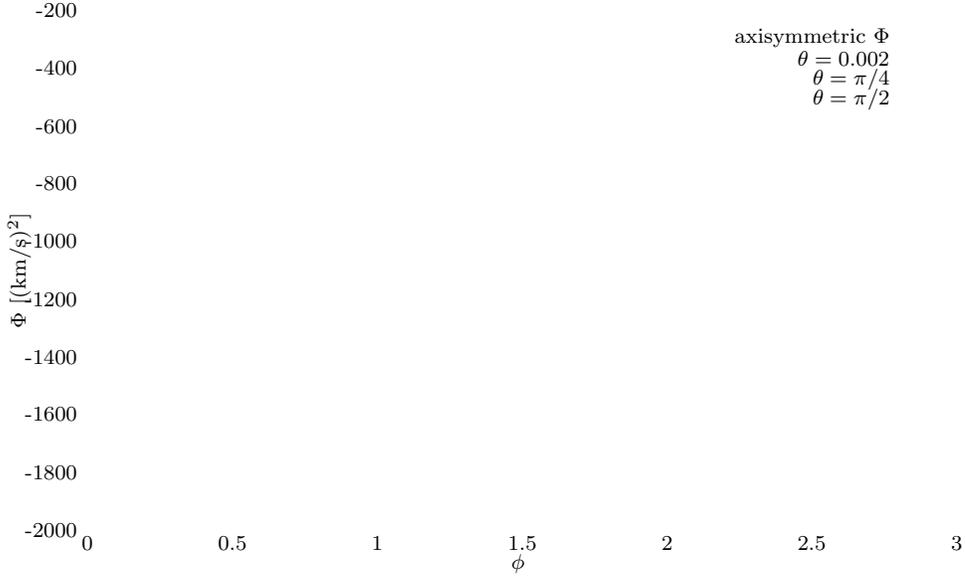}
\caption{The gravitational potential of the bar v. azimuthal angle $\phi$,
as reproduced by summing
(\ref{tw55}) using the diffusion coefficients of (\ref{effdij}).  This plot
is for $r=1.079$pc using the ``E1B'' model.  The flat line is the base
axisymmetric potential that the same stars would have if they were not
assumed to be in the bar.  The curved lines show the bar potential at
different values of polar angle $\theta$, from $\theta=0.002$ for the
least-curved line through $\theta=\pi/4$ and $\theta=\pi/2$ for the
most-curved.  
}
\label{bartest}\end{figure}

The calculation of the bar's diffusion coefficients, as shown in
(\ref{effdij}), was constructed so that the bar potential should describe a
quadrupole while still averaging to the underlying aggregate potential of its
constituent stars.  Figure \ref{bartest} shows that this calculation works as
designed: near the cluster's equatorial plane the bar potential, as
reconstructed from the diffusion coefficients using (\ref{TW55}),
is dominated by the $\Phi_{22}$ term, but towards the pole the
$\theta$ dependence of the spherical harmonic $Y_{22}$ allows $\Phi_{00}$ and
$\Phi_{20}$ to contribute most to the overall bar potential.

%

\section{Merger Losses \& Gains}\label{sect_LG}

\begin{figure}
\input{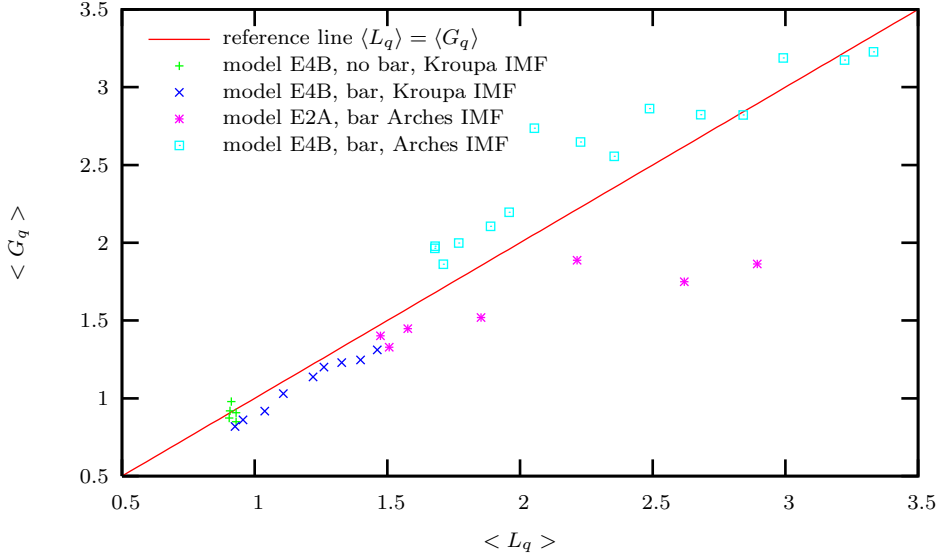}
\caption{Comparison of stellar-merger loss and gain terms $L_q$ and $G_q$
for the primary models studied, averaged over actions $(I,J)$ and stellar-mass
type.  Units are arbitrary, but consistent within a given model.  In general
$\langle L_q\rangle$ and $\langle L_q\rangle$ both increase over time, so the
progression in this plot is left-to-right for a given model.  Full results
for these simulations are given in Chapter \ref{chap_results}.}
\label{mrjmatch}
\end{figure}

The calculations of the rates of mass loss and gain due to stellar mergers,
$L_q$ (as given by (\ref{lm})) and $G_q$ (by (\ref{fullgm})), are done
independently, offering the opportunity to compare them for consistency.
Figure \ref{mrjmatch} shows the results for the three models primarily
studied: the E4B model with Kroupa- or Arches-style IMF, and the E2B model
with Arches-style IMF.  The results shown are taken from the same simulations
described in detail in Chapter \ref{chap_results} and so have upper mass
limits of $125M_\odot$ and either 9 mass bins (for the Arches-style IMF given
by (\ref{ArchesIMF})) or 10 bins (for the Kroupa IMF of (\ref{KrIMF}))
between which merger rates are calculated.

Perfect agreement between the loss and gain calculations would result in
$\langle L_q\rangle=\langle G_q\rangle$.  As can been seen from the figure
the fit is not perfect but is close for the most part, with the largest
deviation being at late times in the E2A model.  Comparing the three
E4B runs shown, it is clear that including a minimal stellar bar in the
model results in a substantial increase in the merger rates, but using an
Arches-style IMF produces much larger merger rates from the start which also
increase more over the simulation time; this will be explored more
in Chapter \ref{chap_results}.

These results, and similar ones for other models, are taken to validate the
overall method of calculating $L_q$ and $G_q$.  However as the fit between
mass loss and gain rates is not perfect, at each timestep the calculated
gain rate $G_q(I,J)$ is multiplied by a correction factor
$\langle L_q\rangle/\langle G_q\rangle$, \ie $G_q(I,J)$ is normalized so
that the effective value of the net rate of mass gain due to mergers
$\langle G_q\rangle$ agrees with the average rate of mass loss
$\langle L_q\rangle$.  ($G_q$ is adjusted to agree with $L_q$ instead of
the other way around because $L_q$ is a much simpler calculation and in
practice shows less fluctuation over simulated time.)

%

\section{Testing the Differencing Scheme}\label{sect_diffscheme}

In order to test the differencing scheme analytic solutions of the
Fokker-Planck equation (\ref{fcfp})
for both constant and varying $\IIJ$ were employed.  Both solutions have the
form $f(x,y,t)=Ae^{k_xx+k_yy+\omega t}$.  In the constant-coefficient case, the
dispersion relation thus produced is
$\omega=\half\II{x}{x}k_x^2+\II{x}{y}k_xk_y+\half\II{y}{y}k_y^2$.
For an analytic solution with non-constant diffusion coefficients, if one takes
$\II{x}{y}=0$, $\II{x}{x}(x)=P_xe^{-k_xx}$ and $\II{y}{y}(y)=P_ye^{-k_yy}$ with
$P_i=\mathit{const}$, then $\omega=0$ and a static system ($\di{f}{t}=0$) results.

\begin{figure}
\input{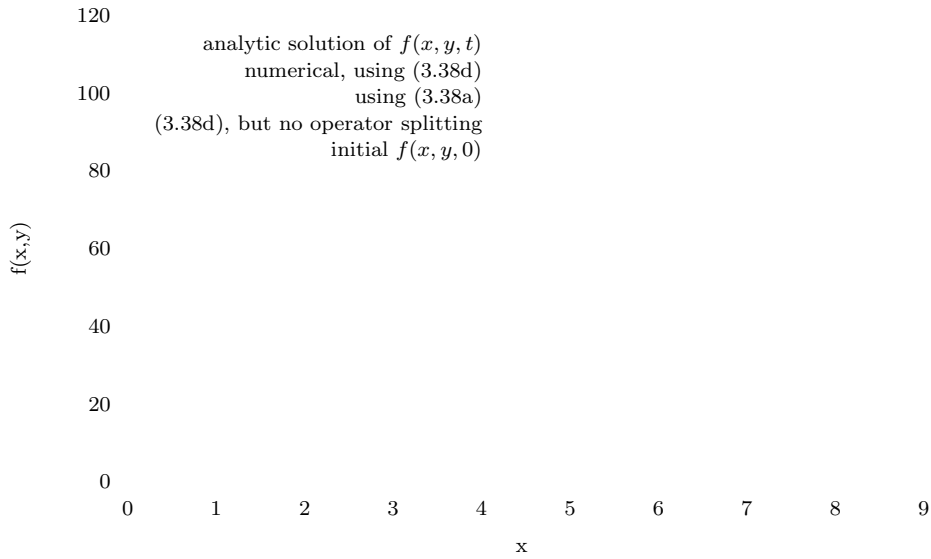}
\caption{Comparison of different finite-differencing and boundary-condition
schemes. The solid line is the analytical solution.
The long- and medium-dashed lines are numerical solutions using
(\ref{natspl}) and (\ref{strika}) respectively.  The dotted line is a
numerical solution also using (\ref{natspl}) but in which (\ref{fd}) is
solved all at once, \ie without the operator-splitting of (\ref{opsplit}).
The dot-dashed line is the $t=0$ starting solution.  The numeric solutions
are shown after 30 timesteps. (Doubling $\Delta t$ and halving the number of
timesteps produced curves indistinguishable from those shown for the
(\ref{natspl}) and (\ref{strika}) cases, and a similar curve for the
no-operating-splitting case that was only slightly different.)
This figure plots a slice midway through the range of $y$ used.}
\label{plot1xy}
\end{figure}

Figure \ref{plot1xy} shows that the numerical boundary condition of
(\ref{natspl}) tracks the true solution closely, only failing to keep up
on the edge at which the per-timstep change is greatest -- this is not surprising
given that (\ref{natspl}) implements a natural spline to deal with the edge
effects and will underestimate changes for which the second derivative is far
from zero.  This is not a large concern in practice, as tests show the
discrepency is worst when the ``crossterm'' diffusion coefficient $D_{xy}$
is of similar magnitude as the smaller of $D_{xx}$ and $D_{yy}$ (as is the case in
Fig. \ref{plot1xy}); in the actual model $D_{xy}$ is typically much smaller
than either $D_{xx}$ or $D_{yy}$ (although it does become of similar size when
all three are close to infinitessimal).  Also, in the simulations the $D_{ij}$
don't increase exponentially near the edges, meaning the natural spline is
a better approximation than it is in these tests.

\begin{figure}
\input{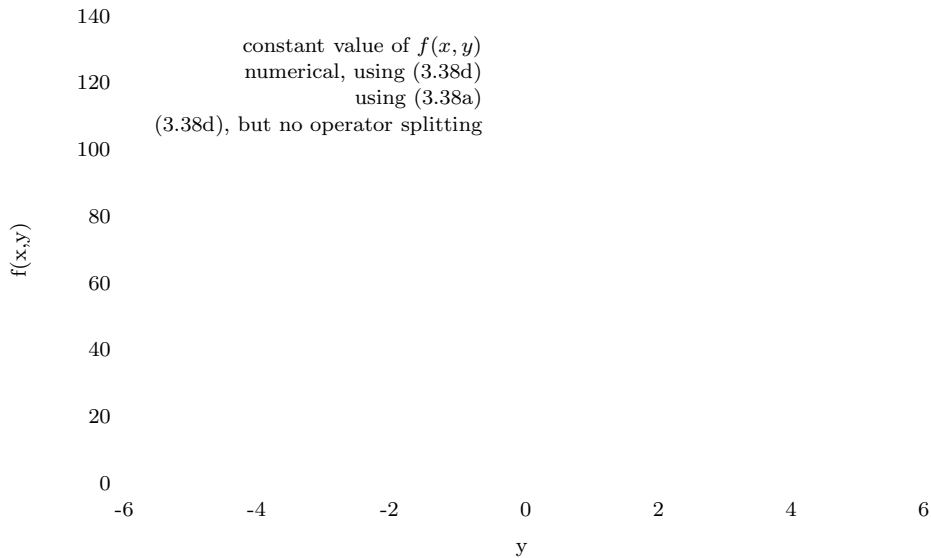}
\caption{As in Fig. \ref{plot1xy} but for the constant-coefficient,
static-solution case (thus the analytic solution is identical to the $t=0$
curve at all times).  Here $f(x,y)$ is plotted for a slice at constant $x$.}
\label{plot2yx}\end{figure}

Figure \ref{plot2yx} is similar to Fig. \ref{plot1xy} except that it plots
the constant-$f$ case and not the constant-coefficient one.

It can been seen that use of the numerical boundary condition of (\ref{strika})
produced a much poorer match to the true solution.  Use of any of the other
conditions of (\ref{striks}) resulted in similar or worse matches. Similarly,
not using a time-splitting scheme was unsatisfactory, as shown by the dotted
line in the figures.  The Courant stability-criterion values \Cite{NR} for the numeric
tests plotted were all $\xi=\frac{1}{4}$, but doubling the timestep and so
doubling $\xi$ made no difference in the time-split case, although it did
make the non-timesplit case even worse than that shown.  Increasing the $\xi$
further (\ie past $\xi\simeq1$) and/or taking a much larger number of timesteps
did start to produce larger discrepencies towards the edges where $f$ is
largest, presumably due to increased inaccuracies caused by the natural-spline
numeric boundary condition more severely underestimating the second derivative
of $f$ there.

\input{Parameters}

\section{Binary Heating}\label{sect_binheat}

As binary stars' orbits harden as a result of encounters with other stars,
the system is effectively heated by the energy tranferred from the binary
system to the cluster as a whole, and potentially disrupted somewhat.  The
physics of binary heating is not incorporated in the simulation code;
however, an upper limit on its possible effect was calculated for a
selection of models using a direct application of the binary-heating
formulation of Quinlan and Shapiro \Cite{QS90}.  The extreme assumption that
all stars in the cluster are in binaries and that all binaries harden
forever (\ie none are soft binaries that do not harden, and none are themselves
disrupted by interactions with other stars) is used to give an absolute
limit on how much binary heating could possibly have occured, had it been
included in the dynamical calculations.

\begin{figure}
\input{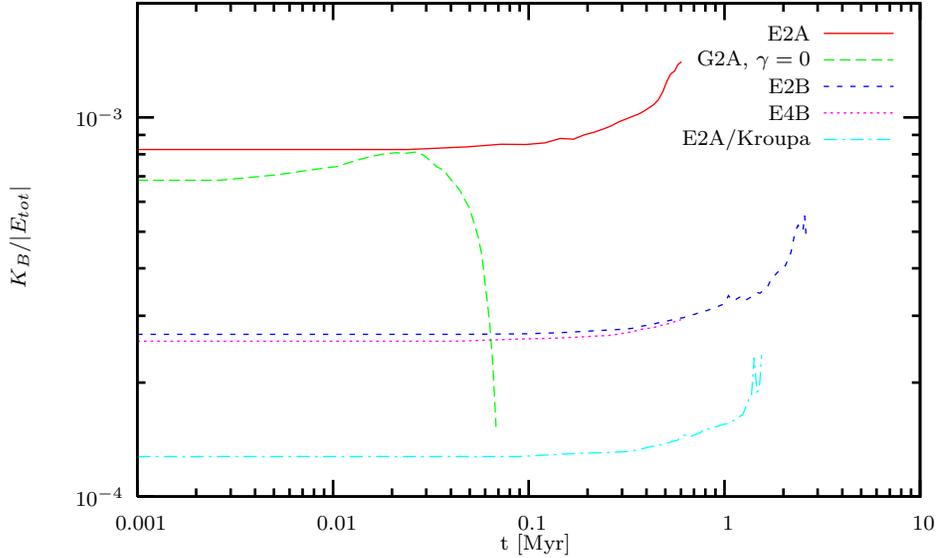}
\caption{Ratio of maximum possible amount of binary heating $K_B$ to overall
(gravitational + kinetic) system energy $|E_{tot}|$ for a variety of models.
All used the Arches-style IMF except for model ``E2A/Kroupa'', and all had an
initial Plummer-sphere distribution except for ``G2A'', which started as a
$\gamma=0$ sphere.  A logarithmic scale is used so that different models' timescales
can be plotted together; exactly flat lines at the start of each model's plot indicate
the first timestep in the calculation, before which binary heating is zero.
}
\label{binheat}\end{figure}

The results are shown in Fig. \ref{binheat}, from which it can be seen that
at all times (except at the very end of one simulation) the upper limit of 
possible binary heating remains a small fraction, $10^{-3}$ or less, of
the overall system energy.  Given that not all stars can be in binaries, that
not all binaries will harden and heat the system, and that some binaries will
be disrupted by encouters and by internal merging, the assumption that
binary heating can be neglected holds.

%% file: Parameters.tex
\newcommand{\lmax}{l_\mathrm{max}}

\section{Model Parameters}

\subsection{Grid size: action space}

\begin{figure}
\input{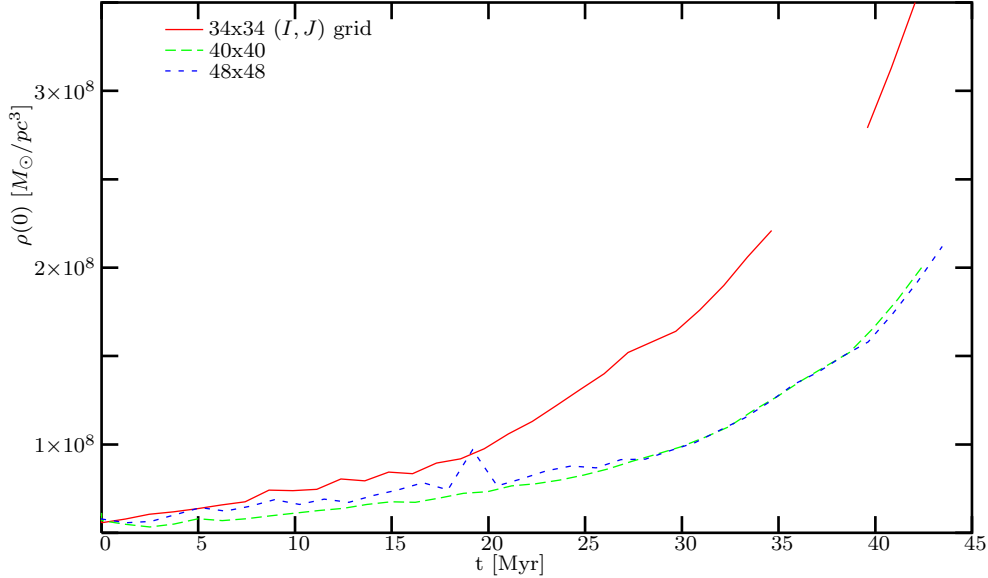}
\caption{Comparison of results of using different-size $(I,J)$ grids.
The break in the otherwise smooth 34x34 curve is a spurious artifact of the
output-calculating procedure and does not affect the model's further
evolution.  All other parameters are the same for each run, and each is
constrained to having the same $r^2$ grid as the others at each timestep.
(These runs used $r^2$ grids with 50 points, a maximum $l$ value of 9 and
a 3-component Miller-Scalo-style IMF.)
}\label{figij}\end{figure}

In order to have confidence in the model's results we must show that the
simulations are stable against changes in parameters which are purely numerical: grid
sizes, timestep sizes \etc  Figure \ref{figij} shows a representative
comparison of using different sizes of action-space grids.  In this and
several other tests, a 40x40 grid was found to give results similar to
those of larger grids.  (In most cases, a 38x38 grid was also sufficient,
but anything smaller would deviate.)  For the full simulations discussed
in Chapter \ref{chap_results} a 40x40 grid was used unless otherwise noted.

Choosing an even larger action-space grid is possible, but at a considerable expense
of computer time: in addition to simply having more gridpoints on which all
calculations need to be performed, in practice a larger grid also required
smaller timesteps in order to remain stable.

\subsection{Grid size: radial coordinate}\label{subsect_gridsize}

The situation for the grid in the radial coordinate $r^2$ is similar to that
for $(I,J)$ in that a too-course grid produces evolution that is unstable to
accumulated errors; however using a too-fine $r^2$ grid results in noisy output
which causes difficulty for the model's mechanism of solving for the new
potential $\Phi(r)$.  In practice a 50-point $r^2$ grid produced a good balance
between these two effects, although sometimes at the cost of an artificially
slower evolution of the model as compared to finer grids.

\begin{figure}
\input{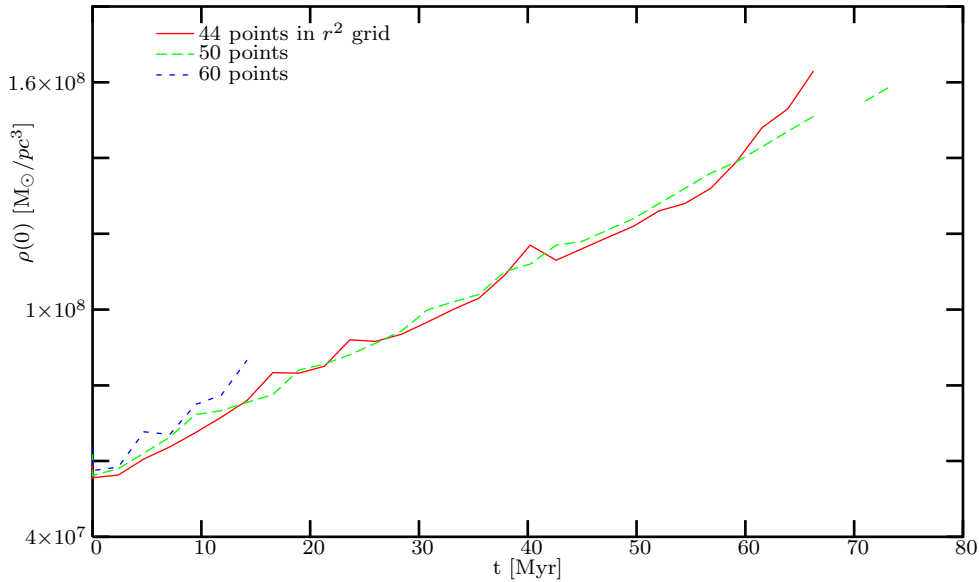}
\caption{
Comparison of runs with $r^2$ grids containing 44, 50 and 60 points; other
parameters (timestep, action-space gridsize \etcend) were identical for each
of the 3 cases.  In order to do a direct comparsion, each run was prevented
from adjusting its $r^2$ grid dynamically.  These runs were for model E4B
with $\lmax=4$ and an initial rotation parameter $\lambda=0.05$.
The mass spectrum was 90\% $1 M_\odot$ and 10\% $2 M_\odot$.
}\label{fignm2}\end{figure}

Figure \ref{fignm2} shows a comparison of use of 44-, 50- and 60-point grids.
The 60-point case evolves slightly more quickly, as there are more gridpoints
near the dense center of the cluster, but it aborts after only a few timesteps.
Unlike the $(I,J)$ grid, which is fixed, in the simulations described in
Chapter \ref{chap_results} the $r^2$ grid is allowed to dynamically
self-update at each timestep, as described in \S\ref{subsect_grid}.
In practice the dynamic updating of the $r^2$ grid helps the finer-grid case
adjust better than shown here: for the runs shown in Fig. \ref{fignm2}
the dynamic updating was disabled in order to allow for a direct comparison
between grid choices.  Even finer 66- and 74-point grids were also tried
and gave results similar to the 60-point case; a 99-point grid was found to
require too-small timesteps with no benefit.  Courser grids of 40 or 36 points
could not track the center of the cluster with sufficient resolution.

\begin{figure}
\input{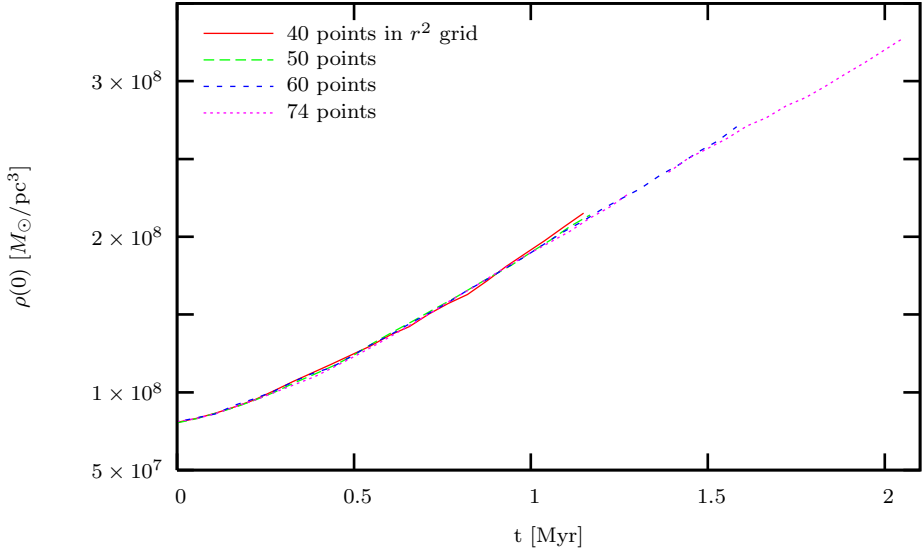}
\caption{
Comparison of runs with $r^2$ grids containing 40, 50, 60 and 74 points using
a 40x40 $(I,J)$ grid, $\lmax=3$ and identical timesteps in each case.
Dynamic $r^2$ grids were enabled, but were similar in all 3 cases.
These runs were for a 7-mass component cluster (mass range $1-32M_\odot$ in
7 bins) with a moderate amount of rotation ($\lambda_0=0.015$), Kroupa-style
IMF and with 1\% of the cluster mass in a stellar bar.
}\label{e4b4rgrid}\end{figure}

A somewhat more typical set of simulations is shown in Fig. \ref{e4b4rgrid},
in which a broader range of stellar masses is used, a stellar bar is included
and the $r^2$ grid is allowed to update itself dynamically at each timestep.
This results in a much smoother evolution in which the 40, 50, 60 and 74-point
$r^2$ grids give almost identical results, although in this case the simulation
using a 74-point grid continued longer than the others.  Note
that the presence of the stellar bar and of a subpopulation of higher-mass stars
caused much more rapid rate of change of the central density compared to that of
the previous example.

Given these findings a 74-point dynamically-updating $r^2$ grid was the default
choice.  In some simulations a 60-point grid was used; those cases are noted.

\subsection{Timestep size}

The choice of timestep size necessarily varied with what cluster parameters
were used as an initial condition; a typical choice was a fraction
$\lesssim0.1$ of the cluster's initial central relaxation time
$t_{rq}(0)$ as given in \S\ref{sect_oafp}: $\Delta t\lesssim0.1 t_{rq}(0)$.
(Stellar mass $m_q$ here is that of the dominant stellar mass in the cluster.)
The goal was to maximize the amount of evolution per timestep in
order to avoid inordinately long computation times, while still having
confidence in the run's results, as demonstrated by selectively performing
similar runs with smaller timesteps and obtaining similar results.

\begin{figure}
\input{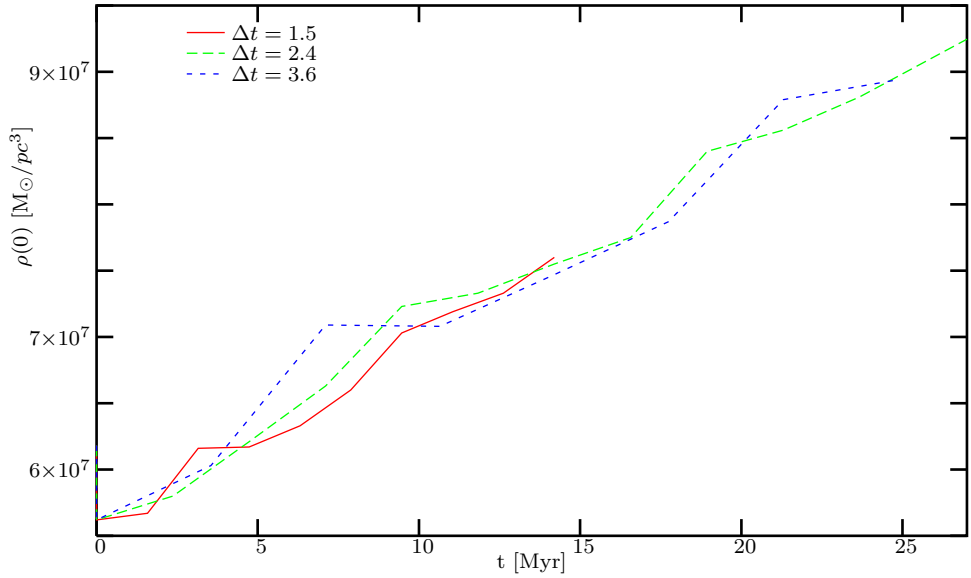}
\caption{
Comparison of the use of different-size timesteps.
Runs are with $r^2$-grids of 50 points; the $\Delta t=2.4$ case continued
until $t\simeq73$, as can be seen in Fig. \ref{fignm2}.  Initial central
relaxation time was $t_{rq}(0)=25.6$ Myr.
}\label{figdt}\end{figure}

Figure \ref{figdt} shows that the simulation model is stable against different sizes of
timestep, although for different choices of timestep size the model may abort
much earlier than for others (usually due to lack of numerical convergence
when iteratively solving for the new $\Phi$ or $\rho$).  When this happens,
sometimes the run can be stably restarted from a slightly earlier time value
but with a smaller timestep, sometimes not; for consistency all simulations
presented are single runs starting at $t=0$.

\begin{figure}
\input{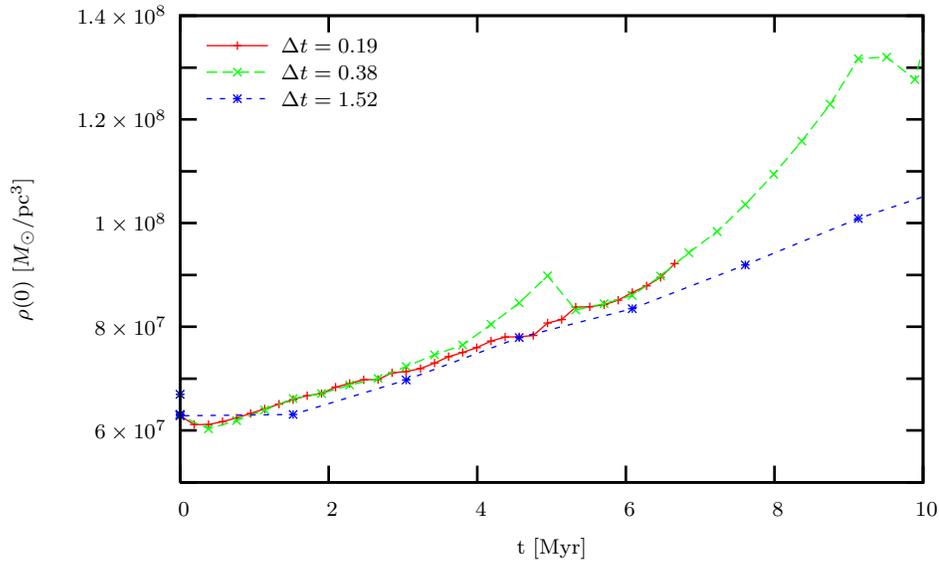}
\caption{Comparison of the use of different-size timesteps.
Runs are of model E4B with an $r^2$-grid of 74 points, a 40x40 $(I,J)$ grid, a
maximum $l$ of 2 and an initial rotation given by $\lambda_0=0.05$ and a Kroupa
IMF with 10 stellar mass bins in the range $1-125M_\odot$.  Initial central
relaxation time for the $1M_\odot$ stars that comprised 51\% of the cluster was
$t_{r1}(0)=158$ Myr; for comparison the $25M_\odot$ stars' mass fraction was
2.4\% with $t_{r7}(0)=0.23$ Myr.
}\label{e4bdt10}\end{figure}

\begin{figure}
\input{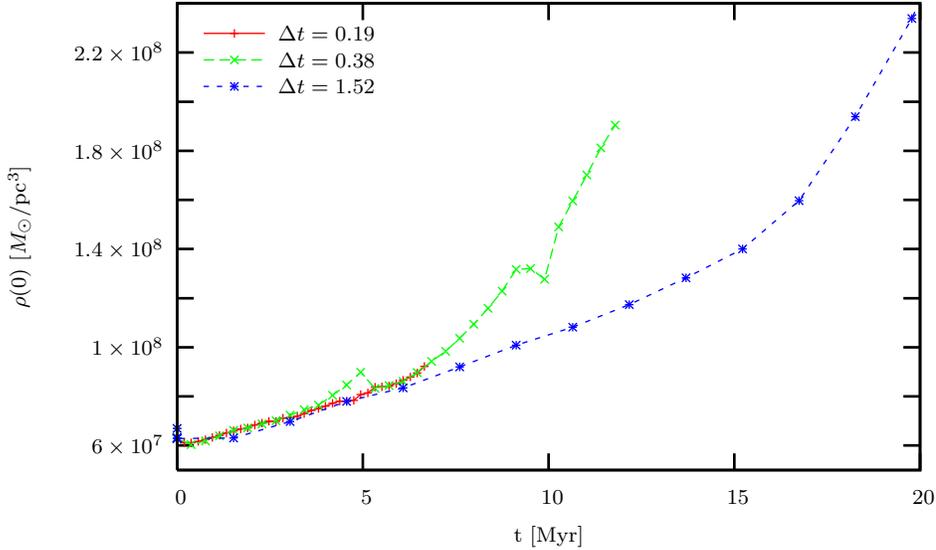}
\caption{Similar to Fig. \ref{e4bdt10} but showing the full range of the
$\Delta t=1.52$ run.
}\label{e4bdt20}\end{figure}

Figure \ref{e4bdt10} displays a set of runs more representative of the actual
simulations presented in Chapter \ref{chap_results}: a wider range of masses is
employed with a realistic IMF, and the $r^2$ grid is dynamically updated as the
system evolves.  Here the largest choice of $\Delta t$, while smooth, clearly
does not track the progression of the central density  as accurately as do the
smaller timestep choices.  The similarity of the two smaller-timestep runs is
interpreted as indicating a reliable outcome, with the extra rise of $\rho(0)$
at $t=4-5$ Myr for the $\Delta t=0.38$ Myr run taken to be a spurious artifact
of the output-calculating procedure as seen earlier.  (Note that in any case
for the upper end of this range of masses the main-sequence lifetime is only
3 Myr \Cite{McZwart}.)  The full span of the largest-$\Delta t$ run is shown in
Fig. \ref{e4bdt20} for comparison, and shows that the $\Delta t=0.38$ run
becomes possibly unstable around $t\gtrsim8$ Myr; if reliable results near
or after $t\simeq7$ Myr had been required, it would have been necessary to
attempt another smaller-$\Delta t$ run as confirmation.

\subsection{Number of expansion terms}

\begin{figure}
\input{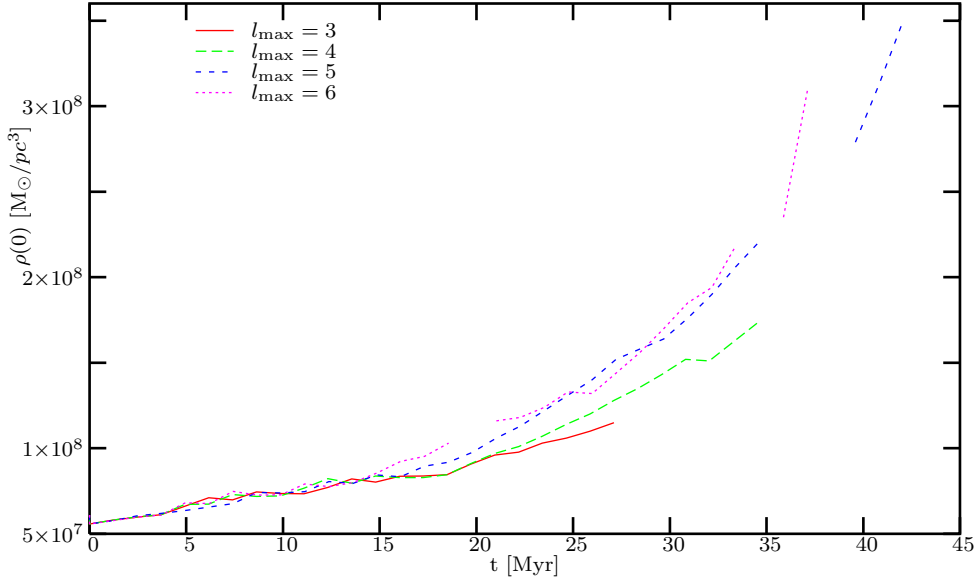}
\caption{
Demonstration of the use of an $\lmax$ value of 3, 4, 5 and 6 in the
diffusion coefficient expansion (\ref{dij}) (and by extension in the
expansion of the potential (\ref{philm})); this corresponds to a total number
of expansion terms of 4, 6, 9 and 12 respectively.  This plot is of a rotating
2-component cluster similar to those of Figures \ref{fignm2} and \ref{figdt}.
Each run was constrained to use the same timestep and the same $r^2$-grid as
the others, to allow for direct comparison.}
\label{fig2nlm}\end{figure}

The main physical quantity that is approximated by a series expansion is the
gravitational potential $\Phi(r)$, which is expressed in term of spherical
harmonics, as shown in (\ref{philm}).  This results in the diffusion
coefficients also being representated in the form of a
series, as in (\ref{dij}); physically, cutting off the series at a certain
maximum spherical harmonic index $l=\lmax$ implies only summing over
orbital resonances of indices equal to or lower than $\lmax$.  Figure
\ref{fig2nlm} shows comparisons of different choices of $\lmax$ value.
The cases with $\lmax$ of 3 (\ie up to the octopole term in the
potential) and 4 do not capture the full effect of the dynamical friction,
which can be seen as starting to converge around $\lmax$ of 5.
(The choices $\lmax=$3, 4, 5 and 6 correspond to a total number of
expansion terms of 4, 6, 9 and 12 respectively when all allowed values of
index $m$ are included.)

\begin{figure}
\input{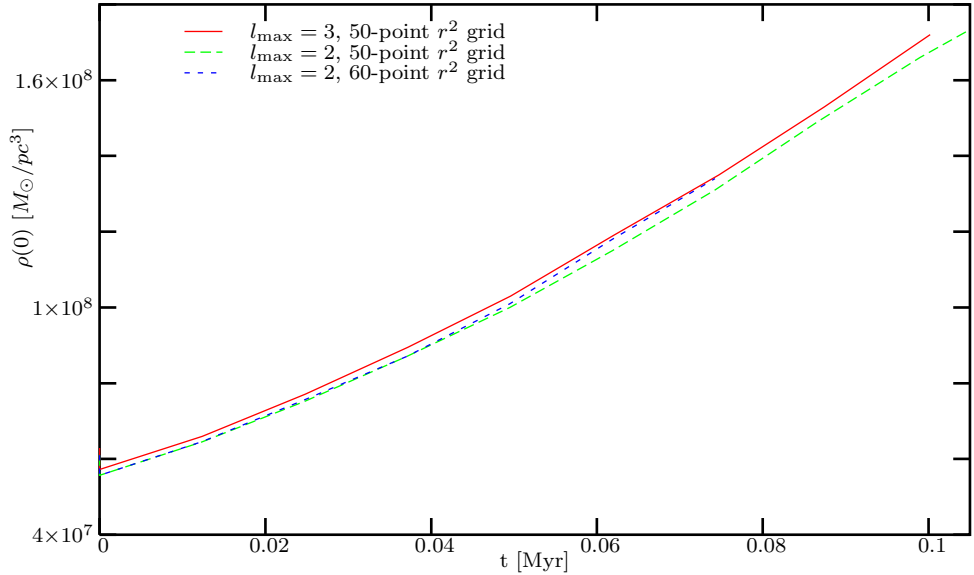}
\caption{
Comparison of different numbers of expansion terms and $r^2$ gridpoints for the
case including a bar perturbation.  The 50-point, quadrupole-only run continued
on a fairly linear path (not shown) until it reached $\rho\simeq2\sn8$ at $t=0.2$.  This
run is for a 2-component cluster with 10\% $2M_\odot$ and 90\% $1M_\odot$ stars,
4\% of which comprised the bar.  The rotational parameter was $\lambda=0.075$.
A run with $\lmax=4$ (\ie 6 expansion terms) was nearly identical to the one
shown using $\lmax=3$, although it ended earlier.
}\label{fig2bar7}\end{figure}

When a strong stellar bar perturbation is incorporated in the potential,
the situation can change dramatically.  As comparing the timescale of Fig.
\ref{fig2bar7} with that of Fig. \ref{fig2nlm} shows, the bar potential
dominates the dynamical friction.  Physically this is because the bar is a
bulk perturbation whose total contribution to the potential is squared (\cf
(\ref{effdij})) whereas ``field'' (\ie non-bar) stars only enter individually
into the $\Psi^2$ factor in (\ref{dij}) before having their contributions
summed over. Because the bar is by construction quadrupole-only and its
dynamical friction timescale is so much shorter than that for field stars,
higher-order resonances become almost irrelevent to the overall cluster
evolution.  So, unless otherwise stated all simulations which incorporate a
bar perturbation have $\lmax=2$ or $3$, which results in only two or four
terms in the expansion (\eg the two terms $m=\pm2$ if $\lmax=2$).

\begin{figure}
\input{nlm/e4bbar-nlm}
\caption{Comparison of different numbers of expansion terms for a
``production'' run: model E4B with a Kroupa IMF, initial rotation parameter
$\lambda_0=0.05$ and a stellar bar mass fraction of 1\%.  The stellar mass
range is $1-125$ $M_\odot$.
}\label{e4bbar4-nlm}\end{figure}

For the ``production'' simulations presented in Chapter \ref{chap_results}
it was found that the choice of $\lmax$ needed to be determined on a
case-by-case basis.  Figure \ref{e4bbar4-nlm} shows the behavior of one of
the main models used, model E4B with a Kroupa IMF and a stellar bar with
mass fraction of 1\%.  In this case $\lmax=2$ gave a fairly steady evolution;
higher values of $\lmax$ hinted at a faster increase in central density
$\rho(0)$ but were not sufficiently numerically stable to be considered
robust; this $\lmax=2$ was used, understanding that it may not quite capture
the full progress of the system.

\begin{figure}
\input{nlm/e2a2-12}
\caption{Comparison of different numbers of expansion terms for a
``production'' run: model E2A with a Kroupa IMF, initial rotation
parameter $\lambda_0=0.05$ and a stellar bar mass fraction of 1\%.  The
stellar mass range is $1-125$ $M_\odot$.
}\label{e2abar-nlm}\end{figure}

Contrasting this is the case shown in Fig. \ref{e2abar-nlm}, of model E2A with
a Kroupa IMF.  Here it is clear that $\lmax=5$ and $\lmax=6$ give almost
identical results, while $\lmax\leq4$ misses a considerable degree of the
increase in central density.  However, the $\lmax\geq5$ runs terminate quite
a bit earlier than the lower-$\lmax$ simulations, and require considerably
more computational time (weeks compared to days) to even get as far as they do.
(Taking $\lmax=4$ produces a total of 6 terms in the expansion, while $\lmax=6$
gives 12 terms).  So in the case of model E2A/Kroupa an $\lmax$ of 4 was used,
but only to give an indication of how the system evolves and not to show
precise quantitative results.

The above two examples are the extreme cases of those studied; in general the
bar perturbation proved dominant enough that $\lmax=2$ or 3 sufficed to
describe the system's evolution while still allowing for numerical stability
and adequately-short computation times.

\subsection{Mass Spectrum: Discretizing the Initial Mass Function}

\begin{figure}
\input{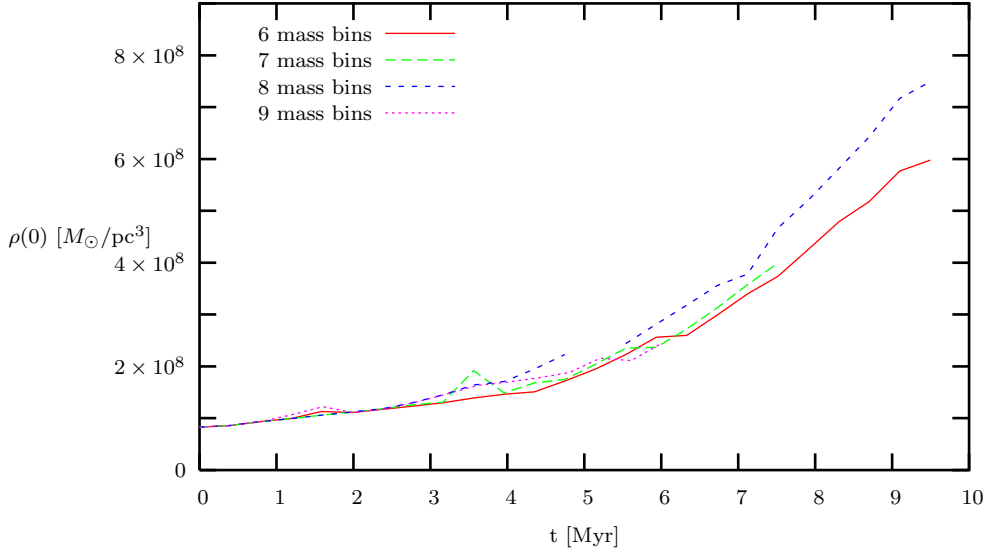}
\caption{Comparison of central density for runs in which the stellar mass
spectrum is split into the number of geometrically-spaced mass bins shown.  All are
model E4B with 50 points in the $r^2$ grid, a 34x34 $(I,J)$ grid, and an overall
mass range of $1-32M_\odot$.  The IMF is Salpeter-like, but with the
uppermost mass bin artifically overpopulated in order to accentuate any
differences in the cases.
}\label{IMFall}\end{figure}

\begin{figure}
\input{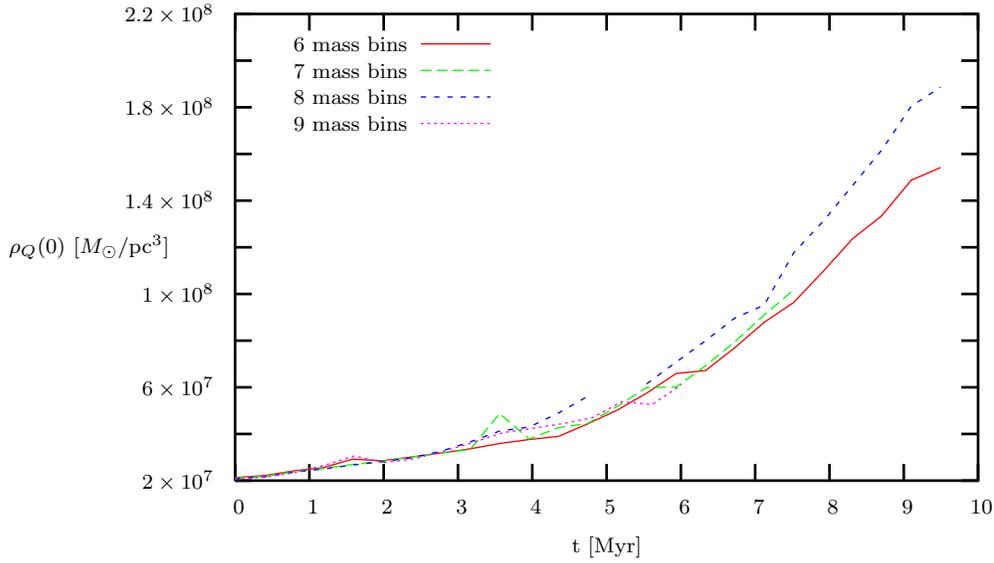}
\caption{Similar to Fig. \ref{IMFall} but showing the central density of the
uppermost ($m_Q=32M_\odot$) stellar mass bin only.
}\label{IMFtop}\end{figure}



\begin{figure}
\input{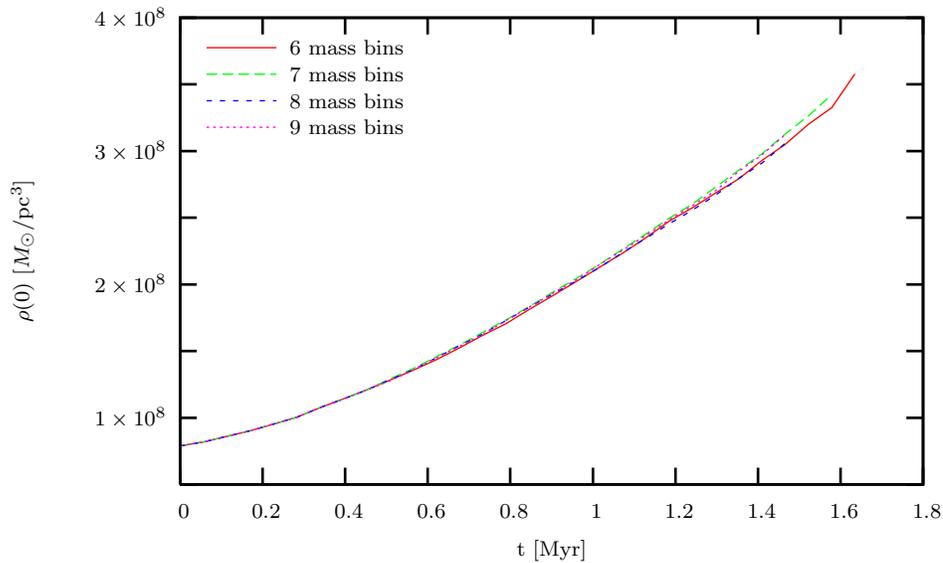}
\caption{Comparison of central density for runs in which the stellar mass
spectrum is split into the number of geometrically-spaced mass bins shown.
All are model E4B with 50 points in the $r^2$ grid, a 34x34 $(I,J)$ grid, and
an overall mass range of $1-32M_\odot$.  The IMF is Kroupa-like, similar to
that used in the later full simulations.  Rotation and a stellar bar are
included, with a bar mass fraction of $1\%$ and an initial rotation parameter
of $\lambda_0=0.016$.
}\label{rotbin}\end{figure}



\begin{figure}
\input{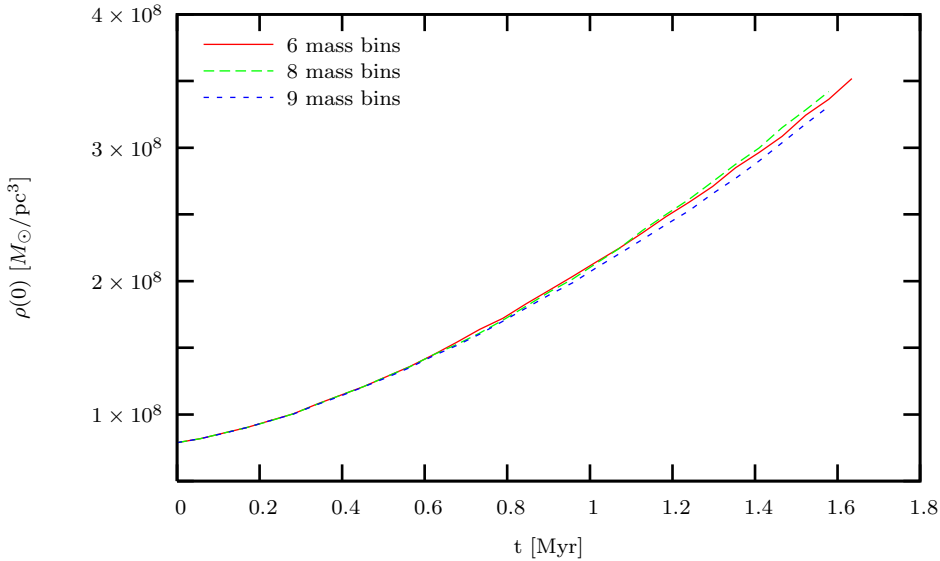}
\caption{Similar to Fig. \ref{rotbin} but with stellar mergers included in
the simulation.  (The 9-bin case shows slightly lower central density than
the others due to it following a somewhat different series of values for the
innermost point of its dynamic $r^2$ grid; this was unusual and did not
occur during the full simulations presented in Chapter \ref{chap_results}
except where noted.)
}\label{e4b4rotmrj}\end{figure}

The initial mass functions described in \S\ref{subsect_IMF} are continuous
functions of the individual stellar mass $m$; in order to separate a given IMF
into discrete bins of discrete stellar mass $m_q$ the
IMF is simply integrated over the range of masses being considered, with the
boundary between bins being the midpoint between them.  The lowest mass bin's
lower limit is taken to be $\half{m_1}$, and the highest-mass $m_Q$ bin is
given the same width as it would have if there was an $m_{Q+1}$ bin with the
same scaling.  As the spectrum of mass bins is geometrically increasing by a
factor $m_{q+1}\lesssim2m_q$, this prescription artifically biases the
numerical IMF slightly towards lower masses as compared to the analytical form
on which it is based.

Figure \ref{IMFall} shows the evolution of the central density for test
runs of model E4B with a stellar mass range between 1 and $32M_\odot$ and a
Salpeter-like IMF in which the $32M_\odot$ bin has been overpopulated, so that
high-mass stars drive the system's evolution more strongly than would occur
naturally.  Figure \ref{IMFtop} simlarly shows results for the same runs, but
for the $32M_\odot$ stellar mass bin only.  Although the curves of the central
density in this artifical case are somewhat rough, a choice of 7 or 8 mass bins
seems to best show the full evolution of the system while avoiding the
numerical instability displayed by the 9-bin run.  For a range of
$1-32M_\odot$, 7 bins corresponds to a mass ratio between adjacent bins of
1.78, and 8 bins gives a ratio of 1.64.  The full simulation runs shown in
Chapter \ref{chap_results}
used adjacent-bin mass ratios of 1.71 for the Kroupa IMF and 1.675 for the
Arches-style IMF.  This is similar to and consistent with the findings of
Amaro-Seoane \Cite{Amaro} who found that an average mass-bin ratio of $\simeq1.72$ was
sufficient to model the Fokker-Planck evolution of clusters of $0.2-100M_\odot$
stars which contained already-formed or primordial massive central objects.

Figures \ref{rotbin} and \ref{e4b4rotmrj} show the central density evolution
using a more-realistic Kroupa-style IMF and incorporating a stellar bar and
a moderate amount of rotation.  It can be seen that both with and without
stellar mergers, the choice of number of stellar mass bins does not strongly
affect the simulation results; as seen in Fig. \ref{e4b4rotmrj} the dynamic
$r^2$ grid can sometimes have a greater, but still small, effect.

%% file: Results.tex
\chapter{Results}\label{chap_results}






\section{General Considerations}
\subsection{Initial Models}

The progression of initial potential-density models studied starts with the
E2A and E2B Plummer spheres which, as listed in Table \ref{tabics} and
described in \S\ref{plummer}, can represent the nuclei of dwarf elliptical or
bulgeless spiral galaxies, as well as very dense globular clusters.  Model E4B
is somewhat more massive and more dense with a higher initial velocity
dispersion, along the lines of the core of a giant elliptical galaxy.  Model
E4A starts out even more dense than E4B, but was found to be numerically
unstable and so could not be simulated.

An alternative to the Plummer sphere is an initial ``$\gamma=0$ sphere'',
which can represent galactic spheroids.  When mapping the masses and core radii
of the above Plummer models onto $\gamma=0$ spheres, only the resulting model G2A (analogous to Plummer model E2A)
has an initial central density and velocity dispersion within the ranges
required for modeling dense astrophysical systems without assuming an
unreasonably high central density.  However, a model G3C, which starts
with the highest initial density studied, was run for comparison purposes,
as was intermediate model G3A; these are described later in this Chapter.

For each of the various potential-density models the distribution of stellar
masses is given by the initial mass function (IMF).  As developed in
\S\ref{subsect_IMF} the base IMF used here is the Kroupa IMF, which is a
good fit for the observed galactic stellar population in general.  The second
is an Arches-style IMF which has recently been determined to specifically
describe the stellar distributions in dense clusters observed at the centers
of galaxies, \ie the objects being simulated here.  Unsurpringly the Kroupa IMF is
more weighted towards lower-mass stars while the Arches IMF has a flatter
distribution of masses, but still with an upper cutoff of $m\lesssim150M_\odot$.

\subsection{Physical Effects}

For sufficient buildup of the cluster's density to occur so that a massive
object can form as the product of collisional mergers of stars, rotational
support against the infall of stellar material to the central region
must be overcome.  As described in Chapter
\ref{chap_intro}, given infinite time such collapse would be expected even in
a nonrotating system if it is sufficiently dense.  However, if it is to
occur before the majority of stars reach the end of their main-sequence
lifetimes -- only 3 Myr for the largest stars in the IMFs considered --
something in addition to interactions of individual stars is expected to be
required.  Fortunately, even though they were not intentionally constructed
to be so, when given a reasonable value for the rotation parameter $\lambda$
the models chosen all satisfy the criterion for being unstable against the
creation of a bar-like perturbation in the stellar potential; this stellar bar
can (hopefully) then
transport angular momentum in bulk from the system's center to its outer regions.

Thus the problem is essentially a race against time: can gravitational
effects, aided by the transport of angular momentum afforded by a stellar bar,
allow the system's central regions to condense quickly enough that collisional
mergers can in turn produce an object of $M>250M_\odot$ which will evolve into
a massive seed black hole, all before the stars outlive their main-sequence
lifetimes.  Simulations for each model and IMF are performed incorporating all
of: an overall rotation of the system, a stellar bar, and
collisional mergers of stars.  In order to examine the various effects some
simulations are also presented in which the system is not rotating, it does
not contain a stellar bar, and/or the stars are not allowed to collide and
merge to produce larger stars.

After a brief summary of what constituted successful simulations and which
models yielded them, the remainder of this chapter reports results for each
choice of initial model and IMF.

\subsection{Overview}

Of the five possible initial models whose density-potential pairs are listed
in Table \ref{tabics} and described in \S\ref{subsect_potdens} (or eight, when
models G2A, G3A and G3C from \S\ref{gamma0} are included), only some yielded
simulations which satisfied all the criteria for stability.  The criteria
included:
\begin{itemize}
\item giving consistent results with different choices of timestep size $\Delta t$;
\item there existing a value of the cutoff $\lmax$ of the potential expansion
high enough that
terms $l>\lmax$ did not contribute significantly 
(as indicated by the central density's increase with time), but low enough to avoid
the numerical instability higher values of $\lmax$ exhibited; and
\item giving consistent results across the Sun, 
Alpha, and AMD/x86 computer architectures.
\end{itemize}
The models which met the above criteria are listed in Table \ref{ICguide}.  For
each model a possible astronomical system to which it best corresponds (if any)
is listed; this expands on the descriptions given in \S\ref{subsect_potdens},
which did not yet take choice of IMF into consideration.
\textit{Table \ref{ICguide} can serve as a convenient reference for the
model-specific sections that comprise the remainder of this chapter.}

\begin{table}
\begin{tabular}{cccc}
Initial Profile &Model Set& IMF & Corresponding Astronomical System\\ \hline\hline
&E2A/E2B&Kroupa&dwarf elliptical or bulgeless spiral galaxy nucleus\\ \cline{3-4}
Plummer
&    &Arches& nuclear cluster \\ \cline{2-4}
sphere
&E4B&Kroupa& core of a giant elliptical galaxy \\ \cline{3-4}
&    &Arches& \textit{none -- inappropriate IMF} \\ \hline
&G2A&Kroupa& galactic spheroid or spiral bulge \\ \cline{3-4}
$\gamma=0$
&    &Arches& \textit{none -- inappropriate IMF} \\ \cline{2-4}
sphere
&G3A&Kroupa& galactic spheroid or spiral bulge \\ \cline{2-4}
&G3C&Kroupa& \textit{none -- initial central density too high} \\ \hline
\end{tabular}
\caption{Models which yielded stable results, along with possible
analogous astronomical systems.}
\label{ICguide}
\end{table}

Simulations 
with 
a Kroupa IMF used a mass range of $1-125M_\odot$; the Arches IMF has a smaller
contribution from low-mass stars (\cf \S\ref{subsect_IMF}) and so employed a
range of $2-125M_\odot$.
The set of possible variations of each model include: whether the system has
initial rotation (with the rotation parameter set near the canonical value
$\lambda\simeq0.05$ as described in \S\ref{subsect_introrot}) or not
($\lambda=0$); if rotating, whether there is a stellar bar or not;
and whether stellar mergers are allowed or not.  For both the Kroupa and the
Arches-style IMF, two models produced stable results in at least some cases:
E2A and E4B.  The E2B model was stable as well but only when using the
Arches IMF.  The results for each model are described in detail below, and
are summarized in Tables \ref{mrjkroupa} and \ref{mrjarches} for the ``full''
runs which included rotation and stellar mergers.

\begin{table}
\begin{tabular}{lcccllcccc}
Model & $\lmax$ & bar & $\rho(0)$ at $t_0$ & $\rho(0)$ at $t_1$ & $t_1$ &
$G_{250}(t_0)$ & $G_{250}(t_1)$ & $S(t_1)$ & $S'(t_1)$\\
\hline\hline
E2A & 4 & 5.3\% & $2.5\sn7$ & $3.0\sn8$ & 1.6 & 0. & 0. & 0.82 & 0.83\\
\hline
E4B & 2 & none & $6.3\sn7 $ & $7.0\sn7$ & 2.5 & 0. & 0. & 1.09 & 1.11\\
    & " & none & "          & $7.2\sn7$ & 3.0 & 0. & 0. & 1.09 & 1.11\\
    & " & 10\% & "          & $3.8\sn8$ & 2.5 & 0. & $6\sn{-5}$ & 1.006 & 1.02\\
\hline
G2A & 4 & none & $1.7\sn7$ & $1.3\sn8$ & .42 & 0. & 0. & 0.95 & 0.96 \\
\hline
G3A & 3 &7.7\%& $4.4\sn7$ & $1.1\sn9$ & .46 & 0. & 0. & $0.87$ & $0.87$\\
\hline
G3C & 2 &7.7\%& $1.4\sn8$ & $3.0\sn9$ & .16 & 0. & 0. & $1.04$ & $1.04$\\
\hline\hline
\end{tabular}
\caption{List of models which yielded stable simulations using the Kroupa IMF,
with central-density and merger-rate results shown at various times $t_1$
in each simulation (usually the endpoint, or chosen for comparison with
another).  Times are in Myr, central densities in $M_\odot/\mathrm{pc}^3$,
and each simulation had a rotation parameter $\lambda\simeq0.05$.
$G_{250}$ is the calculated rate at which 250$M_\odot$ stars would be expected
to be produced via collisional mergers of 125$M_\odot$ stars, in Myr$^{-1}$.
$S$ is the factor by which the central density of the most-massive stars
increased as a result of mass segregation, relative to the overall increase in
central density; $S'$ is the same factor but also including the effects of
stellar mergers.   For the G2A and G3C models, the presence or
absence of a stellar bar had little effect on the system's evolution.
(Model G3A was not run without a stellar bar but it was also expected to have
little effect.)
Note that the central densities listed here are for the initial models with
rotation, and so are lower than those given in Table \ref{tabics}.
Starting time $t_0=0$.
}
\label{mrjkroupa}
\end{table}

\begin{table}
\begin{center}
\begin{tabular}{lcccllcccc}
Model & $\lmax$ & bar & $\rho(0)$ at $t_0$ & $\rho(0)$ at $t_1$ & $t_1$ &
$G_{250}(t_0)$ & $G_{250}(t_1)$ & $S(t_1)$ & $S'(t_1)$\\
\hline\hline
E2A & 3 & 5.3\% &  $2.5\sn7$ & $3.7\sn8$ & .63 & 3.9 & 6.5 & 1. & 1.004\\
\hline
E2B & 2 & none & $8.2\sn6 $ & $1.4\sn7$ & 3.0 & 2.2 & 2.7 & 1. & 1.01\\
    & " & 5.3\% & " & $6.1\sn7$ & 3.0 & 2.6 & 3.3 &1. & 1.01\\
    & 3  & "& " & $9.4\sn7$ & $3.0$ & 2.6 & 3.7  & 1. & 1.01\\
    & "  & 8\% & " & $5.3\sn8$ & $2.7$ & 2.3 & 8.3 & 1. & 1.006\\
\hline
E4B & 2 &  8\% & $6.3\sn7$ & $1.0\sn9$ & 2.2 & 35. & 73.& 1. & 1.015 \\
    & " & 10\% &   "       & $1.5\sn9$ & 2.0 & 33. & 69.& 1. & 1.025 \\
\hline
G2A & 3 & none & $1.7\sn7$ & $1.3\sn8$ & .07 & 1.2&(1.9)*& 1. & 1. \\
    & " & 7.7\%& $1.7\sn7$ & $1.3\sn8$ & .07 & 1.1 & 1.6 & 1. & 1. \\
\hline
\hline
\end{tabular}
\end{center}
\caption{Similar to Table \ref{mrjkroupa}, but for simulations using an
Arches-style IMF.  [*] indicates that the value of $G_{250}(t_1)$ calculated
for model G2A without a bar is considered to be numerically suspect.
}
\label{mrjarches}

\begin{center}
\begin{tabular}{lcccllcccc}
Model & $\lmax$ & bar & $\rho(0)$ at $t_0$ & $\rho(0)$ at $t_1$ & $t_1$ &
$G_{250}(t_0)$ & $G_{250}(t_1)$ & $S(t_1)$ & $S'(t_1)$\\
\hline\hline
E2B & 2  & 5.3\% & $8.2\sn6$ 
& $2.9\sn8$ & 4.3 & 2.6 & 5.1 &1.& 1.01\\
    & 3  & "& " 
& $4.9\sn8$ & 4.3 & 2.6 & 8.7 & 1. & 1.01\\
\hline
\hline
\end{tabular}
\end{center}
\caption{Addendum to Table \ref{mrjarches}, showing results of simulations
allowed to proceed beyond the 3 Myr lifetime of the largest main-sequence stars.
}
\label{mrjarchE2B}
\end{table}

\section{Plummer-sphere Models}
\subsection{Model E4B, with Kroupa IMF: Core of a Giant Elliptical Galaxy
}\label{sect_e4b}

\begin{figure}
\input{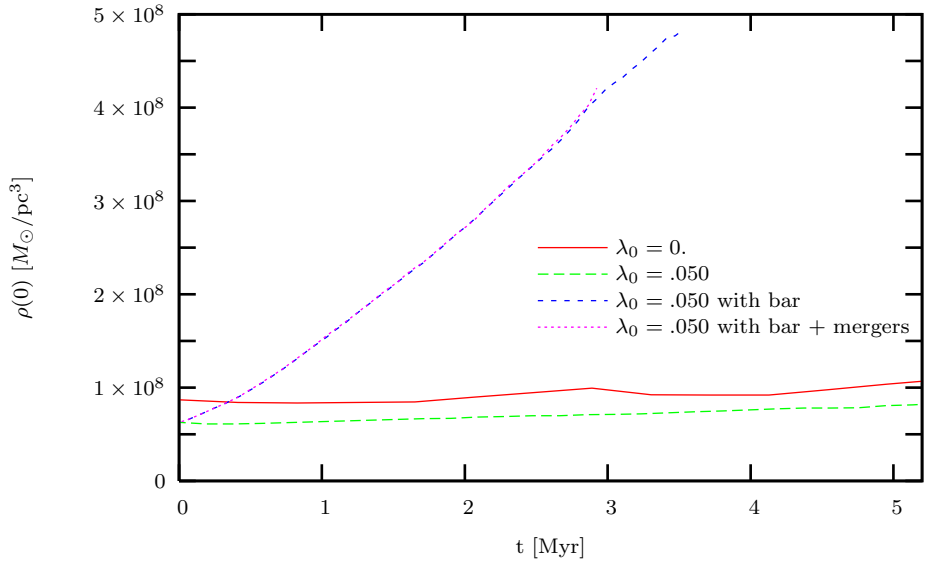}
\caption{Central density v. time for model E4B using the Kroupa IMF with a
stellar mass range of $1-125M_\odot$ 
and $\lmax=2$.
Note that the nonrotating ($\lambda_0=0$) case has an initial $\rho(0)$
slightly higher than the cases that include rotation; this is a product
of the method for introducing rotation into the initial distribution function
of the system.}
\label{e4b0nlm2}
\end{figure}

\begin{figure}
\input{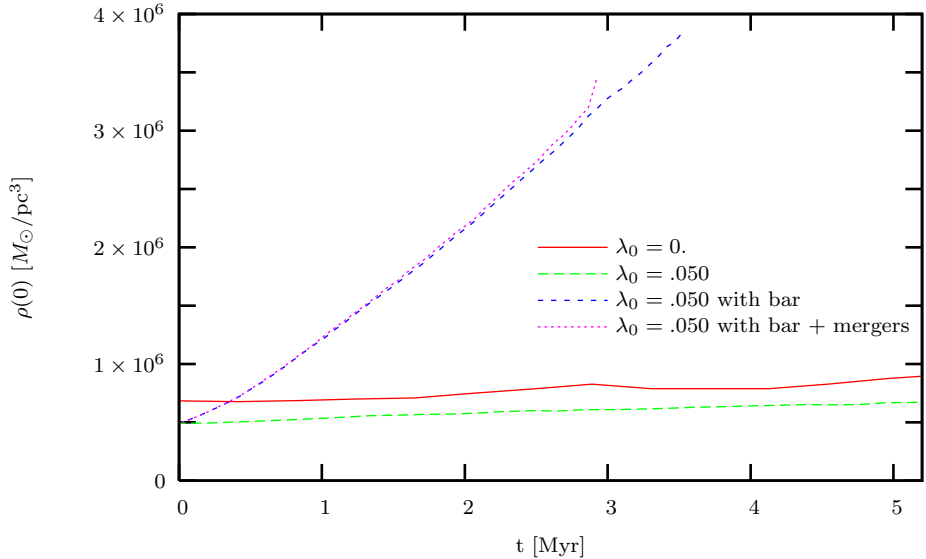}
\caption{Similar to Fig. \ref{e4b0nlm2} but showing only the highest-mass
stars ($m_Q=125M_\odot$).  The upturn at the end of the ``with mergers'' curve
is for the final timestep only and is a result of a numerical instability
which casued the simulation to end early.}
\label{e4bbnlm2}
\end{figure}

With $M_\mathrm{tot}=3.1\sn7M_\odot$ and an initial
$\rho(0)=10^8M_\odot/\mathrm{pc}^3$, Model E4B was the largest overall system
studied with the highest central density.  (Model E4A had an initial
$\rho(0)=3\sn8M_\odot/\mathrm{pc}^3$ but did not yield stable simulations.)
Figure \ref{e4b0nlm2} shows the change in central density $\rho(0)$ with time
for the four cases: (1) nonrotating, (2) rotating, (3) rotating, with a
stellar bar and (4) rotating, with a stellar bar and star-star mergers.
The stellar bar clearly influences the system evolution greatly as it produces
an increase in $\rho(0)$ of more than an order of magnitude in the 3 Myr
main-sequence lifetime of the largest stars in the system, compared to an
increase of $<5\%$ for the cases without a stellar bar.  Contrasting this is
the effect of stellar mergers, which is minimal: at the end of the
simulation that included mergers, $<0.1\%$ of the lowest-mass stars had
undergone a collision and merger, while fewer than 30 of the highest-mass
($125M_\odot$) stars had been created from collisional mergers -- although this
increase is still visible in Figure \ref{e4bbnlm2}.  Table \ref{mrjkroupa}
shows that $G_{250}$, the calculated rate of producing $250M_\odot$ stars by
stellar mergers, is effectively zero even after 3 Myr; similarly, the 
production rate for all stellar masses larger than $200M_\odot$ was also found to be zero.

\begin{figure}
\input{results/e4b2m8rho}
\caption{Density $\rho(r)$ v. radial distance $r$ from the cluster center
for the start, midpoint and end of the simulation for model E4B with a Kroupa
IMF.  Stellar mergers were enabled, as was a 10\% stellar bar.  Rotation
parameter $\lambda=0.05$ and $\lmax=2$.  (This corresponds to the uppermost
plot of Fig. \ref{e4b0nlm2}.)  Note that the decrease in $\rho(r)$ with time for large
$r$ is only exhibited when a stellar bar is present.}
\label{e4brho}
\end{figure}

Looking at the system as a whole, Fig. \ref{e4brho} shows the range of
$\rho(r)$ over the full radial extent of the stellar cluster at the starting
time, the midpoint, and the end of the simulation.  As the core contracts
and gets more dense, the outer regions lose stellar density accordingly.
Interestingly, when the model is run without including a stellar bar no such
change in density is observed in the outer regions; this behavior is seen in
general for all the Plummer-sphere models studied and is attributable to the
stellar bar transporting angular momentum outwards, which rarifies the outer regions
while allowing the inner regions to lose rotational support and contract.

The quantity $S$ in Table \ref{mrjkroupa} gives the relative increase in the
central density of the highest-mass (125 $M_\odot$) stars relative to that of
the overall system, and so indicates how much mass segregation has occured.
$S'$ is a similar measure but also includes the effects of stellar mergers on
the increased density of high-mass stars.  For model E4B without a stellar bar
a moderate ($\lesssim10\%$) relative increase in high-mass stars is seen, with mergers
contributing another couple percent.  (The nonrotating $\lambda=0$ case, plotted
in Figures \ref{e4b0nlm2} and \ref{e4bbnlm2} but not listed in Table
\ref{mrjkroupa}, displayed similar behavior.)  In this measure the presense of
a stellar bar appears to strongly damp the relative mass segregation to a large
extent, but that is merely due to how much more the overall central density
increases with the bar (by a factor of $\simeq6$ with the bar as compared to
$\simeq1.1$ without).  Thus a small relative mass segregation in the with-bar
case still corresponds to a larger absolute increase in the central density of
the highest-mass stars.  Still, it also means that the bar is dominating the
dynamics and the different stellar-mass populations evolve less distinctly
than in the no-bar case.



\subsection{Model E4B, Arches-style IMF}\label{Ae4b}

\begin{figure}
\input{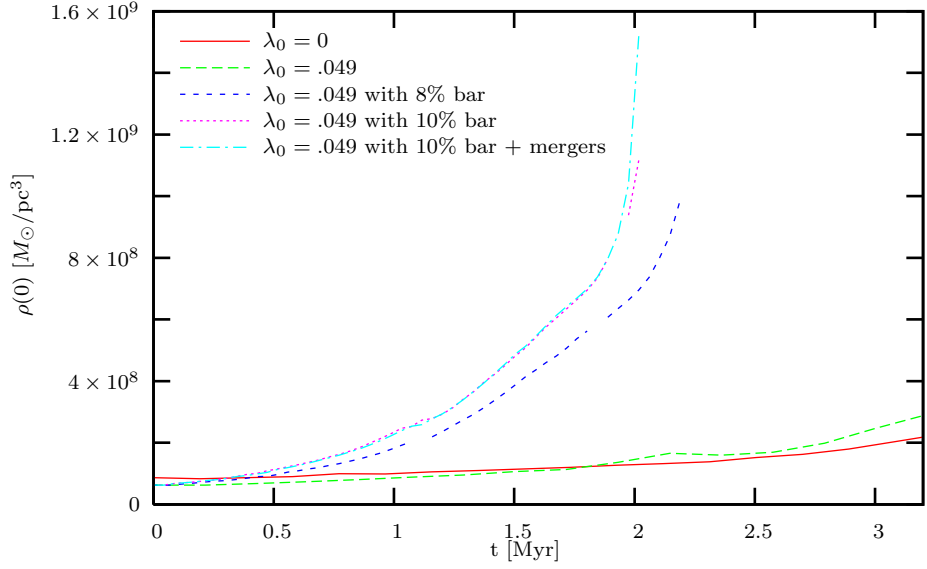}
\caption{Central density v. time for model E4B using the Arches IMF with a
stellar mass range of $2-125M_\odot$ 
and $\lmax=2$.
}
\label{Ae4b0nlm2}
\end{figure}

\begin{figure}
\input{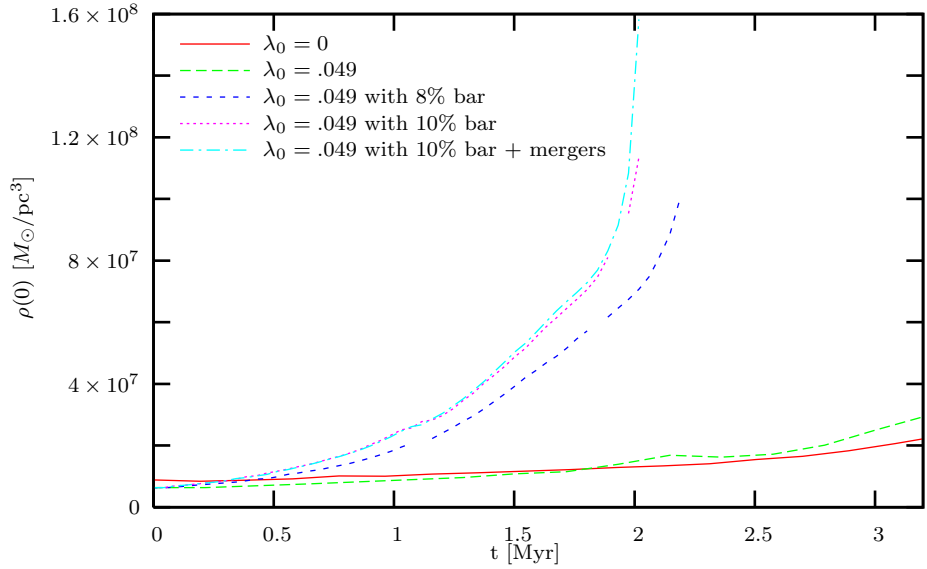}
\caption{Similar to Fig. \ref{Ae4b0nlm2} but showing only the highest-mass
stars ($m_Q=125M_\odot$).}
\label{Ae4banlm2}
\end{figure}

When studying model E4B with the Kroupa IMF replaced by an Arches-style IMF the situation changes
quite a bit.  As shown by Figures \ref{Ae4b0nlm2} and \ref{Ae4banlm2} both the
overall central density and that of the largest-mass stars increase much more
rapidly even in simulations without a stellar bar: by 3 Myr, $\rho(0)$ is up by
a factor of $\gtrsim3-4$ for the nonrotating and rotating cases, even without
the presence of a stellar bar.  More dramatically, the systems which included a
bar perturbation showed an increase in $\rho(0)$ of $\simeq20\times$ or more,
and in a shorter amount of simulated time
(by 2 Myr when using a bar of strength 10\%, by 2.2 Myr for an 8\%
bar\footnote{In model E4B the strength-10\% bar case used the default
1\% mass-fraction bar, as described in \S\ref{sect_barimpl}; the strength-8\%
bar corresponded to a mass fracton of 0.8\%, below which the code had
difficulty converging.}), after which the simulations were not able to track the
evolution stably.  This even more rapid increase is consistent with the Arches
IMF being weighted more towards larger stellar masses than is the Kroupa IMF.
The figures show that unlike in the Kroupa-IMF case, with an Arches IMF the
stellar system achieves core collapse before the 3 Myr stellar-lifetime limit,
with the central density $\rho(0)$ increasing exponentially: the run with a 10\%
bar and mergers was able to track somewhat into the core collapse stage before
aborting, while the other with-bar runs reached the turning point of $\rho(0)$.
Once core collapse begins, the central core of the cluster decouples dynamically
from the outer regions -- a situation the Fokker-Planck model is not constructed
to deal with.

The stellar merger situation is also qualitatively different with an Arches
IMF: as seen in Table
\ref{mrjarches}, even at $t=0$ the rate at which $250M_\odot$ stars were being
produced from collisions of $125M_\odot$ stars was $\simeq33\,\mathrm{Myr}^{-1}$,
increasing to $\simeq69\,\mathrm{Myr}^{-1}$ by the end of the simulation
with a 10\% stellar bar.  Thus
the Arches IMF was sufficient to produce $250M_\odot$ stars from
the outset, although the presence of a stellar bar enhanced the production.
It was also found that by $t=2.0$ Myr, approximately 500 stars of mass
125$M_\odot$ had been produced by collisional mergers of lower-mass stars.

Interestingly, the flatter Arches IMF seems to strongly damp any mass
segregation when compared to the Kroupa-IMF case above: the relative
increase in central density $\rho_Q(0)$ of the highest-mass stars tracks that of
the overall system $\rho(0)$ to several significant figures, regardless of the
presence or strength of the stellar bar.  And again, the nonrotating $\lambda=0$
case plotted in Figures \ref{Ae4b0nlm2} and \ref{Ae4banlm2} (but not listed in
Table \ref{mrjkroupa}) displayed a similar complete lack of mass segregation;
$S\equiv1$ for all E4B/Arches cases.  There is only a slight relative increase in
$\rho_Q(0)$ due to stellar mergers, as shown by $S'(t_1)=1.025$.

\begin{figure}
\input{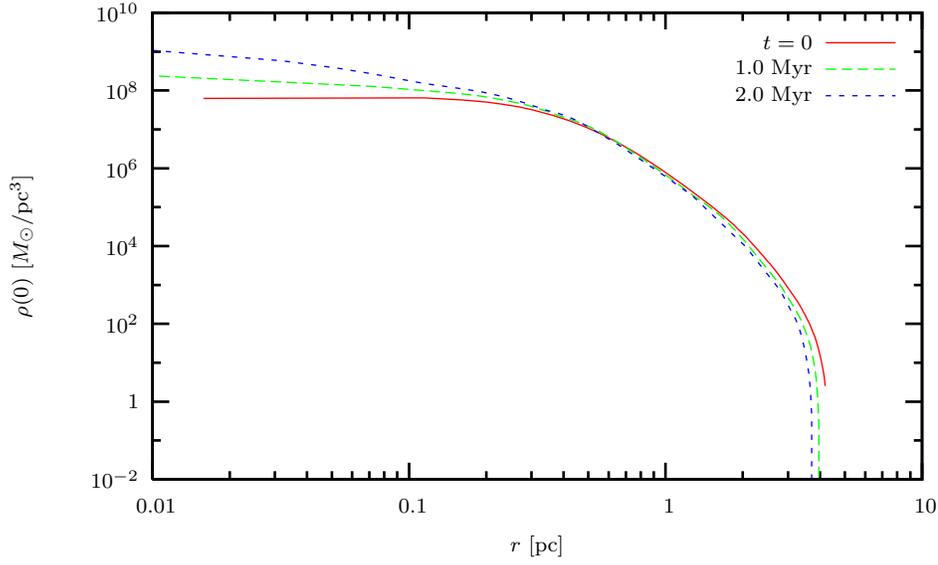}
\caption{Density $\rho(r)$ v. radial distance $r$ from the cluster center
for the start, midpoint and end of the simulation for model E4B with an
Arches-style IMF.  Stellar mergers were enabled, as was a 10\% stellar bar;
rotation parameter $\lambda=0.049$ and $\lmax=2$. (This corresponds to the
leftmost plot of Fig. \ref{Ae4b0nlm2}.)}
\label{Ae4b2m3rho}
\end{figure}

\begin{figure}
\input{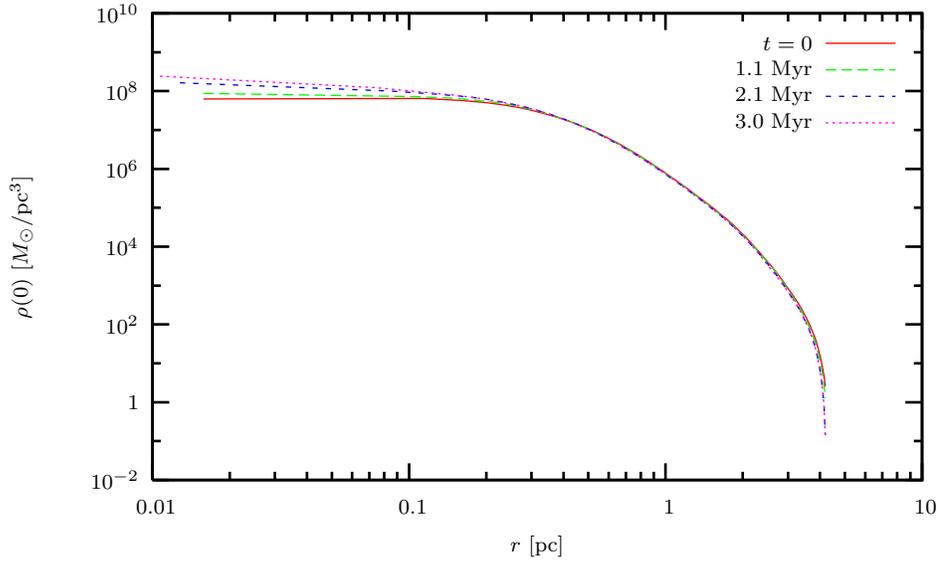}
\caption{Density $\rho(r)$ v. radial distance $r$ from the cluster center
for the start and end of the simulation for model E4B/Arches, as well as at
two similar times as those shown in Fig. \ref{Ae4b2m3rho} for comparison.
Neither stellar mergers
nor a stellar bar were enabled (in contrast with Fig. \ref{Ae4b2m3rho}).
Rotation parameter $\lambda=0.049$ and $\lmax=2$.}
\label{Ae4brho}
\end{figure}

Figure \ref{Ae4b2m3rho} plots the stellar mass density $\rho(r)$ from the
center of the cluster outwards.  As in the Kroupa-IMF case, the stellar bar
transports angular momentum from the inner regions to the outer, allowing core
contraction and a corresponding decrease in stellar density at large $r$.
A similar simulation but which does not include a stellar bar is plotted in Fig.
\ref{Ae4brho}.  Without a bar to transport angular momentum
outwards, there is very little decrease in stellar density in the outer
parts of the cluster.  Note that Figures \ref{Ae4b2m3rho} and \ref{Ae4brho}
are plotted on the same scale for easier comparison.

\subsubsection{Other quantities: bar composition, velocity dispersion and relaxation time}
\label{subsect_otherqs}

For the ``with bar'' runs presented in this chapter, the default
was to populate the bar with stars from the lowest-mass stars considered in
the simulation, \eg 2$M_\odot$ stars when using the Arches IMF.  As a test,
the E4B/Arches ``with bar'' simulation was also run with a bar comprised of
1$M_\odot$ stars.  It gave identical results to the default case, which is
expected since for purposes of its perturbing potential the bar is treated as
a single bulk object and not as a collection of individual stars.

\begin{figure}
\input{results/Ae4bm_vms0}
\caption{Central velocity dispersion $\vrms(0)$ v. time of the lowest-mass
(2$M_\odot$) stars in model E4B with an Arches IMF, for nonrotating,
rotating and with-bar cases.  (The case shown with a stellar bar
also had stellar mergers enabled, but a run without mergers was similar.)}
\label{Ae4bvms0}
\end{figure}

\begin{figure}
\input{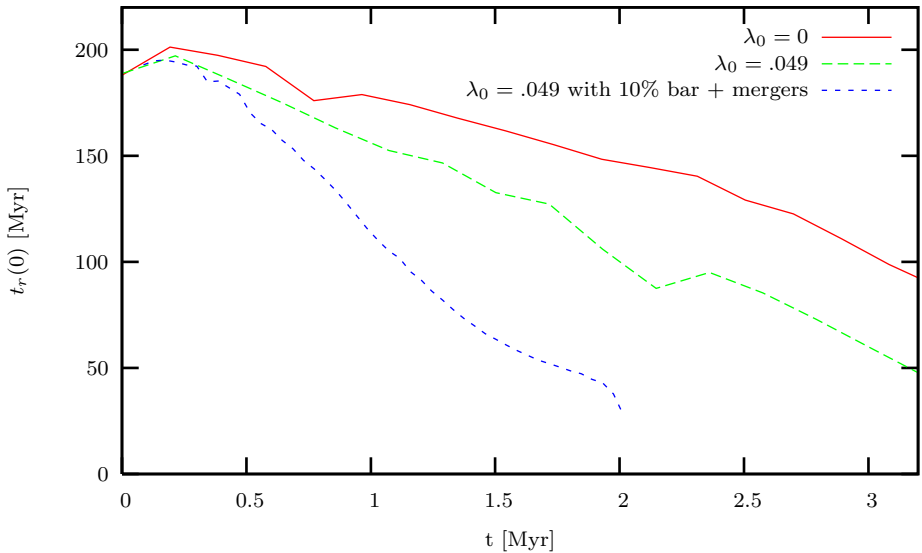}
\caption{Median central relaxation time $t_r(0)$ v. time of the
lowest-mass (2$M_\odot$) stars in model E4B with an Arches IMF, for
nonrotating, rotating and with-bar cases.  (The case shown with a stellar bar
also had stellar mergers enabled, but a run without mergers was similar.)}
\label{Ae4btr0}
\end{figure}

In addition to stellar density, several other quantities are tracked by the
simulation code.  Two of the most
interest physically are the velocity dispersion $\vrms$ and the
relaxation time $t_r$.  Model E4B/Arches can be used as an example to show the
results for these properties, which evolved similarly for all models.

Figure \ref{Ae4bvms0} shows the change with time of $\vrms(0)$ in the various
cases studied.  The systems which are initially rotating start at a lower value of
$\vrms(0)$ because the method of incorporating rotation into the initial
distribution function $f(I,J,0)$ necessarily ``cools'' the system by increasing
the amount of order present in the distribution of angular momentum $J$.
The two runs which do not include a stellar bar have a slowly
increasing $\vrms(0)$ as the system contracts and the central core gains
entropy.  Contrasting this is the case with a stellar bar, which collapses
much more quickly and has a corresponding rapid increase in the central
velocity dispersion.  Of note, the expected initial value of $\sigma_o(0)$ for
model E4B without rotation as given by Table \ref{tabics} is 400 km/s; all
models had a calculated initial $\vrms(0)$ which was similarly lower than would
be expected if isotropy of the velocity dispersion (as given by (\ref{sigma2})
with $\delta\equiv0$) strictly held.  To make the calculated initial $\vrms(0)$
values agree with those of $\sigma_o(0)$ in Table \ref{tabics}, an anisotropy
parameter $\delta\simeq-0.3$ is required; fortunately no derived quantities in
the calculation depend strongly on an assumption of isotropy.

The median central relaxation time $t_r(0)$ for the same set of cases is
shown in Fig. \ref{Ae4btr0}.  The initial value of $t_r(0)$ is also slightly
different than that listed in Table \ref{tabics}, which is expected as the full
calculation of $t_r$ applies to a given test star, not to a specific point in
space (\cf the footnote in \S\ref{sect_oafp}).  Thus only a median value of
$t_r$ across all stars which traverse the system center can be determined, and
any slight error in $\vrms(0)$ does propogate to the deduced $t_r(0)$, via
its dependence on the cube of $\vrms$.  This being the case, it can be seen that the
presence of a stellar bar does cause a rapid decrease in the central relaxation
time of the system, in this case by close to an order of magnitude within 2 Myr.

\subsection{Model E2A, Arches-style IMF: Nuclear Cluster}

\begin{figure}
\input{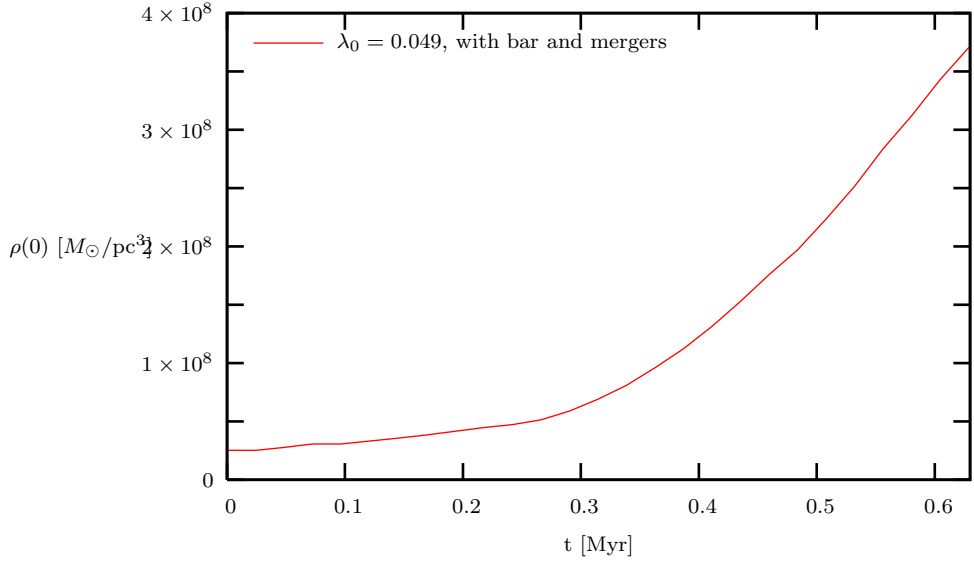}
\caption{Central density v. time for model E2A with Arches-style IMF, $\lmax=3$
and a stellar mass range of $2-125M_\odot$.
}
\label{Ae2a0nlm4}
\end{figure}

Model E2A is somewhat different than E4B: the total mass, initial central
density and velocity dispersion are all lower for E2A, as is the initial
relaxation time (\cf Table \ref{tabics}).  When using the Arches IMF the
simulations which did not include a stellar bar were not numerically consistent
for different choices of the timestep size, and so only the ``full'' case that
included a stellar bar and collisional mergers is shown in Fig. \ref{Ae2a0nlm4}.

Despite these differences, the increase in $\rho(0)$ by a factor of $\simeq14$
before the simulation ended was similar to that of model E4B prior to its core
collapse.  Model E2A did not reach core collapse before aborting -- although for E2A the
simulation ended much earlier, at $t=0.63$ Myr, consistent with the shorter
initial value for $t_r$ and smaller $\vrms$.  And similarly to as was found for
model E4B, there was sufficient merging of 125$M_\odot$ stars initially to
create 250$M_\odot$ stars -- in this case, at a rate of
$3.9\,\mathrm{Myr}^{-1}$, which increased to $6.5\,\mathrm{Myr}^{-1}$ by the
simulation's end.  Approximately 15 stars of mass 125$M_\odot$ had been created
via collisional mergers of lower-mass stars as well by the simulation's end,
a small number attributable to the short amount of simulation time.  The mass
segregation behavior, with no segregation observed (\ie $S=1$) due to dynamical
effects and only a small amount ($S'\gtrsim1$) from stellar mergers, was also
similar to that of the E4B/Arches model, as was to be expected given the short
simulated time of evolution of the system.

\subsection{Model E2B, Arches-style IMF: Nuclear Cluster}\label{E2Barches}

\begin{figure}
\input{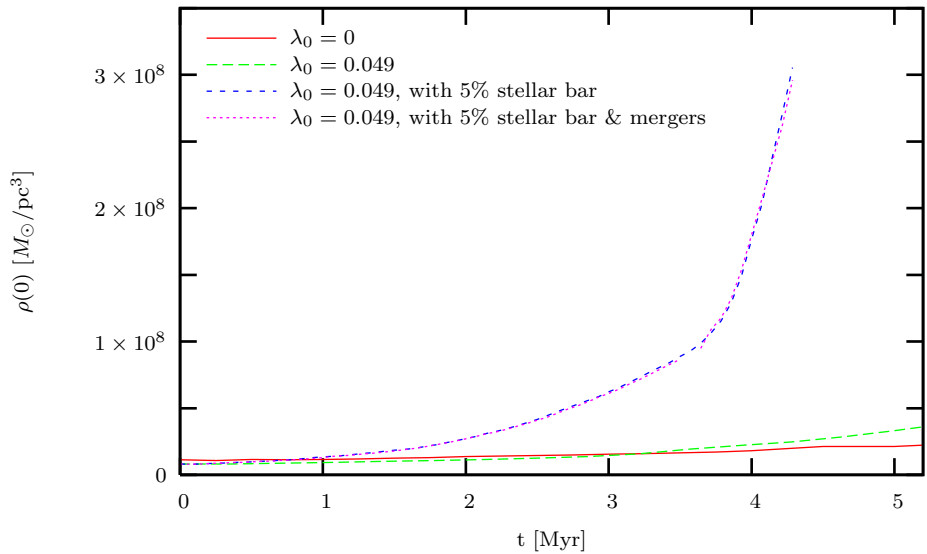}
\caption{Central density v. time for model E2B with $\lmax=2$ and using the
Arches IMF with a stellar mass range of 2-125$M_\odot$. 
Note that the main-sequence lifetime of the most massive stars is 3 Myr
and so the region of the graph beyond $t=3$ is nonphysical.}
\label{Ae2bt5}
\end{figure}

\begin{figure}
\input{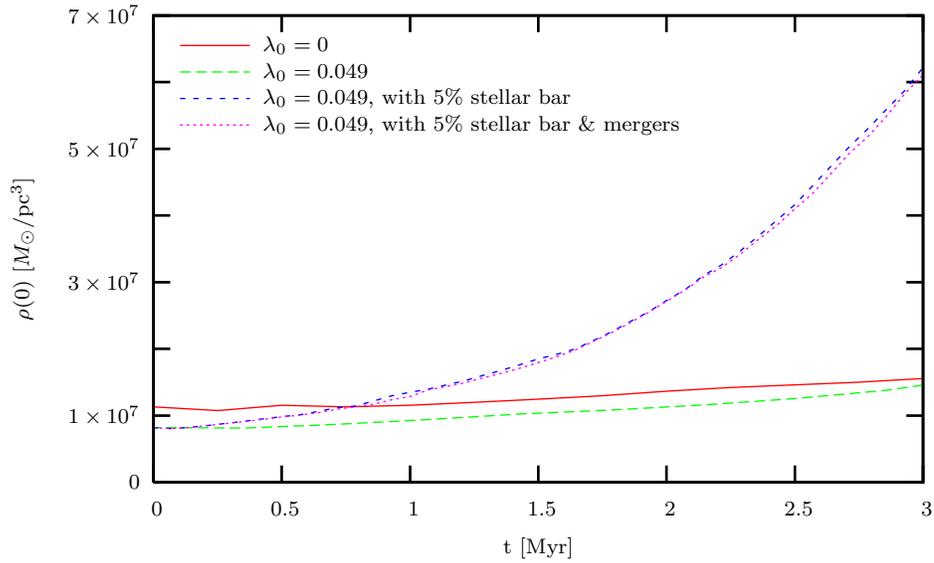}
\caption{Detail of the first 3 Myr of Fig. \ref{Ae2bt5}.
}
\label{Ae2bt3}
\end{figure}

\begin{figure}
\input{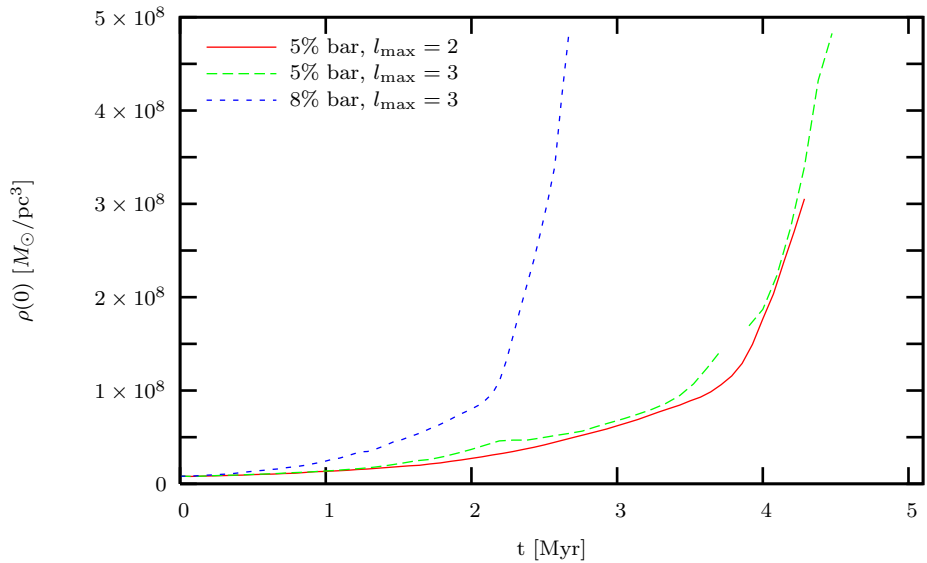}
\caption{Similar to Fig. \ref{Ae2bt5} but including $\lmax=3$ and an 8\% bar
case.  All have $\lambda=0.049$ and do not include stellar mergers.  (Simulations
performed with mergers enabled yielded almost-identical runs of $\rho(0)$ to
those shown in this figure.)  Note that the ``5\% bar'' plot here is the same as
that shown in Fig. \ref{Ae2bt5}.  The gap in one plot is due to a glitch in the
output routine, as described in Chapter \ref{chap_tests}.
}
\label{Ae2b4}
\end{figure}

\begin{figure}
\input{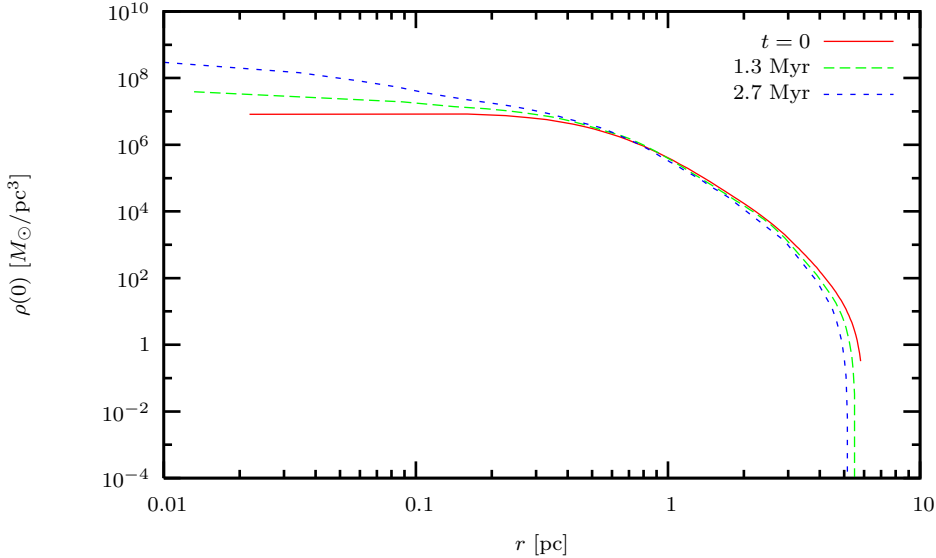}
\caption{Density $\rho(r)$ v. radial distance $r$ from the cluster center for
the start, midpoint and end of the simulation for model E2B with an Arches-style
IMF.  Stellar mergers were enabled, as was an 8\% stellar bar; rotation
parameter $\lambda=0.049$, and $\lmax=3$.  This corresponds to the leftmost
plot of Fig. \ref{Ae2b4}.  The decrease in $\rho(r)$ with time for large
$r$ is only seen when a stellar bar is present.
}
\label{Ae2brho}
\end{figure}

Model E2B, when given an Arches-style IMF, shares many properties with model
E4B/Arches described in \S\ref{Ae4b}.  Figure \ref{Ae2bt5} shows that the
nonrotating case evolves very slowly, as does the rotating case which does
not have a stellar bar; in contrast, a modest 5.3\% stellar bar induces core
collapse and a $\simeq60\times$ increase in central density $\rho(0)$ -- but
only at $t\simeq4$ Myr of simulation time, well beyond the 3 Myr main sequence
lifetime of the largest stars.  As seen in Fig. \ref{Ae2bt3}, the evolution
prior to $t=3$ Myr is much more modest.
Stellar mergers have little affect on $\rho(0)$, and Table \ref{mrjarches}
shows that with a 5.3\% bar, by $t=3$ Myr the rate of massive star production
increases modestly from $G_{250}=2.6$ Myr$^{-1}$ to 3.7 Myr$^{-1}$.  (\ie a small
number of massive stars are produced collisionally at the start, and a small
number are produced at the end of the stellar lifetimes.)  Table
\ref{mrjarchE2B} shows that if the system was able to continue on to
$t\simeq4$ Myr, core collapse would result in a somewhat larger increase in
the rate of massive-star production.


A rise over time in the rate of producting $250M_\odot$ stars via mergers can
be attributed either to an increase in overall density $\rho$, or to the
presence of a greater number of massive stars due to previous collisional
mergers.  A comparison to determine how much each of the above factors
contributed was performed by running a simulation in which stellar mergers were
shut off at first and then turned on at $t=3$ Myr.  The result was a value of
$G_{250}\simeq3.4$ Myr$^{-1}$ at $t=3$, compared to 2.6 Myr$^{-1}$ at $t=0$,
and to 3.3 Myr$^{-1}$ at $t=3$ when mergers were allowed from the start.
Hence, at least for model E2B with a 5.3\% bar, there is little cumulative
effect of collisional mergers on the later merger rate of larger stars.

Allowing for a somewhat stronger bar gives a significant effect: Figure
\ref{Ae2b4} shows that with an 8\% bar (still well within the typical range of
$5-10\%$) core collapse is accelerated and shifted earlier to $t\simeq2.5$ Myr;
likewise the rate of $250M_\odot$ star production increases somewhat, now to
8.2 Myr$^{-1}$ -- still not as high as for model E4B/Arches but notably more
than at $t=0$ in this model.  As well, ultimately $\sim60$ stars of 125$M_\odot$
were created by mergers of lower-mass stars by the time the simulation ended at
$t=2.7$ Myr.  And also similarly to model E4B/Arches, no mass segregation is
seen before the simulations terminate -- which again was also the case for a
simulation performed with $\lambda=0$ (\ie without rotation).

Finally, the full range of stellar mass density $\rho(r)$ plotted at various
times in Fig. \ref{Ae2brho} again shows the core of the cluster contracting
and becoming denser, while the outer regions lose density over time.  As was
also seen for
model E4B, a simulation performed without a stellar bar did not display any
similar effects of angular momentum transport to the very outer regions of
the cluster as did one which included a bar.  (The non-bar case is not plotted here, but compares similarly
to Fig. \ref{Ae2brho} as do Figures \ref{Ae2b4} and \ref{Ae4b2m3rho} for E4B.)



\subsection{Model E2A, Kroupa: Bulgeless Spiral or Dwarf Elliptical Nucleus}

The simulations of model E2A with the Kroupa IMF behaved somewhat strangely:
for very similar cases -- \eg two runs both with rotation and a stellar bar,
but only one with stellar mergers (which had a small effect when using the Kroupa
IMF) -- the dynamic $r^2$ grid would evolve rather differently for the various
cases, making direct comparisons difficult.  Also, finding a satifactory value
for the potential-expansion parameter $\lmax$ proved elusive: $\lmax=5$
appeared to be required to capture the bulk of the interaction strength, but
was even more unstable than $\lmax\leq4$. Still, as shown in Table
\ref{mrjkroupa} this model gave results consistent with model E4B/Kroupa: the
stellar bar produced an increase in $\rho(0)$ of an order of magnitude in a
short time compared to the main-sequence lifetime of the most massive stars,
but the rate of collisional mergers of those massive stars was negligible.
The ``reverse mass segregation'' indicated by $S<1$ is unexpected.  However,
test runs using a lower value of $\lmax=2$, while not capturing the full extent
of the model's evolution, also showed a similar trend towards $S<1$ through
$t=1.6$ Myr -- but at later times $S$ turned around and began increasing again.
Even so, why there would be a delay in mass segregation being exhibited for
this model remains unexplained.


\subsection{Effect of Collisional Merger Rates}\label{upfactor}

\begin{table}
\begin{center}
\begin{tabular}{lccllcccc}
IMF & $factor$ & $\rho(0)$ at $t=0$ & $\rho(0)$ at $t_1$ & $t_1$ &
$G_{250}(t=0)$ & $G_{250}(t_1)$ & $S$ & $S'$\\
\hline\hline
Kroupa & 1  & $6.3\sn7$ & $4.2\sn8$ & 2.9 & 0. & 0. &0.98&  1.04\\
       & 5  &    "      & $4.4\sn8$ & 2.9 & "  & 0. &  " &  1.12\\
       & "  &    "      & $4.8\sn8$ & 3.0 & "  & 0. &  " &  1.13\\
\hline
Arches & 1  &    "      & $4.0\sn8$ & 1.6 & 33.& 58. & 1. & 1.01\\
       & 5  &    "      & $5.0\sn8$ & 1.6 &  " & 68. & " &  1.07\\
       & "  &    "      & $9.1\sn8$ & 2.1 &  " & 87. & " &  1.10\\
\hline
\hline
\end{tabular}
\end{center}
\caption{Effect of artificially increasing the collisional merger rate by a
factor of 5 in order to account for the effect of tidal-capture binaries, as
described in \S\ref{upfactor}. 
Listed are results
for model E4B with $\lmax=2$ and a 10\% stellar bar.  The $factor=1$ entry has
slight differences from the values shown in Table \ref{mrjarches} due to
different timestep sizes being used for the simulations listed here.}
\label{mrjfact}
\end{table}

\begin{figure}
\input{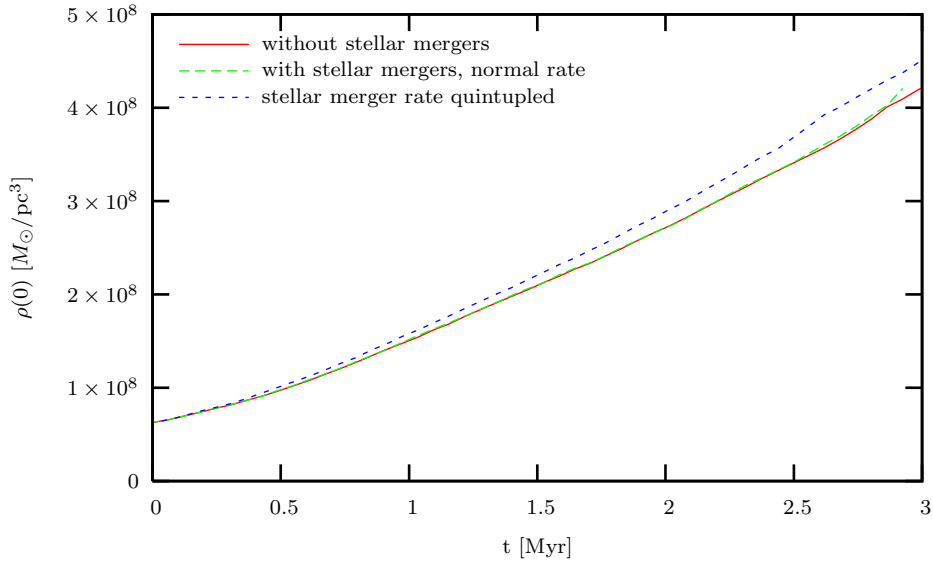}
\caption{Effect of artificially increasing the collisional merger rate.
Plotted is model E4B using a Kroupa IMF with $\lmax=2$ and a 10\% stellar bar.
Also see Table \ref{mrjfact}.}
\label{e4bn2b7f5}
\end{figure}

The development of the collisional merger rates in \S\ref{mergerrates} assumed
only direct 2-body collisions.  However, indirect collisions -- due to the
creation of tight binaries formed via tidal capture which then merge on a short
timescale -- can enhance the merger rate by a factor of 3 to 5 in some situations
\Cite{Freitag}.  (Other factors which may enhance the expected rate include
that low mass [$5-10M_\odot$] stars do not to contract to their long-term main
sequence radius immediately and so have a larger cross section initially,
and a second-order effect in that when those lower-mass stars collide
with more massive stars they form a disk, which increases the new star's effective
cross section.) To test how an enhanced merger rate would affect the overall
results, model E4B was re-run with a the loss and gain coefficients of (\ref{lm})
and (\ref{fullgm}) increased by a factor of 5, the effect of which is shown in
Fig. \ref{e4bn2b7f5} and Table \ref{mrjfact}.  (Tabulated but not plotted are
the results for model E4B with the Arches IMF; while stable, the $r^2$ grid of
that case's $factor=5$ run sufficiently diverged from that of the base $factor=1$
simulation to preclude a direct visual comparison.)

While in both the Kroupa- and Arches-IMF cases an enhanced merger rate resulted
in a faster evolution of the overall system, little qualitative difference was
observed.  The central density $\rho(0)$ and merger-induced mass segregation
ratio $S'$
both increased more quickly with the additional factor of 5 in the merger rate
-- as did the rate of production of massive stars $G_{250}(t_1)$ when using the
Arches IMF.  But still no massive stars were created when using the Kroup IMF,
and core collapse did not occur any more quickly with the Arches IMF.  As 
model E4B
has the highest initial $\rho(0)$ and is already the fastest-evolving of the
various models studied, other models are expected to also show little
qualitative change when given an enhanced merger rate.

\section{$\gamma=0$ Sphere Models}\label{gamm0}

As described in \S\ref{gamma0}, of the $\gamma=0$ models which could be made
using same values of total mass $M$ and core radius $r_\mathrm{core}$ to the
Plummer-sphere models described above, only the ``G2A'' model has reasonable
values for the initial central density $\rho(0)$ and velocity dispersion
$\vrms(0)$.
An extreme model ``G3C'' in which the half-mass radius was the same as that
for model E2B was also run and is described here, as is an intermediate model
``G3A''.

\begin{figure}
\input{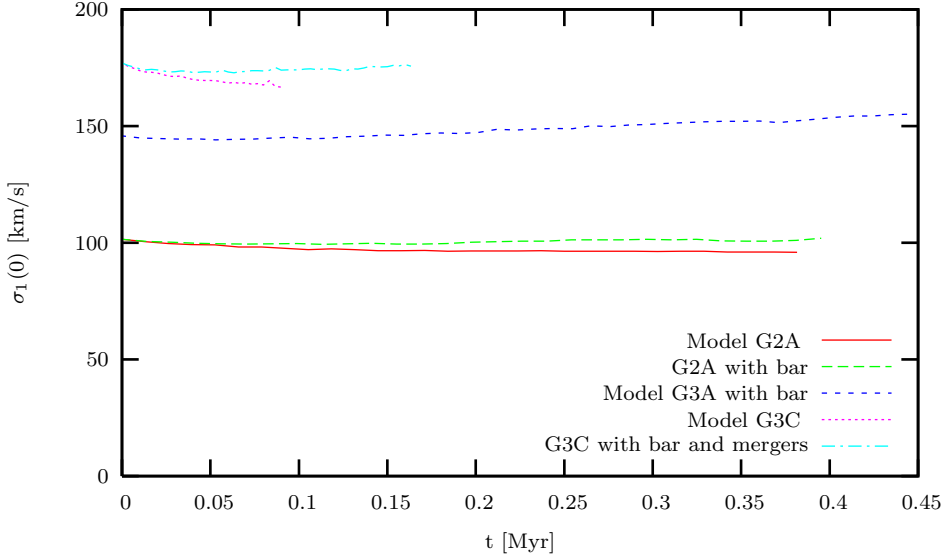}
\caption{Central velocity dispersion v. time for the lowest-mass (1$M_\odot$)
stars in the $\gamma=0$ sphere models with a Kroupa IMF.  All have
$\lambda\simeq0.05$.  Plots of runs with and without stellar mergers were
identical to each other; differences within each model are due to the
presence or absence of a stellar bar.}
\label{g2vms}
\end{figure}

\begin{figure}
\input{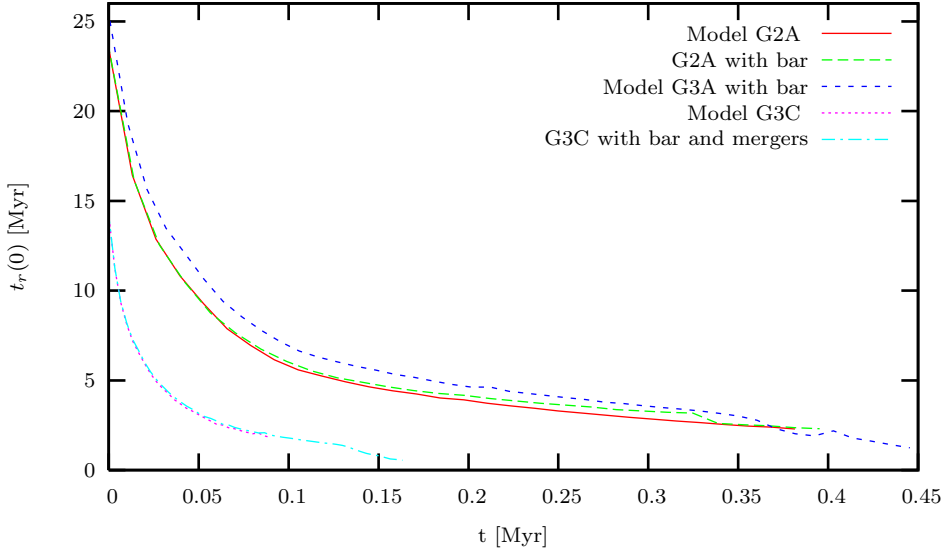}
\caption{Central relaxation time for the lowest-mass (1$M_\odot$) stars in
the $\gamma=0$ sphere models with a Kroupa IMF.  All have $\lambda\simeq0.05$.}
\label{g2tr}
\end{figure}

Figures \ref{g2vms} and \ref{g2tr} show $\vrms(0)$ and the central
relaxation time $t_r(0)$ for various cases of models G2A, G3A and G3C, all
with a Kroupa IMF.  Unlike as was seen for the Plummer sphere models, the
stellar bar had little effect on the $\gamma=0$ 
models' evolution, as can be seen from the contrast of Fig. \ref{g2vms} with
Fig. \ref{Ae4btr0}.  This trend was common to all the
$\gamma=0$ sphere results and can at least partially be attributed to the
short amount of simulation time before the models ended, as indicated by the
small values on the plots' horizontal axes.  Relaxation times were much shorter than
for the Plummer sphere models, and so the bar had little time to transport
angular momentum -- and mergers had little time to produce larger stars as well.
Even so, in some cases the short $t_r(0)$ values did allow for rapid overall evolution as
is described in the following sections for each specific model.

\subsection{Model G2A, Kroupa IMF: Galactic Spheroid / Spiral Bulge}

\begin{figure}
\input{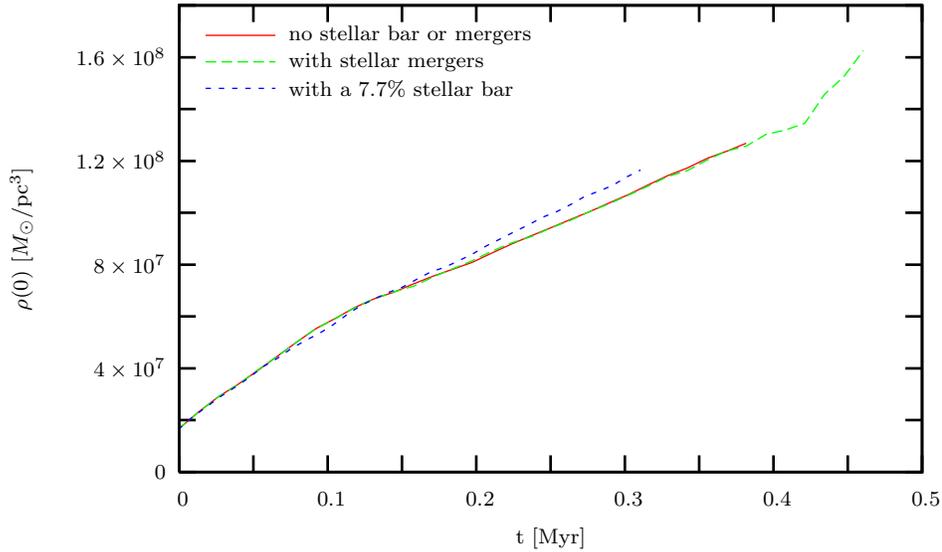}
\caption{Central density v. time for $\gamma=0$ sphere model G2A with a Kroupa
IMF, $\lmax=4$ and $\lambda=0.051$.  The nonrotating ($\lambda=0$) case gave
similar results but was less stable and not plotted.}
\label{g2a6}
\end{figure}

\begin{figure}
\input{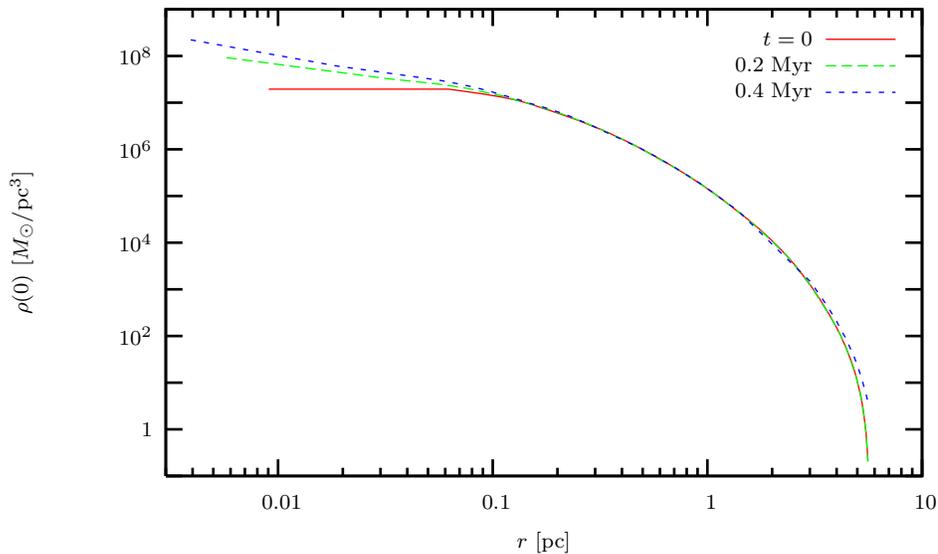}
\caption{Density $\rho(r)$ v. radial distance $r$ from the cluster center
for the start, midpoint and end of the simulation for model G2A with a Kroupa
IMF.  Neither rotation nor stellar mergers were enabled, and $\lmax=4$; this
corresponds to the solid-line plot of Fig. \ref{g2a6}.  
The behavior of $\rho(0)$ v. $r$ for the other two cases from Fig. \ref{g2a6}
is almost identical to that plotted here and are so not shown.}
\label{g2a6rho}
\end{figure}

Model G2A with a Kroupa IMF showed only an almost linear increase in $\rho(0)$
before ending due to numerical issues; Figure \ref{g2a6} shows that this was
the situation for all cases regardless of whether either mergers or a bar were
included.  
(The slight upturn in the ``with mergers'' line in the plot was not
reproducable and is likely a numeric artifact.)  Figure \ref{g2a6rho} gives
the full density run $\rho(r)$ at various times in the simulation;
interestingly the stellar density rises with time at both small and large $r$, and drops
slightly with time in between; this compares to the Plummer sphere behavior
in which the stellar density dropped with time for the largest values of $r$,
\eg as in Fig. \ref{Ae4b2m3rho}.  The difference could be due to the $\gamma=0$ sphere
having relatively more mass to absorb angular momentum from the bar at intermediate
values of $r$, to the slower pattern speed of the bar
($\Omega_b\simeq160$ Myr$^{-1}$, compared to $\Omega_b\simeq400$ Myr$^{-1}$ for
model E2A), or simply to the short evolution time of the simulation not
allowing for angular momemtum to be transported all the way to the cluster's
edge.

As was the case for all the Plummer sphere
models, the calculated rate of 250$M_\odot$ star production remained nill
by the end of the simulation, and no stars of mass 125$M_\odot$ had been created
via collisional mergers.
Table \ref{mrjkroupa} shows that model G2A exhibited reverse mass segregation
($S<1$); a test run with $\lmax$ lowered from 3 to 2 showed that
$S$ initially decreased below 1 and then increased again at times later than
were reached by the ``full'' $\lmax=4$ simulation, indicating that the apparent
reverse mass segregation is a temporary phenomenon.
Again, this behavior was also as was found for model E2A.


\subsection{Models G3A (Larger Galactic Spheroid) and G3C, Kroupa IMF}

\begin{figure}
\input{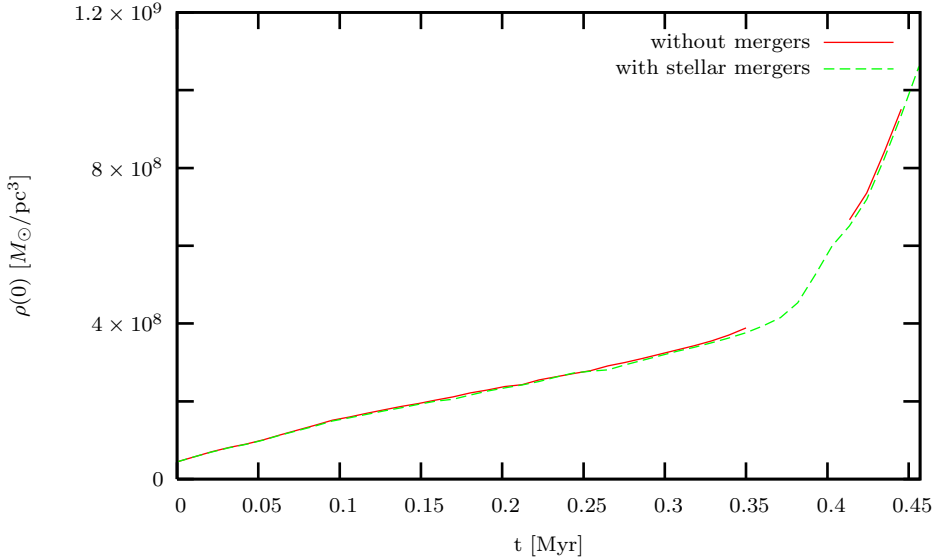}
\caption{Central density v. time for $\gamma=0$ sphere model G3A with a Kroupa
IMF, $\lmax=3$, $\lambda=0.046$ and a 7.7\% stellar bar.  The gap in the
no-merger plot is due to a glitch in the output routine.  Runs without
a stellar bar did not yield stable simulations and are not plotted.}
\label{g3a4}
\end{figure}

\begin{figure}
\input{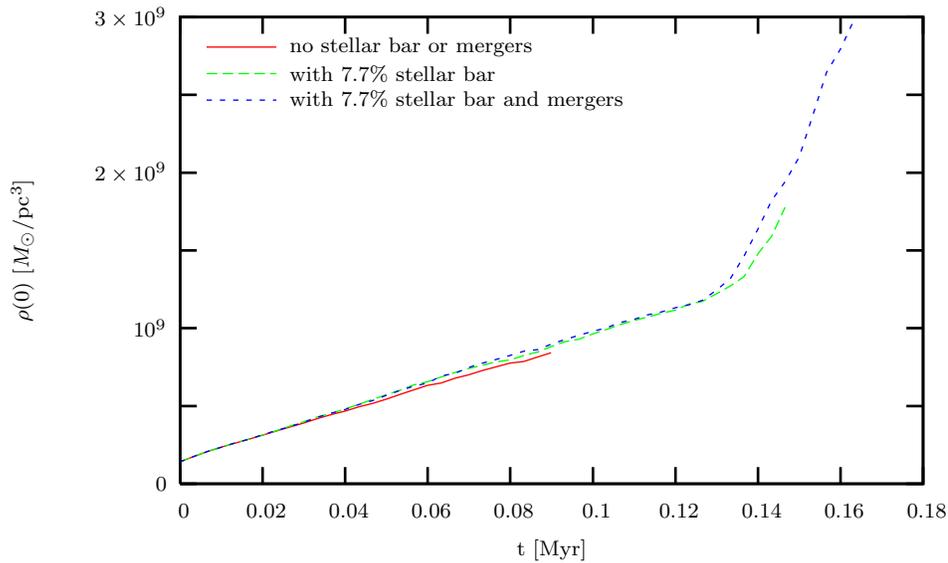}
\caption{Central density v. time for $\gamma=0$ sphere model G3C with a
Kroupa IMF, $\lmax=2$, and $\lambda=0.046$.  The nonrotating ($\lambda=0$)
case did not yield a stable simulation and is not plotted.}
\label{g3c2}
\end{figure}

Model G3A started with a somewhat higher density than model G2A did, and had
a corresponding faster and greater amount of evolution of the system; model
G3C continued this trend.  In Figures \ref{g3a4} and \ref{g3c2} the runs
without a stellar bar were either unstable or ended early, but the case with
a bar case continued long enough to reach the beginnings of core collapse;
these were the only Kroupa-IMF models studied to do so.  Even so, the density of
high-mass stars remained low enough, and the total simulated time was short
enough, that no 125$M_\odot$ stars were created by the simulations' end. (The
calculated value was $\sim0.6$ new stars of mass 125$M_\odot$ for both models,
and the Fokker-Planck code ignores stellar population numbers that are less
than 1.)  Similarly the calculated rate of 250$M_\odot$ star production
remained nill, as given in Table \ref{mrjkroupa}.  A modest amount of
mass segregation did take place even in the short simulation time.

Despite the shorter simulation time, the run of $\rho(r)$ given in Fig.
\ref{g3c2rho} displays an increase in density for model G3C at small and large
$r$ values (and a decrease for intermediate $r$) more clearly than does Fig.
\ref{g2a6rho} for model G2A or Fig. \ref{g3a4rho} for G3A.  This is consistent
with G3C's much-shorter relaxation time, \eg as shown in Fig. \ref{g2tr}.
However, while model G3A may be seen as representing a larger version of the
galactic spheroid modeled by G2A, it is doubtful that model G3C corresponds
to a realistic astronomical system, and is more useful here as a demonstration case.

\begin{figure}
\input{results/g3a4rho}
\caption{Density $\rho(r)$ v. radial distance $r$ from the cluster center
for the start, midpoint and end of the simulation for model G3A with a Kroupa
IMF, $\lambda=0.046$, a 7.7\% stellar bar, and stellar mergers enabled.
This corresponds to the dashed plot of Fig. \ref{g3a4}.}
\label{g3a4rho}
\end{figure}

\begin{figure}
\input{results/g2c2rho}
\caption{Density $\rho(r)$ v. radial distance $r$ from the cluster center
for the start, midpoint and end of the simulation for model G3C with a Kroupa
IMF, $\lambda=0.046$, a 7.7\% stellar bar, and stellar mergers enabled.
This corresponds to the short-dashed plot of Fig. \ref{g3c2}.}
\label{g3c2rho}
\end{figure}

\subsection{Model G2A, Arches IMF}

\begin{figure}
\input{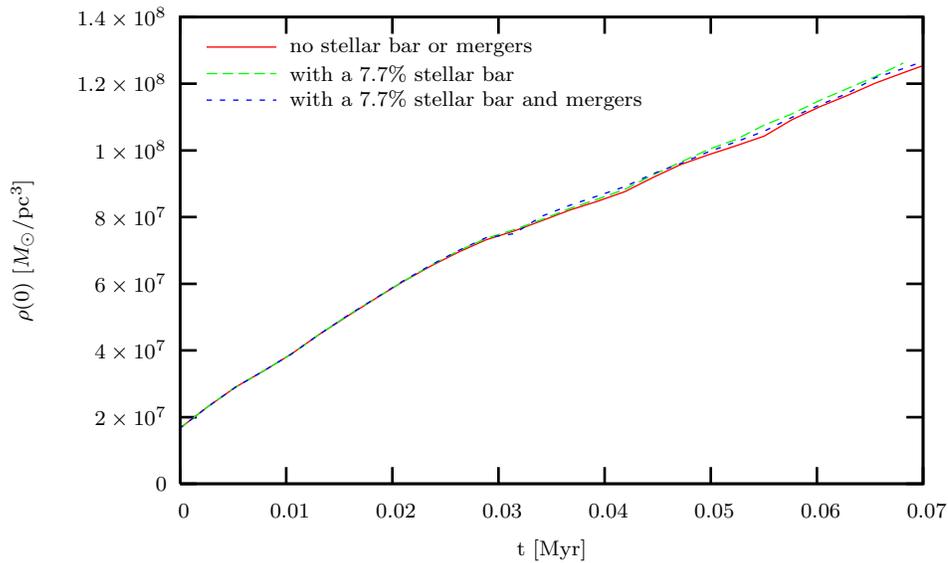}
\caption{Central density v. time for $\gamma=0$ sphere model G2A with an
Arches-style IMF, $\lmax=3$ and $\lambda=0.051$.  Two of the runs continued
beyond $t=0.07$ Myr but became unstable; those portions are not shown.}
\label{Ag2a4}
\end{figure}

Model G2A with an Arches-style IMF behaved much like model G2A/Kroupa but
with a more rapid evolution.
Figure \ref{Ag2a4} shows the central density $\rho(0)$ increasing roughly
linearly with time but only a factor of $\simeq8$ before the simulation ended
due to numerical difficulties.
The presence or absence of either a stellar bar or collisional mergers again
had very little effect on the overall evolution of the system.  Table
\ref{mrjarches} shows that the system starts with just a high enough rate
of collisions of massive stars $G_{250}=1.1$Myr$^{-1}$ to produce a small
number of $250M_\odot$ stars, with the rate increasing modestly bebfore the
simulation ends.  And in the short simulation time exactly one star of mass
$125M_\odot$ was created through collisional mergers of smaller stars.



This chapter has presented the simulation results for each model studied
in turn.  In the next chapter these results will be discussed with respect
to what astrophysical effects were observed for which models, how that
compares to what would be expected from timescale and density arguments,
and what conclusions can be reached regarding the production of massive
central objects in actual stellar systems.

%% file: Discussion.tex
\chapter{Discussion}\label{chap_disc}


According to B{\"o}ker, nuclear clusters -- massive, dense stellar clusters in the nuclei of
galaxies -- are observed to be nearly ubiquitous and share a similar relation
to the host galaxy as do active galactic nuclei \Cite{Boeker}.  He closes with
the following quote:
\begin{quote}

It has recently been proposed by [Ferrearse \etal 2006] that
nuclear clusters extend the well-known scaling relation between the mass of
a galaxy and that of its central super-massive black hole (SMBH) to lower
masses. This has triggered speculation about a common formation mechanism of
nuclear clusters and SMBHs, being governed mostly by the mass of the host
galaxy. The idea put forward is that nuclear clusters and SMBHs are two
incarnations of a central massive object which forms in every galaxy. In
galaxies above a certain mass threshold ($\simeq10^{10}M_\odot$), galaxies
form predominantly SMBHs while lower mass galaxies form nuclear clusters.

{\ldots} 
Is a nuclear cluster possibly a
pre-requisite for the formation of a SMBH? Is the formation of a BH (not
necessarily a super-massive one) a logical consequence of the high stellar
densities present in nuclear clusters? Progress along these lines will
require a better understanding of the formation of pure disk galaxies in the
early universe, as well as improved models for the evolution of extremely
dense stellar systems.

\end{quote}

This study is an attempt at just such an improved model of dense stellar
systems, linking what the above reference calls nuclear clusters with the
eventual supermassive black holes in the centers of active galaxies.  This
chapter begins with two brief discussions, first of the timescale on which
mass segregation alone would be expected to result in core collapse, and then
of what central density is required before either runaway accretion or runaway
collisional mergers might occur.  The main body of the chapter then discusses
the physical effects exhibited by the various simulations and what they imply
for astronomical systems.  Finally a brief closing considers what future
observations may support the assumptions and results of these simulations,
and what further improvements of the simulation method are indicated.

\section{Timescale Arguments and Expectations}

\subsection{Mass Segregation}\label{massseg}

For mass segregation to occur, energy equipartition across different mass
populations must be an unstable situation.  Spitzer in 1969 formulated an
analytic condition for preventing the mass-segregation instability: for a system
of total mass $M_1+M_2$, with $M_1$ contained in stars of individual mass $m_1$
and $M_2$ in stars of individual mass $m_2>m_1$ as
\beq\left(\frac{M_2}{M_1}\right)\left(\frac{m_2}{m_1}\right)^{3/2}<0.16\eeq
which for realistic mass spectra is never satisfied, \ie there are enough heavy
stars that they decouple from the dynamics of the system as a whole and
experience mass segregation \Cite{GFR}.  For a system with multiple masses,
the analogous condition has been numerically determined to be \Cite{Amaro}
\beq\left(\frac{N_h}{N_l}\right)\left(\frac{m_h}{m_l}\right)^{2.4}<0.16\eeq
in which $N_l$ is now the total number of stars of individual mass $m_l$ and
$N_h$ is the total number mass $m_h>m_l$.  For the Kroupa IMF used here,
the value of the above relation is 13; for the Arches IMF, 209.  Even if the
Kroupa IMF is extended down to $0.2M_\odot$, the value is still 1.4.  So
indeed for all IMF studied here there are sufficient heavy stars so that mass
segregation should at least be possible.

Given the plausibility of mass segregation, one can ask if it would be
expected to foster core collapse.  The timescale for segregation-driven core
collapse has been variously estimated to be $\sim10\%$ of the initial
half-mass relaxation time \Cite{Amaro} or $\sim15\%$ of the initial mean-mass
central relaxation time \Cite{GFR}, both independent of the IMF used. For the
models studied only the extreme model G3C has a resulting expected
segregation-collapse timescale which falls under the 3\ Myr stellar-lifetime
limit, being in the $1-2$\ Myr range for the two conditions given above.
(Figure \ref{g2tr} shows model G3C having a very short relaxation time in
general.) The runs of model G3C all ended before 0.5\ Myr of simulated time
and so would not be expected to exhibit segregation-driven collapse.  In
addition, other models were allowed to run beyond the 3\ Myr limit as a test
but none reached its expected segregation-collapse time  -- the closest being
model E4B with an Arches IMF and no stellar bar, which had an expected
collapse time of $t_c\simeq15$\ Myr and which ended at $t\simeq10$\ Myr
showing no signs of collapse.

\subsection{Critical Density}

Whether or not core collapse is achieved, the central question is whether,
given the $\sim150 M_\odot$ upper mass limit of the IMF, a more-massive object
can be formed which will evolve to be a seed black hole.
Once a critical local density of
$\rho\gtrsim\rho_\mathrm{crit}=5\sn9(\frac{\vrms}{100\mathrm{km/s}})M_\odot/\mathrm{pc}^3$
is achieved, a $\sim100 M_\odot$ star can directly accrete $10^3 M_\odot$ of
material in $\sim5$\ Myr, creating an IMBH \Cite{McZwart}.  As all
models used here have a central velocity dispersion $\vrms\gtrsim100$\ km/s,
Tables \ref{mrjkroupa} and \ref{mrjarches} show that none of the models meet
that requirement.  Model G3A/Kroupa may eventually reach it once deeper
into core collapse than the simulation was able to track: Fig. \ref{g2vms}
shows its $\vrms\simeq150$\ km/s and remaining roughly steadly even as core
collapse starts at $t\simeq0.4$\ Myr in Fig. \ref{g3a4}. Model E4B/Arches reaches
an even higher central density during its core collapse (Fig. \ref{Ae4b0nlm2}), but
it has a central velocity dispersion 
approaching 500\ km/s and still increasing (Fig. \ref{Ae4bvms0}).  However
neither model is expected to represent a typical astronomical system.  As
described in \S\ref{sect_ICs}, G3A has a somewhat  high density for the
galactic spheroids that the $\gamma=0$ spheres best model.  Likewise the Arches
IMF used in model E4B/Arches is determined from stellar clusters, while E4B's
velocity dispersion places it in the range of galaxy cores, not individual
clusters.

Other than accretion, collisional mergers are another avenue for forming
massive objects from smaller stars.  For massive ($125M_\odot$) stars, a local
density of $\rho_{_{125}}\simeq1.2\sn8 M_\odot/\mathrm{pc}^3$ is required in
order for the average star to experience 1 collision per main sequence lifetime
\Cite{Freitag}.  None of the models started with such a high central density,
and only model E4B/Arches with the maximally strong bar achieved it by the end
of the simulation.  While by no means is it necessary for a collision rate of
1 per 125$M_\odot$ star be reached in order for a massive object to form --
such a rate would produce an abundance of 250$M_\odot$ objects, while previous
simulations of young clusters with an upper mass limit of 120$M_\odot$ have
found that more than one very massive star is never formed \Cite{FRB} -- it
does give an indication of the density range required.  As will be seen in the
following discussion of detailed simulation results, for a broad range of
initial models using the Arches IMF (observed in stellar clusters),  250$M_\odot$
objects are calculated to have been formed, while with the Kroupa IMF
(representing the field star population) no such massive objects are created.

\section{Summary of Simulation Results}

The models studied fall into two main classes.  The majority are Plummer-sphere
initial models, which to be consistent with Quinlan and Shapiro \Cite{QS90} are labeled with an
initial ``E''.  Models E2A and E2B can represent either dense globular
clusters or the nuclei of dwarf elliptical or bulgeless spiral galaxies,
and E4B the core of a giant elliptical galaxy.

The other class of model is the ``$\gamma=0$ sphere'', labeled with a ``G'',
which fit the surface brightness of galactic spheroids.  Model G2A has 
identical mass and core radius as model E2A, while models G3A and G3C
have the same mass as model E2B but with smaller core radii.  G3C is an
extreme case which does not likely represent a realistic astronomical system.

For each model two possible initial mass functions were employed: a Kroupa
IMF based on general field star populations, and an Arches-style IMF similar
to that observed in dense clusters specifically.  Note that the Arches IMF
may not be a realistic IMF for either model E4B or the $\gamma=0$ spheres, as
those models best represent galactic centers while the Arches IMF is derived
from observations of individual stellar clusters.  However, the simulations of
those models serve to complete the overall picture of the physical evolution
of stellar systems at galactic centers.  More complete details of the
individual models and IMFs, and how rotation is introduced into them, is given
in \S\ref{sect_ICs}.  The remainder of this section describes the physical
effects exhibited by the various simulations, first for the Plummer models and
then for the $\gamma=0$ spheres.

\subsection{Plummer-Sphere Simulations}

The simulations which started with a Plummer Sphere distribution (the ``E''
models) give a clear picture.  As each model's results show, the
presence of a moderate-strength stellar bar can dramatically raise the rate
of central density increase of a dense stellar cluster, by a factor of 1-2
orders of magnitude.  However, the IMF also plays a crucial role: with the
standard Kroupa IMF the rate of increase of central density $\rho(0)$ remains
roughly linear with time through the main-sequence lifetime of the largest
stars in a cluster, even for the most-dense model studied (E4B) -- as shown in
Fig. \ref{e4b0nlm2} -- whereas with the cluster-specific Arches-style IMF core
collapse can take place near to the 3 Myr stellar lifetime limit
(\eg model E2B with a 5\% bar, shown in Fig. \ref{Ae2bt5}),
or even within it (model E4B with an 8 or 10\% bar, Fig. \ref{Ae4b0nlm2}).

More detailed discussion of the various physical aspects is given below.

\subsubsection{Rotating v. Non-rotating Clusters}

As described in \S\ref{massseg}, none of the Plummer models was expected to
experience mass-segregation-driven core collapse, and none was observed in the
simulations.  For all of models E4B/Kroupa (Fig. \ref{e4b0nlm2}), E4B/Arches
(Fig. \ref{Ae4b0nlm2}) and E2B/Arches (Fig. \ref{Ae2bt5}) both the nonrotating
and rotating cases showed only slow 
growth in the central density by simulation's end, if no stellar bar was
incorporated.  Interestingly, rotation was not a strong impediment to this
essentially secular increase in $\rho(0)$, \eg the E4B/Kroupa cases' density
plots track each
other with the only difference being the lower initial $\rho(0)$ value of the
rotating cluster, while in the E4B/Arches and E2B/Arches models the rotating
case actually has a somewhat larger rate of central density increase.  While
counterintuitive, rotation serving to enhance the rate of collapse has been
observed before in Fokker-Planck simulations \Cite{KYLS}, possibly due to
individual stars being able to interact with a larger (or at least different)
range of neighboring stellar orbits.

\subsubsection{Effects of a Stellar Bar Perturbation}

When a moderate stellar bar -- 5\% to 10\% strength, in line with observed bars
-- was included, however, all models showed rapid increase in central density.
Core collapse was achieved in both E4B/Arches and E2B/Arches, and in
less than the 3 Myr stellar-lifetime limit when the bar strength was $\simeq8\%$
or greater. Figure \ref{Ae2b4} shows the dramatic effect a greater bar strength
can have on reducing the time to core collapse, going from $t\simeq4$ Myr with a $5\%$
bar to slightly over 2 Myr with an 8\% bar.  Comparing Figures \ref{Ae2a0nlm4}
and \ref{Ae2bt5} it appears that bar-driven core collapse would also be expected
in model E2A/Arches had its simulation been able to remain numerically stable
for a longer amount of simulated time.

\subsubsection{Collisional Stellar Mergers: Effect on Cluster Dynamics}

In general stellar mergers had little effect on the overall dynamics in the
Plummer-sphere simulations.  Figures \ref{e4b0nlm2} for E4B/Kroupa and
\ref{Ae2bt5} for E2B/Arches show only a very slight extra increase in $\rho(0)$
when stellar mergers are enabled, as is also born out in comparing values of $S$ and
$S'$ in Tables \ref{mrjkroupa} and \ref{mrjarches}.  Even in the most
extreme Plummer-sphere model studied, E4B/Arches with a 10\% stellar bar,
the main effect was that core collapse was achieved slightly earlier with
stellar mergers, as shown in Fig. \ref{Ae4b0nlm2}.  That figure does hint that
as core collapse progresses, collisional mergers may begin to drive the
evolution of the central regions.  However, this computational model cannot
track into that regime, and it remains that the core collapse that is seen
here is consistently bar-driven.

\subsubsection{Collisional Stellar Mergers: Massive Star Formation}

When the buildup of massive ($250M_\odot$) stars via collisional stellar mergers
is considered, the accelerated central density increase brought about by the bar
perturbation is insufficient to produce even a single massive star within 3 Myr
using the Kroupa IMF, as seen in Table \ref{mrjkroupa}.  But the Arches IMF is
already sufficiently populated with high-mass (up to $125M_\odot$) stars to allow
for the collisional production of massive $250M_\odot$ stars, and the accelerated
collapse induced by the bar only increases the rate of production, as seen in the
$G_{250}(t_1)$ column in Table \ref{mrjarches}.  So as far as the formation of
massive stars is concerned it is the IMF which determines whether or not
$250M_\odot$ or larger objects are created at all.  The overall rate of formation
due to collisional mergers is primarily set by the initial conditions of the
cluster, \ie its size and density, as can be seen by comparing the $G_{250}(0)$
values for models E2A, E2B and E4B in Table \ref{mrjarches}.  However,
the table also shows that the presence and strength of any stellar bar can
have a large effect on how much the rate of massive star formation $G_{250}$
increases in the period before core collapse occurs.

Complementing the above discussion of the calculated rate of massive object
(250$M_\odot$) formation, one can look at the numbers for the largest stars
in the IMFs used.  The physical upper limit was 150$M_\odot$, which was
implemented numerically as a mass bin centered on and associated with stars of
mass 125$M_\odot$.  In the Kroupa IMF models, no simulation yielded an
increase in the number of $125M_\odot$ stars due to collisional mergers,
which is consistent with a null rate for $250M_\odot$ object formation.
As detailed in Chapter \ref{chap_results}, the Arches-IMF models ranged from
creating 15 stars of mass 125$M_\odot$ stars in model E2A with a 5.3\% bar,
to 500 stars of mass 125$M_\odot$ stars in model E4B with a 10\% bar.  Noting
the short amount of simulation time for which model E2A ran, these numbers are
commensurate with the calculated rates of 250$M_\odot$ object formation.

\subsubsection{Collisional Stellar Mergers: Possibility of Runaway Merging}

The question of how massive-star production due to collisional mergers might
be self-reinforcing is more difficult to answer.  It would be expected that as
more low-mass stars merge and so produce more intermediate-mass products, then
more intermediate-mass merger candidates would be available to again
collide and produce even more-massive stars.  As discussed in \S\ref{E2Barches}
model E2B/Arches was allowed to run until time $t=3$ Myr without stellar
collisions enabled, and the rate of massive-star production due to mergers was
at that point
calculated to be within 3\% of the value found when mergers had been allowed
{\sl ab initio} (and both were $\gtrsim25\%$ larger than the rate at $t=0$).
Thus the amplification of the rate of massive star production due to the
collisional merging of smaller stars is not significant when compared to the
increase in the merger rate which results from the overall central density
increase afforded by the stellar bar, at least in the pre-core-collapse period.

\subsection{``$\gamma=0$ Sphere'' Simulations}

The $\gamma=0$ sphere models have similar density profiles to their Plummer
sphere counterparts, although they are somewhat less centrally concentrated.
A major distinction is that the $\gamma=0$ spheres have a much shorter
initial relaxation time, as seen by comparing Figures \ref{g2tr} and
\ref{Ae4btr0}.  This results in a very different physical evolution, as
described next.

\subsubsection{Rotating v. Non-rotating Clusters, Effect of Bar Perturbations}

Contrasting the situation for the Plummer-sphere clusters, simulations of
initial $\gamma=0$ spheres showed no significant differences in evolution
between nonrotating systems, rotating systems without a stellar bar, and
rotating systems with a bar.  Figures \ref{g2a6} for model G2A/Kroupa 
and \ref{Ag2a4} for G2A/Arches show this with respect to the central density
$\rho(0)$.  In model G2A/Kroupa the with-bar case exhibits a somewhat more
rapid increase in $\rho(0$), but it is extremely slight compared to the effect
the bar has in the Plummer sphere simulations.  In Fig. \ref{Ag2a4} model
G2A/Arches shows even less difference, although it ran for a very short
simulated time.  

Even though the simulations of $\gamma=0$ sphere models 
were not able to track the system as far into their evolution as some of the
Plummer-sphere simulations were able to do, Figures \ref{g2a6} (for G2A/Kroupa)
and \ref{g3c2} (for G3C/Kroupa)
show a trend for the $\gamma=0$ sphere models with no stellar bar
to evolve almost identically to the same models with a bar, in stark contrast
to how the Plummer sphere models behaved.  Thus for the $\gamma=0$ sphere
models, overall system relaxation dominates over any bar- or collision-driven
collapse, and again rotation does not obviously support the system against an
increase in the central density.

\subsubsection{Collisional Stellar Mergers}

Stellar mergers played a similar role in $\gamma=0$ sphere simulations as 
they did for the Plummer-sphere models.
Figures \ref{g2a6} and \ref{g3a4} show that the central density $\rho(0)$, in
cases in which stellar mergers are enabled, exactly tracks $\rho(0)$ in the
corresponding no-mergers simulations for models G2A and G3A with a Kroupa IMF.
Mergers have only a slight effect on $\rho(0)$ for model G2A/Arches (Fig.
\ref{Ag2a4}).  Only in the extreme model G3C of Fig. \ref{g3c2} do stellar
mergers affect the evolution, and then only by producing a slightly more rapid
start to core collapse.

In terms of numbers of high-mass stars formed through collisions, the $\gamma=0$
situation is slightly different, in that for both models G3A/Kroupa and
G3C/Kroupa the calculated number of 125$M_\odot$ stars formed via collisions of
lower-mass stars was between 0.5 and 1.  Given the short simulation time of
each model, a few additional 125$M_\odot$ would be expected to be created
within the 3\ Myr limit.  While not interesting astrophysically, this
contrasts with the situation for Plummer spheres in which no Kroupa IMF
simulation produced any 125$M_\odot$ stars.  However, Table \ref{mrjkroupa}
still shows that no 250$M_\odot$ objects at all would be created through
mergers even in these cases -- and Table \ref{mrjarches} shows that only a
small number are formed even in model G2A with a more top-heavy Arches IMF.

What remains unclear is whether model G3A/Kroupa will reach a regime in
which massive star formation becomes plausible: as seen in Fig. \ref{g3a4} it
does achieve the beginnings of core collapse, unlike any Plummer model with
a Kroupa IMF.  It does so in a short amount of simulation time, leaving
$\sim2.5$ Myr for post-core collapse evolution to occur before the stellar
population begins to evolve off the main sequence.  Model G3C/Kroupa achieves
core collapse even more rapidly, but it is doubtful that it represents a
realistic astronomical system.

\subsection{Summary}

Before performing the simulations it was expected that rotational support
against collapse would be countered by the outward transport of angular
momentum afforded by a stellar bar, allowing core collapse to proceed even
in a rotating system.  While the effect of the bar was observed in all Plummer
sphere simulations it did not result in the anticipated core collapse in all
models, nor in an ensuing runaway of collisional mergers creating a massive
object.  For the $\gamma=0$ spheres the overall system relaxation dominated over
the effect of the bar.

In the end the initial mass function turned out to be the
determining factor: with the possible exception of model G3A no simulation
performed using a Kroupa IMF, representative of a general galactic stellar
population, reached core collapse, and none obtained a large enough stellar
merger rate to create a massive object.  In contrast, all models which
employed the Arches-style IMF, which is specific to dense stellar clusters,
started with a merger rate already sufficient to produce massive
250$M_\odot$ objects -- although the presence of a stellar bar did make the
difference in whether or not core collapse was reached, as well as in producing
a large increase in the massive-object formation rate.

The above description is fairly robust:
\begin{itemize}
\item with a Kroupa IMF only the most extreme models (G3A and G3C) reached core collapse;
\item with an Arches IMF all models reached core collapse (or for E2A, showed indications of it);
\item no $\gamma=0$ sphere model was dominated by the effect of a bar;
\item even with an Arches IMF, a bar was required in order for any Plummer model
to reach core collapse; and
\item runaway mergers were not observed in any simulation, although a large
increase in high-mass merger rates followed from core collapse in barred models.
\end{itemize}
Of note is that a cluster's evolution did not change qualitatively even when
the merger rate was artificially increased by a factor of 5, as described in
\S\ref{upfactor} -- so it does not appear that a collisional-merger runaway
is likely to occur for any pre-core-collapse system of main sequence stars.


Thus the results indicate two possible paths for formation of a massive object
within a dense stellar system, one clear and one somewhat tentative.  For
cluster-sized systems similar to models E2A or E2B -- including dwarf
elliptical galaxy nuclei or bulgeless spiral galaxies -- with an Arches-style
IMF, the expected degree of rotation is sufficient to support a stellar bar
perturbation which in turn leads to core collapse through the outward transport
of angular momentum.  However, even without core collapse, through stellar
mergers the system will produce massive objects in less than the main-sequence
stellar lifetime sufficient to seed supermassive black hole growth.  For larger
systems, \eg galactic spheroids similar to model G3A, core collapse can be
obtained even with a Kroupa IMF, although massive object formation through
stellar mergers is not expected to be achieved within main-sequence lifetimes
and so will require interactions of collapsed relativistic objects, which
was outside the scope of this study.

\section{Future Observations, Future Work}

A basic assumption in this study is that stars preceded quasar black holes in
the early universe.  Two future space missions currently being considered by
the European Space Agency in its {\it Cosmic Vision 2015-2025} program will go
a long way towards addressing this assumption by observing the earliest luminous
objects whose surrounding clouds of dense gas and dust obscure them from
current instruments. First, the Far Infrared Interferometer (FIRI) will perform
high-resolution imaging spectroscopy to resolve the creation of the first
luminous objects in the universe, separating out the formation of stars and
the growth of black holes \Cite{FIRI}.  Complementing FIRI, the ESA's XEUS
X-ray observatory will detect the earliest quasars as they form.  Together
these two missions, or similar ones, could answer the question of whether
stars or quasar black holes formed first \Cite{FIRM}.  They or similar
observations may also address the main question remaining about the current
study's results: whether or not an Arches-style IMF, with its weighting
towards higher stellar masses, is in fact a better representation of the
actual IMF that existed in early-universe stellar clusters than the Kroupa
IMF is. 
If so, then the main result of these simulations -- that individual nuclear
stellar clusters are likely to form seed objects for massive black holes --
holds.  If not, then one must resort to post-core-collapse evolution to obtain the
seed objects, and likely in galactic spheroids instead of in nuclear clusters.



As with all simulations, the results presented here have some limitations
that could be addressed in future work.  
\begin{itemize}
\item \textbf{Finite stellar lifetimes.}  A conceptually straightforward but
technically nontrivial extension would be to track the system's dynamical
behavior as the more-massive stars move off the main sequence, which would
involve the appropriate moving of stars from their present distribution functions
to new ones representing white dwarfs, neutron stars and stellar-mass black holes
-- as represented by terms $B_q$ and $R_q$ in the general Fokker-Planck
equation of \ref{fullfp}, but not yet implemented in the simulations.
Corresponding different collisional cross sections would also be required, as has
been done by Quinlan and Shapiro in modeling one-dimensional, nonrotating
Plummer sphere systems (\Cite{QS90}, \Cite{QS89}).
\item \textbf{Rotation-induced flattening.}  As described in \S\ref{sect_gravpot},
the provision for a non-spherical gravitational potential is
already implemented in the simulation code, but it is not yet debugged and
shown to be numerically stable; once that is achieved it may be possible to
add another degree of realism to the simulations of rotating systems.  For
globular clusters at least, flattening of the system has been found to
correlate with rotation \Cite{FSK}, and to increase merger rates somewhat
\Cite{Arabadjis}, which could strengthen the results presented here.
\item \textbf{Dark Matter.}  No provision has been made for there to be a
background component in the gravitational potential, \eg due to the presense
of collionless dark matter.  This is consistent with all previous Fokker-Planck
studies of stellar systems -- in recent works Fiestas \etal (\Cite{FSK}),
Kim \etal (\Cite{KYLS}, \Cite{KLS}) and Takahashi (\Cite{TakaI}, \Cite{TakaIII})
all ignore dark matter in their globular cluster simulations, while Arabadjis
\Cite{Arabadjis} and Quinlan \& Shapiro (\Cite{QS89}, \Cite{QS90}) look to the
collisionally-produced massive objects in their simulations as \textit{being}
dark matter.  Amaro-Seoane \Cite{Amaro} points out that the current state of
astrophysical modeling ``comes to its limits'' in embedding simulated systems
in collisionless dark halos.  Still, it is conceivable that introducing an
\adhoc offset to the overall gravitational potential, or an additional
collisionless component to the system's matter-density profile, could serve
as a toy model of the effect of a smooth dark matter background.
\end{itemize}

The early evolution of quasars and host galaxies is not simple -- we have
only observed the final $\simeq30\%$ of quasar evolution, but even that is
enough to indicate galactic spheroids and massive black holes do not grow
together \Cite{FIRM}.   
The results presented here show that even when rotation is accounted for,
dense clusters of main-sequence stars in early galaxy cores can provide a
possible mechanism for forming the seed objects that eventually become 
quasar black holes.

%% file: Appendix.tex
\appendix
\chapter{Non-Spherical Gravitational Potentials}
\renewcommand{\theequation}{A.\arabic{equation}}
\input{Homeoidal}
\input{Nonspherical}
\input{symbols}

%% file: Homeoidal.tex
\section{Potential Calculations in Homeoidal Coordinates}
\label{subsect_nonsph}

For the more general, but rarely needed case in which there is an overall
and variable ellipticity $e(a)$ to the potential, a somewhat more complex
procedure than that given in \S\ref{sect_gravpot} is required.  One still
starts with the inverted Poisson equation
as given in (\ref{cohn24}) but now the derivation of the inverted Laplacian
$\mathcal{L}^{-1}$ is more involved, as developed below.

To describe the non-spherical potential $\Phi$ when the isodensity
surfaces of $\rho$ are ellipsoids, homeoidal coordinates are called for
\Cite{BT}.  Consider the potential $d\Phi$ of a thin ellipsoid of mass
$dM$, semimajor axis $a$ and ellipticity $e$ at homeoidal radius $u_o$.
Defining $\Delta\equiv ae$, the 
homeoidal radius $u$ is related to the standard cylindrical coordinates by
\beq\frac{R^2}{\cosh^2u}+\frac{z^2}{\sinh^2u}=\Delta^2\label{homeo}.\eeq
(There is also an angle-like coordinate $v$ which does not come into play.)
Then the potential $d\Phi(u)$ at $u$ due to the ellipsoid at $u_o$ is
\beq\label{bt2-67}d\Phi(u)=-\frac{G\,dM}{\Delta}\left\{
   \begin{array}{ll}\sin^{-1}e,&\mbox{$u<u_o$}\\
            \sin^{-1}(\sech u),&\mbox{$u\geq u_o$}\end{array}\right.\eeq
where the mass element $dM=\rho dV$ for the ellipsoidal shell\footnote{Here it
is implictly assumed that the ellipticity $e$ is a slowly-enough varying
function of radial coordinate $a$ so as to not affect the differential
$dV$; this approximation is examined in the Appendix.} is
\beq\label{rhodv}\rho dV=4\pi\rho a^2(1-e^2)^{1/2}\,da\eeq
and the total potential is found by integrating over mass shells at all
values of $u_o$ as will now be shown.
Let us define $d\Phi_e\equiv-\frac{G\,dM}{\Delta}\sin^{-1}e$ and
$d\Phi_u\equiv-\frac{G\,dM}{\Delta}\sin^{-1}(\sech u).$
Having no explicit dependence on $u$, the integration of $d\Phi_e$ is
straightforward: \beq\Phi_e(u)=-4\pi 
G\int_{a(u)}^\infty(1-e^2)^{1/2}\rho\,\frac{\sin^{-1}e}{ae}\,da.\eeq
For the contribution to $\Phi(u)$ due to shells with $u_o<u$ we need a
relation for $u(a)$.  Noting that in the $z=0$ plane, $R=a$ by definition,
we accomplish this by solving (\ref{homeo}) for $u$; the resulting quadratic
equation in $\sech^2u$ yields
\newcommand{\rzd}{(R^2+z^2+\Delta^2)}\beq\label{sech2u}\sech^2u=
\frac{1}{2R^2}\left[\rzd\mp\left(\rzd^2-4R^2\Delta^2\right)^{-1/2}\right]\eeq
where the necessity for $\sech^2u\leq1$ implies that only the upper sign is
allowed (\eg taking $R=z$ makes this clear).  Then the contribution to the
total potential due to mass shells interior to $u$ is obtained by
integrating (\ref{bt2-67}) using (\ref{rhodv}):
\beq\label{phiu}\Phi_u(u)=-4\pi
G\int_0^{a(u)}(1-e^2)^{1/2}\rho\,\frac{\sin^{-1}(\sech u)}{ae}\,da.\eeq
Equation (\ref{sech2u}) may be expanded, yielding approximations necessary for
use when $R\rightarrow0$.  (When $\Delta\rightarrow0$ the simpler
spherically-symmetric procedure can be used.)  The total gravitational
potential at an arbitrary value point $u(R,z)$ is thus given by
\beq\Phi^\new(u)=\Phi_u+\Phi_e.\label{phitot}\eeq
Note that with use of (\ref{sech2u}) in
(\ref{phiu}), the only reference to the homeoidal coordinate $u$ in the
calculation of $\Phi(u)$ is in locating the transition from $\Phi_u$ to
$\Phi_e$, despite $u_o$ being the (implicit) variable of integration.
Note also that the coordinate transformation of (\ref{homeo})
is referenced to the integration variable $u_o$, and not to the location of
measurement $u$; thus each homeoidal transform is defined by the mass shell
responsible for the element of $d\Phi_u$ currently contributing to
$\Phi_u=\int d\Phi_u$.  Hence $a=a(u_o)$ and $e=e(u_o)$; neither $a$ nor
$e$ has an explicit dependence on $u$.  Likewise, the $\Delta$ in (\ref{sech2u}) is $\Delta=a(u_o)e(u_o)$ even
though the coordinate being transformed is $u$ and not $u_o$, and so
(\ref{homeo}) constitutes a ``running coordinate transformation'' which
is continuously evolving across the range of integration of (\ref{phiu}).

%% file: Nonspherical.tex
\section
{Laplacian Solution in Ellipsoidal Coordinates: Numerical Aspects}\label{sect_num}

A novel aspect of this study is that $e$ is allowed to vary over the range
of $a$.  This is different from the fixed-ellipiticy methods (\eg as described
by Tremaine and Weinberg \Cite{TW}) but was found to be both advantageous in that it allows
potentially more accurate modeling, and necessary to avoid a pitfall of the
fixed-ellipiticy schemes: physically interesting interactions will primarily
occur in high-density regions, \ie the core of the distribution, but these tend
not to be the high-ellipticity regions which are usually far from the center of
the cluster -- and so taking a single value of $e$ for the entire cluster can
artifically increase the effects of non-sphericity.  This was found to be the
case in practice, and was what motivated the allowance for a variable $e$.
(Provision for fixing $e$ at a predetermined value is also built in however.)

One may worry that a changing $e(a)$ may induce troublesome shell-crossing.
Numerically, this is only a concern if the $a^2$ grid is so fine and $e(a)$
changes so rapidly that the semiminor axes $b=a(1-e)^{1/2}$ of adjacent shells
on the grid cross (\ie if $a_{r+1}(1-e_{r+1})^{1/2}<a_r(1-e_r)^{1/2}$ for
gridpoints $a^2_r$ and $a^2_{r+1}>a^2_r$).  In practice the variability of
$e(a)$ is not large enough for this to occur, given the courseness of the
$a^2$ grids.  And in principle, possible shell-crossing is not a fundamental
concern anyway, as the ellipsoidal isodensity surfaces are merely smoothed
representations of the underlying discrete stellar distribution.

To see how much effect the variability of $e(a)$ has on the entire
$\rho$---$\Phi$ scheme, consider what effect it has on the mass element $dM$,
which was not taken into account in (\ref{rhodv}).  This is not a strict calculation of what effect variable
$e$ has on the calculus, but it does give an idea of the size of its effect:
\beq dM=\rho\,dV=\rho\,d\left[\frac{4\pi}{3}a^3(1-e^2)^{1/2}\right]=4\pi\rho
a^2(1-e^2)^{1/2}\left[1-\frac{ae}{3(1-e^2)^{1/2}}\frac{de}{da}\right]da\eeq
which shows that the effect of the dependence of $e$ on $a$
is of $O(e^2)$ smaller than the dominant term.
Using the more straightforward form of (\ref{phiu}) greatly simplifies the
numerical calculation without explicitly affecting the results, as the
$e(a)$ run is still allowed to converge on a consistent set of values along
with $\rho(a)$ and $\Phi(a)$ during the iterative solution for all of these at
each new timestep, with resulting new $f$.

Thus the assumption of ellipsoidal isodensity surfaces, along with $f_n$,
determines the potential $\Phi$.  The required integrals described above,
however, are too computationally expensive to be performed each time the
knowledge of potential is needed in the Fokker-Planck coefficient
calculation.  After testing interpolation schemes and analytic
approximations, only the Clutton-Brock self-consistent field (``SCF'') method
as described by Hernquist and Ostriker \Cite{HernJPE} proved adequate in both accuracy and speed.
For the SCF ``core radius'' $\rcore$ the value at which
$\Phi(\rcore)=\Phi(0)/\sqrt2$ is used.  This is also 
the core radius of the Plummer potential with central potential $\Phi(0)$.
The field method is tested at each timestep, and if sufficient accuracy
cannot be achieved with a reasonable number of expansion terms, the code
falls back on direct integration.  For interpolating over the other,
one-dimensional grids of quantities discussed in Chapter \ref{chap_pot}, simple
polynomial or cubic spline schemes are employed.

%% file: symbols.tex
\chapter{Tables of Symbols}
\begin{table}
\begin{tabular}{@{\hspace{0pt}}l@{\hspace{5pt}}l@{\hspace{5pt}}l@{\hspace{5pt}}l}
& {\bf Quantity} & {\bf Symbol} & {\bf Example} \\ \hline
\S\ref{sect_dyn}
& radial component                  & $r$ or $1$              & $v_r$, $I_1$\\
& tangential component              & $T$ or $2$              & $v_T$, $I_2$\\
& azimuthal compoment               & $z$ or $3$              & $J_z$, $w_3$\\
& orbital endpoints                 & $p$ and $a$         & $r_p$, $\tr_a$\\
& dummy component variable & $i$ or $j=1$, $2$, or $3$  & $\Omega_j$, $\IIJ$\\
\S\ref{subsect_orbfreqs} 
& ``circular'' value or component   & $c$                     & $r_c$\\
\S\ref{subsect_I}
&maximum dynamically-allowed value&max&$J_\mathrm{max}$, $\beta_\mathrm{max}$\\
\S\ref{subsect_rhoI}
& velocity-space coordinates        & $v$        & $v$, $\theta_v$, $\phi_v$\\
& a specific mass ``bin''  & $q$; later also $q'$ \& $\tM$ & $m_q$, $f_{q'}$\\
\S\ref{orbinc}
& total value of a dynamical quantity & tot & $J_\mathrm{tot}$\\
& minimum dynamically-allowed value & min         & $J_\mathrm{min}$ \\
\S\ref{subsect_vdisp}
& average over a given quantity $x$,\\&~~~~possibly with a weighting function
                             &$\left<\cdot\right>_x$&see text for definitions\\
\S\ref{sect_ICs}
& core (radius)                     & core       & $\rcore$ \\
\S\ref{subsect_introrot} & amount of a dynamical quantity \\
&~~~ due to rotation             & rot & $J_\mathrm{rot}$, $T_\mathrm{rot}$\\
\S\ref{subsect_pert}
& spherical harmonic indices & $l$, $m$ & $Y_{lm}$\\
& resonance indices & $knm$; later $\ell_{1}\ell_{2}\ell_{3}$&$\Psi_{knm}$\\
& an individual object (whether\\
&~~~~a single star or the stellar bar) & $*$ & $\Phi_*$\\
\S\ref{sect_fieldpert}
& the stellar bar   & $B$ & $f\barb$, $\Omega\barb$\\
\S\ref{coeffs_coeffs}
& repeated summation index & [any repeated index] &
$\ell_p\Omega_p\equiv\sum_p\ell_p\Omega_p$\\

\end{tabular}
\caption{List of frequently-used subscripts.}

\end{table}

\begin{table}
\begin{tabular}
{@{\hspace{0pt}}l@{\hspace{5pt}}l@{\hspace{5pt}}l}
& {\bf Quantity} & {\bf Symbol}\\ \hline
\S\ref{sect_units}& Newton's constant of gravitation& $G$\\
\S\ref{sect_dyn} &radial component of orbital action & $I_{1}$ or $I$; later also $\ti$\\
& tangential component of orbital action & $I_{2}$ or $J$ or $J$; later also $\tj$\\
& vertical component of orbital action & $J_{z}=J\cos\beta$\\
& orbital energy per unit mass & $E$ \\
& orbital position coordinates & $r$ , $\theta$, $\phi$\\
& radial \& tangential orbital frequencies & $\Omega_{1}$, $\Omega_2$\\
& gravitational potential per unit mass & $\Phi$\\
& canonical orbital angle coordinates & $w_{1}$, $w_{2}$, $w_{3}$; later also $\tw_1$ \etc\\
& orbital inclination (Euler angle) & $\beta$\\
& orbital azimuth (Euler angle) & $\psi$\\
& orbital radial endpoints & $r_p$, $r_a$; later also $\tr_p$, $\tr_a$\\
\S\ref{subsect_endpts}
& radial component of orbital velocity &$v_r\equiv\sqrt{\vrsq}$\\
\S\ref{subsect_orbfreqs}
& ``circular'' orbital radius &$r_c$ where $\partial v_r(r_c)/\partial r=0$\\
\S\ref{subsect_r}
& individual stellar mass & $m$ \\
\S\ref{subsect_rhoI}
& tangential component of orbital velocity & $v_T=J/r$ \\
\S\ref{sect_vrot}
& average rotational velocity & $\vrot(r)$ \\
\S\ref{orbinc}
&orbital polar angle &$\alpha\equiv\halfpi-\beta$\\
\S\ref{subsect_grid}
&''ellipsoidal'' radial coordinate & $a(r,e(r))$\\
& cylindrical radial and vertical coordinates & $R$, $z$\\
\S\ref{subsect_potdens}
& one-dimensional (nonrotating)  \\
&~~~~distribution function for stars of mass $m_q$ & $\evenf(E)$\\
\S\ref{subsect_introrot}
&total cluster gravitational potential energy & $W_\mathrm{grav}$\\
&total cluster rotational angular momentum & $J_\mathrm{rot}$\\
&cosmological rotation parameter&
 $\lambda=\frac{J_\mathrm{rot}|W_\mathrm{grav}|^\frac{1}{2}}{GM^{2.5}}$\\
\end{tabular}
\caption{Symbols first introduced in Chapter \ref{chap_pot}, listed by section
in which each was first introduced.   These denote properties of individual
stars in the system, not bulk properties.
Bulk properties based on summing over the individual stars' values are denoted
by subscripts, for example total energy of the stellar system $E_\mathrm{tot}$
or the rotational contribution to the total angular momentum $J_\mathrm{rot}$.
Only symbols that are used in more than one section are included.}
\end{table}

\begin{table}
\begin{tabular}
{@{\hspace{0pt}}l@{\hspace{5pt}}l@{\hspace{5pt}}l}
& {\bf Quantity} & {\bf Symbol}\\ \hline
\S\ref{sect_dyn}& gravitational potential per unit mass & $\Phi$\\
\S\ref{subsect_r}
& stellar mass density & $\rho$ \\
& total cluster mass & $M$ \\
& stellar distribution function in position space & $\ftd(\mathbf{r},\mathbf{v})$\\
& stellar distribution function \\
&~~~~in energy/angular momentum space & $\ftd(E,J)$\\
\ref{subsect_I}
& orbit-averaged stellar distribution function & $f(I,J)$\\
\ref{orbinc}
&stellar distribution function parameterized \\
&~~~~in terms of orbital inclination&
   $f_\beta(I,J;\beta)\equiv h(\alpha(\beta))f(I,J)$\\
&even ($g$) and odd ($\Theta$) components\\
&~~~~of inclination parametization &
   $h(\alpha)\equiv g(|\alpha|)+\Theta(\alpha)$\\
& total angular momentum of all stars of  energy $E$ & $J_\mathrm{tot}(E)$\\
\S\ref{subsect_grid}
&ellipticity of stellar mass distribution&$e(r)$\\
\S\ref{subsect_vdisp}
& one-dimensional \textit{``rms''} stellar velocity dispersion&  $\vrms$\\
& anisotropy parameter & $\delta$\\
& generalized stellar velocity dispersion& $\sigma_o\equiv\sqrt{(1-\delta)\vrms^2}$\\
\end{tabular}
\caption{Symbols first introduced in Chapter \ref{chap_pot},
listed by section in which each was first introduced.
These are bulk properties of the system or of a subpopulation thereof.
Only symbols that are used in more than one section are included.}
\end{table}

\begin{table}
\begin{tabular}{@{\hspace{0pt}}l@{\hspace{2pt}}l@{\hspace{2pt}}l}
& {\bf Quantity} & {\bf Symbol}\\ \hline
\S\ref{sect_oafp}& Fokker-Planck drift coefficient &$\J(I,J)$\\
& Fokker-Planck diffusion coefficient& $\IIJ(I,J)$\\
& \adhoc merger loss and gain terms\\
&~~~~in F-P equation & $L_q(I,J)$, $G_q(I,J)$\\
\S\ref{subsect_pert}
& gravitational potential of individual perturber \\
&~~~~(whether a single star or the stellar bar) & $\Phi_*$\\
& spherical harmonics & $Y_{lm}(\theta,\phi)$\\
& radial coefficients\\
&~~~~ of perturbation potential expansion & $\Phi_{lm}(r)$\\
& coefficients of perturbation potential\\
&~~~~ expansion in action space & $\Psi_{knm}(I,J)$\\
& Slater rotation coefficient & $V_{lnm}(\beta)$\\
& resonance strength coefficient & $W_{klnm}(I,J)$\\
& resonance indices & $(k,n,m)$; later $(\ell_{1},\ell_{2},\ell_{3})$\\
\S\ref{coeffs_orbinc}
& any quantity $y(\beta)$, weighted-averaged\\
&~~~~over orbital inclination&
  $\avgb{y}=\frac{\pi}{\beta_\mathrm{max}}\int d\beta\,h(\beta)y(\beta)$\\
& any quantity $y(\beta)x(\theta)$, weighted-averaged\\
&~~~~over polar angle $\theta(\psi,\beta)$&
  $\kaplan{y}{x}=\frac{2}{\beta_\mathrm{max}}\int d\beta\,h(\beta)y(\beta)
     \,d\psi\,x(\theta)$\\
& average-squared perturbation\\&~~~~ coefficient strength &
  $\overline{\Psi^2}_{knm}\equiv\avgb{\Psi_{knm}(\beta)^{2}}$\\
\S\ref{coeffs_coeffs}
& perturbation "turning on" parameter & $\eta\gtrsim0$\\
& (complex) perturbation orbital frequency & $\omega=m\Omega_*+i\eta$\\
\S\ref{sect_e4b}
& Relative density increase of highest-mass stars &
    $S=[\rho_{_Q}(0)/\rho(0)]_{t_1}/[\rho_{_Q}(0)/\rho(0)]_{t_0}$\\
& {\sl As above, but including effect of stellar mergers} & $S'={\ldots} $ \\
& Rate of massive ($250M_\odot$) star creation & $G_{250}$\\
\end{tabular}
\caption{Symbols first introduced in Chapters \ref{chap_coeffs},
\ref{chap_tests} and \ref{chap_results}, listed by section in which each was
first introduced.  Only symbols that are used in more than one section are included.}
\end{table}
